%% file: M51_Schirm.tex
\documentclass[usenatbib]{mn2e}

\usepackage{graphicx}
\usepackage{epstopdf}
\usepackage{amssymb}
\usepackage[draft]{hyperref}
\usepackage[toc,page]{appendix}
\usepackage{tablefootnote}
\usepackage[version=3]{mhchem}

\newcommand{\unit}[1]{\ensuremath{\, \mathrm{#1}}}
\newcommand{\mol}[1]{\ensuremath{\mathrm{#1}}}

\newcommand{\molH}{\ensuremath{\mathrm{H_2}}}
\newcommand{\CO}{$\mol{CO}$}
\newcommand{\CI}{$\mol{[CI]}$}

\newcommand{\Tkin}{$T_{\mol{kin}}$}
\newcommand{\nH}{$n(\molH)$}
\newcommand{\NCO}{$N_{\mol{CO}}$}
\newcommand{\NCI}{$N_{\mol{[CI]}}$}
\newcommand{\FF}{$\Phi_A$}

\newcommand*{\measComp}{_all_CI_13CO}
\newcommand*{\oneComp}{_all7down_CI}
\newcommand*{\coldComp}{_7downjcut4_CIcold_n9}
\newcommand*{\warmComp}{_7downjcut4_CIcold_n9}

\newcommand{\apjs}{ApJS}
\newcommand{\aap}{A\&A}
\newcommand{\apjl}{ApJL}

\begin{document}
\title[Herschel SPIRE-FTS M51]{Probing the cold and warm molecular gas in the Whirlpool Galaxy: \emph{Herschel\footnote{Herschel is an ESA space observatory with science instruments provided by European-led Principal Investigator consortia and with important participation from NASA}} SPIRE-FTS Observations of the central region of M51 (NGC 5194)}
\author[M.R.P Schirm et al.]{M.R.P. Schirm$^1$,  C.D. Wilson$^1$,  J. Kamenetzky$^{2,3}$, T.J. Parkin$^1$, J. Glenn$^4$, \newauthor P. Maloney$^4$, N. Rangwala$^5$, L. Spinoglio$^6$,  M. Baes$^7$, A. Boselli$^8$, A. Cooray$^9$, \newauthor I. De Looze$^{7, 10}$, J. A. Fern\'andez-Ontiveros$^6$, O. \L. Karczewski$^{11}$, R. Wu$^{12}$ 
\\ 
$^1$Department of Physics and Astronomy, McMaster University, Hamilton, ON L8S 4M1 Canada \\
$^2$Steward Observatory, University of Arizona, 933 North Cherry
Avenue, Tucson AZ 85721, USA \\
$^3$Westminster College, 1840 S 1300 E
                Salt Lake City, UT 84105, USA  \\
$^4$Center for Astrophysics and Space Astronomy, 389-UCB, University
                of Colorado, Boulder, CO, 80303, USA \\
$^5$Space Science and Astrobiology Division, NASA
                Ames Research Center, Moffet Field, CA 94035, USA \\
$^6$Istituto di Astrofisica e Planetologia Spaziali, INAF-IAPS, Via Fosso del Cavaliere 100, I-00133 Roma, Italy \\
$^7$Sterrenkundig Observatorium, Universiteit Gent, Krijgslaan 281 S9, B-9000 Gent, Belgium \\
$^8$Aix Marseille Universit\'e, CNRS, LAM (Laboratoire d'Astrophysique de Marseille), UMR 7326, F-13388, Marseille, France \\
$^9$Department of Physics \& Astronomy, University of California, Irvine, CA 92697, USA\\
$^{10}$Department of Physics and Astronomy, University College London, Gower Street, London, WC1E 6BT, UK \\
$^{11}$Laboratoire AIM, CEA/DSM--CNRS--Universit\'{e} Paris Diderot, IRFU/Service d'Astrophysique, CEA Saclay, 91191 Gif-sur-Yvette, France. \\
$^{12}$JSPS International Research Fellow, Department of Astronomy, the University of Tokyo, Bunkyo-ku, Tokyo 113-0033, Japan \\
}

\maketitle

\begin{abstract}

We present Herschel SPIRE-FTS intermediate-sampled mapping
observations of the central $\sim 8 \unit{kpc}$ ($\sim 150''$) of M51,
with a spatial resolution of $40''$. We detect 4 $^{12}$\CO
\ transitions ($J=4-3$ to $J=7-6$) and the \CI \ \ce{^3P2 - ^3P1} and
\ce{^3P1 - ^3P0} transitions. %in the nucleus and centre regions. 
We
supplement these observations with ground based observations of
$^{12}$\CO \ $J=1-0$ to $J=3-2$ and perform a two-component non-LTE
analysis. 
We find that the molecular gas in
the nucleus and centre regions has a
cool component ($T_{kin} \sim 10 - 20 \unit{K}$) with a moderate but
poorly constrained  density
($n(\mol{H_2}) \sim 10^{3} - 10^{6} \unit{cm^{-3}}$), as well as 
significant 
molecular gas in a warmer ($T_{kin} \sim 300 - 3000 \unit{K}$),
lower density ($n(\mol{H_2}) \sim 10^{1.6} - 10^{2.5}
\unit{cm^{-3}}$) component.  
We compare our \CO \ line ratios and calculated densities along with
ratios of \CO \ to total infrared luminosity to a grid of photon
dominated region (PDR) models and find that the cold molecular gas
likely resides in PDRs with a field strength of $G_0 \sim
10^{2}$. The warm component likely requires an additional source
of mechanical heating, 
% most likely 
from supernovae and stellar winds or possibly shocks produced in the
strong spiral density wave. 
When compared to similar two-component models of other
star-forming galaxies published as part of the Very Nearby Galaxies
Survey  (Arp 220, M82 and NGC 4038/39), M51 has the lowest density
for the warm component, while having a warm gas mass fraction that
is comparable to those of Arp 220 and M82, and significantly higher
than that of NGC 4038/39.

\end{abstract}

\begin{keywords}

galaxies: individual(NGC 5194), galaxies: ISM, ISM: molecules

\end{keywords}

\input{Introduction.tex}

\input{Observations.tex}

\input{RadTrans.tex}

\input{PDR.tex}

\input{Discussion.tex}

\input{Conclusions.tex}

\section*{Acknowledgments}

We thank the referee  and the editor for constructive comments that improved
the paper. This research was supported by grants from the Canadian Space Agency and the Natural Sciences and Engineering Research Council of Canada (PI: C. D. Wilson). PACS has been developed by a consortium of institutes led by MPE (Germany) and including UVIE (Austria); KU Leuven, CSL, IMEC (Belgium); CEA, LAM (France); MPIA (Germany); INAF-IFSI/OAA/OAP/OAT, LENS, SISSA (Italy); IAC
(Spain). This development has been supported by the funding agencies BMVIT (Austria), ESA-PRODEX (Belgium), CEA/CNES (France), DLR (Germany), ASI/INAF (Italy) and CICYT/MCYT (Spain). SPIRE has been developed by a consortium of institutes led by Cardiff University (UK) and including Univ. Lethbridge (Canada); NAOC (China); CEA, LAM (France); IFSI, Univ. Padua (Italy); IAC (Spain); Stockholm Observatory (Sweden); Imperial College London, RAL, UCL-MSSL, UKATC, Univ. Sussex (UK); and Caltech, JPL, NHSC, Univ. Colorado (USA). This development has been supported by national funding agencies: CSA
(Canada); NAOC (China); CEA, CNES, CNRS (France); ASI (Italy); MCINN
(Spain); SNSB (Sweden); STFC (UK); and NASA (USA). HIPE is a joint
development by the Herschel Science Ground Segment Consortium,
consisting of ESA, the NASA Herschel Science Center, and the HIFI,
PACS and SPIRE consortia. This research has made use of the NASA/IPAC
Extragalactic Database (NED) which is operated by the Jet Propulsion
Laboratory, California Institute of Technology, under contract with
the National Aeronautics and Space Administration. This research made use of the python plotting package matplotlib \citep{hunter2007}. This research made use of APLpy, an open-source plotting package for Python hosted at http://aplpy.github.com. 
We would like to thank Mark Wolfire for providing the PDR model grids used in this paper. This work made use of HERACLES, `The HERA CO-Line Extragalactic SurveyÕ \citep{leroy2009}. This research was supported in part by the Grant-in-Aid for Scientific Research for the Japan Society of Promotion of Science (140500000638). IDL gratefully acknowledges the support of the Flemish Fund for Scientific Research (FWO Vlaanderen).

\bibliographystyle{mn2e}
\bibliography{M51_Schirm}

\newpage 

\input{appendix.tex}

\end{document}

%% file: Introduction.tex
\section{Introduction}

M51 (NGC 5194) is a well-studied, relatively normal, nearby spiral
galaxy.  Its recent interaction with the nearby lenticular galaxy NGC
5195 has led to triggered star formation throughout the galaxy
\citep{nikola2001}, and this  interaction may be responsible for the prominent spiral arms of M51 \citep{zartisky1993, dobbs2010}. \cite{rose1982} first suggested the presence of a non-stellar nuclear source of radiation at the centre of M51: it has been classified as a Seyfert type 2 galaxy without a hidden broad-line region detected in polarized light \citep{tran2001} and as a Low Ionization Nuclear Emission Region (LINER) galaxy (e.g., \citealt{satyapal2004}).  The Seyfert-2 activity may have been triggered as a result of the interaction \citep{kouloridis2014}.

 %These latter report mid-infrared spectroscopy of M51, among a sample of LINER, with the detection of the high ionisation line of [OIV]26µm. However the [OIV] luminosity of M51, as compared to its total IR luminosity, is fully consistent with the emission of a sample of ÒpureÓ starburst galaxies \citep{bernard-salas2009}. Therefore the presence of the low luminosity AGN in M51 is probably not responsible for the excitation of [OIV].

M51 is an excellent source in which to study both cold and warm molecular gas, due to its nearly face-on orientation, the prominence of its spiral arms, and its recent interaction with NGC 5195. Observations of \molH  \ rotational lines have found that the ratio of warm  ($T = 100 - 300 \unit{K}$) to hot ($T=400-1000 \unit{K}$) molecular gas varies across the system, which may suggest a varying excitation mechanism \citep{brunner2008}. \cite{roussel2007} found that the \molH \ is generally excited in photon dominated regions (PDRs, also known as photodissociation regions). Recently, \cite{parkin2013} modelled PDRs in M51 using various transitions of $\mol{[OI]}$, $\mol{[CII]}$ and $\mol{[NII]}$, along with the total infrared luminosity, and found that the far-ultraviolet field strength necessary to reproduce PDRs in M51 varies between $G_0 \sim 10^{1.5} - 10^{4.0}$, with the highest values occurring in the nucleus. 

The cold molecular gas has been studied predominantly through
observations of $^{12}$\CO \ (hereafter \CO),
and its isotopologues $^{13}$\CO \ and $\mol{C^{18}O}$. M51 has been
observed using ground-based single-dish telescopes in \CO \ $J=1-0$
\citep{scoville1983, garciaburillo1993,nakai1994, kramer2005,
  koda2009}, $J=2-1$ \citep{garciaburillo1993, kramer2005, israel2006,
  schuster2007, leroy2009}, $J=3-2 $ \citep{israel2006, vlahakis2013}
and $J=4-3$ \citep{israel2006}, and in $^{13}$\CO \ $J=1-0$
\citep{kramer2005}, $J=2-1$ \citep{kramer2005, israel2006} and $J=3-2$
\citep{israel2006}.  In addition, \cite{israel2006} presented
observations of \CI \ in the $^3P_1 - ^3P_0$ (hereafter $J=1-0$)
transition at $492 \unit{GHz}$, which has also been proposed as a molecular gas tracer (e.g. see \citealt{papadopoulos2004} and \citealt{offner2014}). 

Higher-resolution interferometric observations of \CO \ $J=2-1$,
$^{13}$\CO \ $J=1-0$ and $\mol{^{12}C^{18}O}$  $J=1-0$ in M51 have
been performed using the Owens Valley Radio Observatory (OVRO) by
\cite{schinnerer2010}. These observations were limited to two regions
within the spiral arms of M51.  Using non-local thermodynamic
equilibrium (non-LTE) excitation models  with an escape probability
formalism, they found that the temperature of the molecular gas in the
observed giant molecular clouds (GMCs) is $T_{kin} \sim 20
\unit{K}$. This temperature is similar to  clouds in the Milky Way when observed at the same resolution ($\sim 180\unit{pc}$).

More recently, M51 was observed at arcsecond resolution in \CO
\ $J=1-0$ and $^{13}$\CO \ $J=1-0$ as part of the Plateau de Bure
Interferometer (PdBI) Arcsecond Whirlpool Survey (PAWS,
\citealt{schinnerer2013, pety2013, hughes2013, meidt2013, hughes2013b,
  colombo2014, colombo2014b}). These observations were corrected for
short-spacings using single dish observations. \cite{colombo2014}
detect 1507 objects in \CO \ $J=1-0$, and find that the mass
distribution, brightness and velocity dispersion of GMCs vary across
the different environments in M51.  Some of
these differences  seem to be dynamically driven \citep{meidt2013}. 
\cite{pety2013} detect extended \CO \ $J=1-0$ emission that resides in a thick molecular
disk with a scale height $\sim 200 \unit{pc}$, and accounts for $\sim
50$ percent of the total \CO \ $J=1-0$ emission. 
\cite{pety2013} suggest that this thick, extended disk could be the result of galactic fountains or chimneys due to the ongoing star formation. 

In this paper, we present observations of M51 using the Spectral and
Photometric Imaging Receiver (SPIRE; \citealt{griffin2010}) Fourier
Transform Spectrometer (FTS; \citealt{naylor2010}) on board the ESA
Herschel Space Observatory (\emph{Herschel}; \citealt{pilbratt2010}).
The SPIRE-FTS is a low spatial and spectral resolution imaging
spectrometer covering a spectral range from $194 \unit{\mu m}$ to $671
\unit{\mu m}$ ($\sim 450 \unit{GHz} - 1545 \unit{GHz}$). At the
redshift of M51 ($z \sim 0.002$), this spectral range includes a total
of 10 \CO \ transitions ($J=4-3$ to $J=13-12$), 10 $^{13}$\CO
\ transitions ($J=5-4$ to $J=14-13$) and 2 \CI \ transitions (\ce{^3P1
  - ^3P0} and \ce{^3P2 - ^3P1}, hereafter $J=1-0$ and $J=2-1$,
respectively) all of which trace molecular gas. The SPIRE-FTS
$\mol{[NII]} 205\unit{\mu m}$ data were previously published by
\cite{parkin2013}; in this paper we present the detected \CO \ and \CI
\ transitions for the first time. We adopt a distance  to M51 of $9.9 \pm 0.7
\unit{Mpc}$ \citep{tikhonov2009}, based on observations of the tip of
the red giant branch.

These observations were performed as part of the Very Nearby Galaxies
Survey (VNGS; PI: C.D. Wilson) whose primary goal is to study the
interstellar medium (ISM) of very nearby galaxies using both
SPIRE and the Photoconductor Array Camera and Spectrometer (PACS;
\citealt{poglitsch2010}). From the sample of 13 galaxies in the VNGS,
SPIRE-FTS \CO \ data have been published for five: Arp 220
\citep{rangwala2011}, M82 \citep{kamenetzky2012}, NGC 1068
\citep{spinoglio2012}, NGC 4038/39 \citep{schirm2014}, and M83
\citep{wu2014}. We present the observations and data reduction in
Section \ref{obsSect}. In Section \ref{radTransSect}, we present the
non-LTE analysis of our detected \CO \ and \CI
\ transitions, while in Section \ref{pdrSect} we present models of
photon dominated regions (PDRs) in M51. We discuss the implications of
the solutions of our non-LTE models and our PDR models in
Section \ref{discSec}, along with a comparison of the
results for M51 with previously studied galaxies within our sample.

%Throughout the universe, molecular gas and star formation is intrinsically linked: stars form from molecular gas from within the confines of giant molecular clouds (GMCs), and these stars proceed to inject energy back into the surrounding interstellar medium (ISM) which in turn can quench star formation and heat the surrounding gas. Multiple mechanisms exist with which stars can heat the nearby gas, including photon dominated regions (PDRs \citethis), along with supernovas and stellar winds \citethis. In addition to these mechanisms, other heating mechanisms exist which can heat molecular gas, including x-ray dominated regions (XDRs) \citethis, shocks \citethis and turbulent heating \citethis. Which mechanism dominates the heating depends largely on galaxy type: interacting and merging galaxies, for example, are often heated primarily by mechanical heating mechanisms, such as turbulent heating, shocks, supernova and stellar winds \citethis. In more normal spiral galaxies, PDRs are typically the dominant source of heating \citethis. 

%Interactions and mergers between galaxies play a major role in the evolution of galaxies \citethis along with the structure within our universe (e.g. \citealt{springel2005}). 

%Below this we will have rough notes from the PAWS colaboration. We will include a paragraph on the work done by PAWS highlighting the interesting aspects of it

%% file: Observations.tex
\section{Observations} \label{obsSect}

\subsection{FTS Data Reduction} \label{FTSdatared}

M51 was observed using the SPIRE-FTS on OD 438 (July 25th, 2010) in
intermediate sampling mode, with 32 repetitions per jiggle position
(Observation ID 1342201202). The observation is centered at
($13\unit{h}29\unit{m}52.71\unit{s}, +47^{\circ}11'42.60''$), covering
a region roughly $\sim 160'' \times 180''$  with a total integration
time of 17603 seconds ($\sim 5 \unit{hours}$). These data were reduced
with the Herschel Interactive Processing Environment (HIPE) version
11.0 and SPIRE calibration 11.0. We used a modified version of the
standard mapping pipeline \citep{swinyard2014}, with the primary
difference that we skip the map making step, instead saving each
individual jiggle position as a level 1 spectrometer point source (SPS) product. 

The standard mapping pipeline assumes either that the source is a
point source or that the source is fully-extended, filling the entire
beam uniformly. As with many of the sources in the VNGS sample, M51
cannot be characterized as either a point-source or a fully-extended
source relative to the FTS beam. 
%instead, we characterize the source as being semi-extended. 
In addition, the beam size and shape of the SPIRE-FTS varies with
frequency, with the size varying from $\sim 17 ''$ to $\sim 43''$. In
previous works where we had fully Nyquist sampled maps of our sources
(e.g. see \citealt{kamenetzky2012, spinoglio2012, schirm2014}), we
convolved our point-source calibrated integrated intensity maps using
custom convolution kernels. The same technique cannot be used here as
our map is not Nyquist sampled.  Instead, we match the beam size
across the entire spectrum using the recently developed semi-extended
correction tool (SECT) in HIPE version 11.0 
%We discuss the highlights of the tool here while the method used by
%the tool is described in more detail in 
\citep{wu2013}.  

When a source is semi-extended, correcting for the FTS beam requires
correcting for the source-beam coupling at every frequency. The SECT
corrects for the source-beam coupling by assuming that the
distribution of the emitting gas, whether it is $\mol{[CI]}$-
and $\mol{CO}$-emitting molecular gas or $\mol{[NII]}$-emitting
ionized gas, follows the same spatial distribution as the selected reference
image \citep{wu2013}.  It first calculates the source-beam coupling in the form of a
forward coupling efficiency, $\eta_f(\nu, \Omega_{source})$, for each
bolometer at a given jiggle position using derived FTS beam profiles
and the normalized reference map. This source-beam coupling is
frequency-dependent, and so must be calculated at every frequency. It
then multiplies the intensity at each frequency by this factor. The
resulting data cube has an equivalent beam size and shape of a $40''$
Gaussian beam. For our reference image, we opt to use the PACS
$70\unit{\mu m}$ image, which correlates
strongly with star formation \citep{calzetti2010} and also has the
advantage of having a   significantly higher 
  resolution than any of the FTS data (beam
size $\sim 6 ''$). 
%$70\unit{\mu m}$ correlates
%strongly with star formation \citep{calzetti2010}, and 
We expect that
the molecular gas traced by $\mol{CO}$ and $\mol{[CI]}$ correlates
well with the ongoing star formation in a relatively normal galaxy
like M51. We perform this correction on the level 1 SPS products at
each jiggle position using the SECT, with the correction varying
between $\sim 0.5$ and $\sim 1.6$. The correction is dependent upon
the frequency and the location of the bolometer. 

\begin{figure*} 
\includegraphics[width=\linewidth]{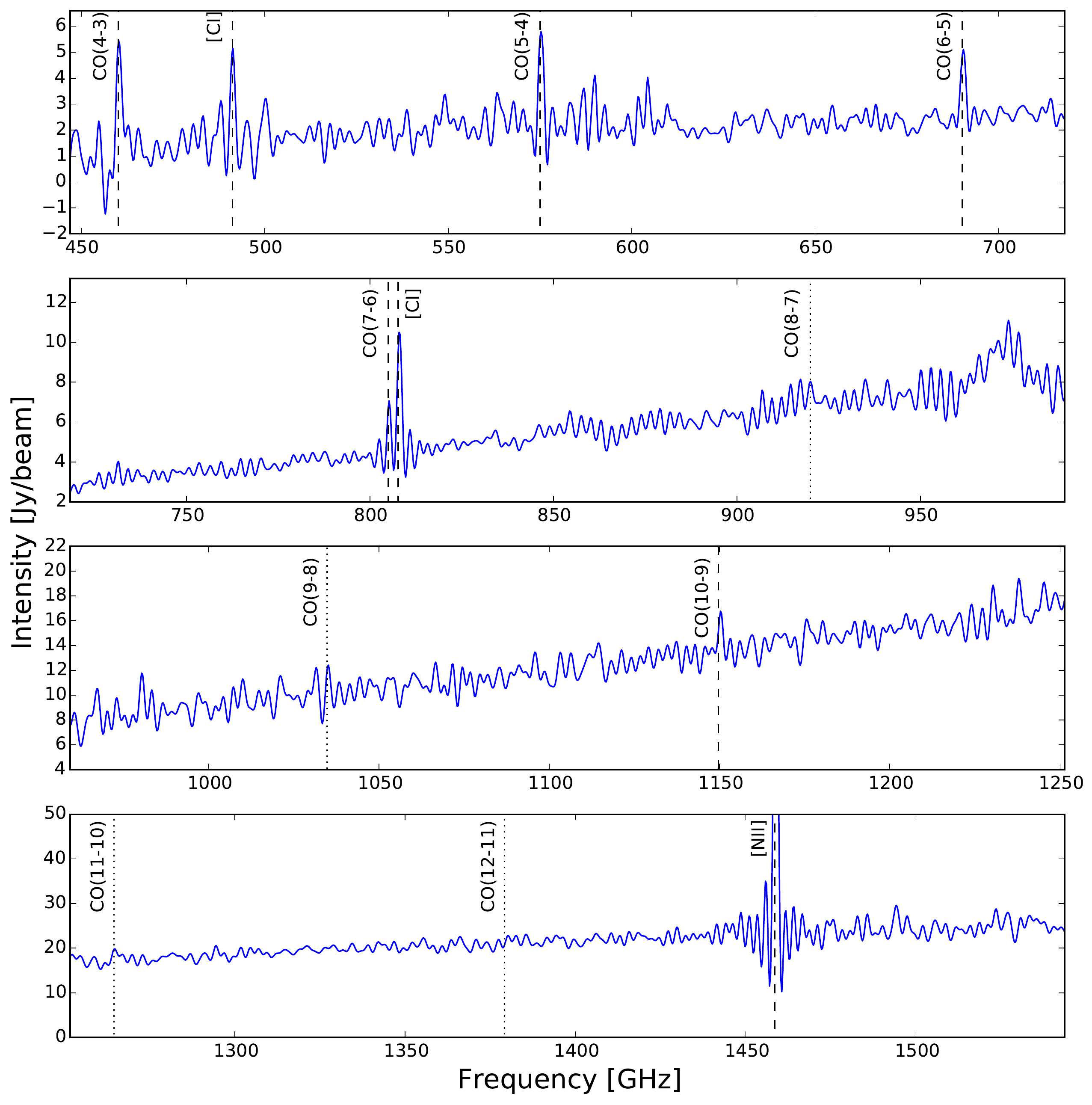} 
\caption[]{FTS spectrum for the nucleus of M51 in units of $\mol{Jy
    \ beam^{-1}}$. All of the detected atomic and molecular
  transitions are indicated by a dashed line, while undetected
  $\mol{CO}$ transitions are indicated by a dotted line at the
  expected location. This spectrum has been corrected for the
  semi-extended nature of the emission (see Section \ref{FTSdatared}
  for details).  The peak of the $\mol{[NII]}$ line is at $\sim 100
  \unit{Jy \ beam^{-1}}$, beyond the scale of the plot. Note that \CO
  \ $J=10-9$ is detected only in the nucleus.}
\label{centerFTS}
\end{figure*}

We create a level 2 data cube for each set of detectors, the SPIRE Long Wavelength Spectrometer Array (SLW) and the SPIRE Short Wavelength Spectrometer Array (SSW), from the semi-corrected level 1 products using the \emph{spireProjection} task. We chose a pixel size of $10''$ for both cubes. It is important to note that the pixel size will have no significant effect on our data cubes, provided we limit the pixel size such that only one detector is assigned to an individual pixel for each of the SLW and SSW level 2 data cubes.  The resulting SLW and SSW data cubes contain 28 and 68 pixels with spectra, respectively.  The complete semi-extended corrected FTS spectrum for the centre of M51 is shown in Figure \ref{centerFTS}.

\subsubsection{Regions} \label{regSect}

\begin{figure}
	\centering
	$\begin{array}{@{\hspace{-0.2in}}c@{\hspace{0in}}c}
		\includegraphics[width = 0.5\linewidth]{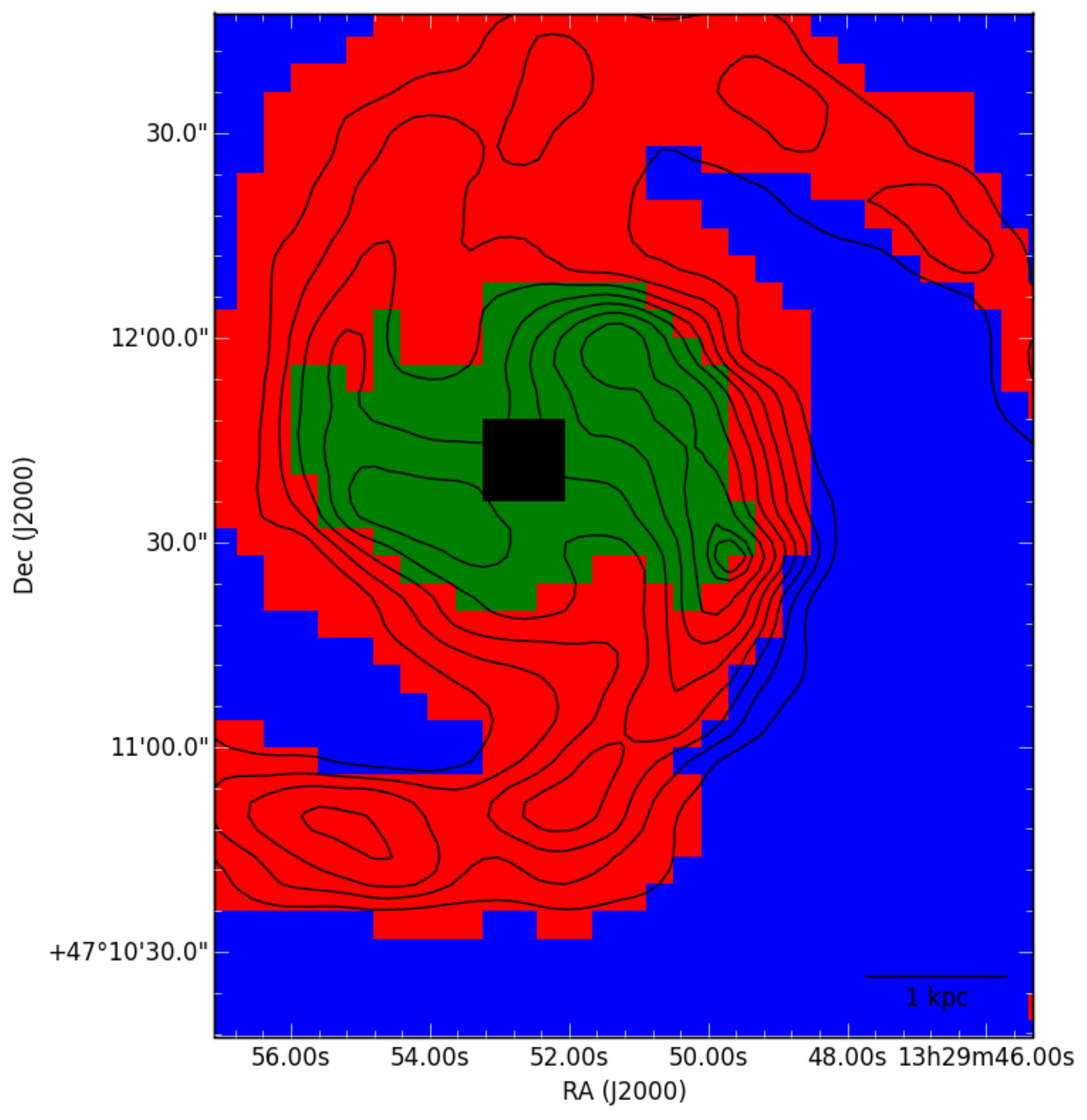} &
		\includegraphics[width = 0.5\linewidth]{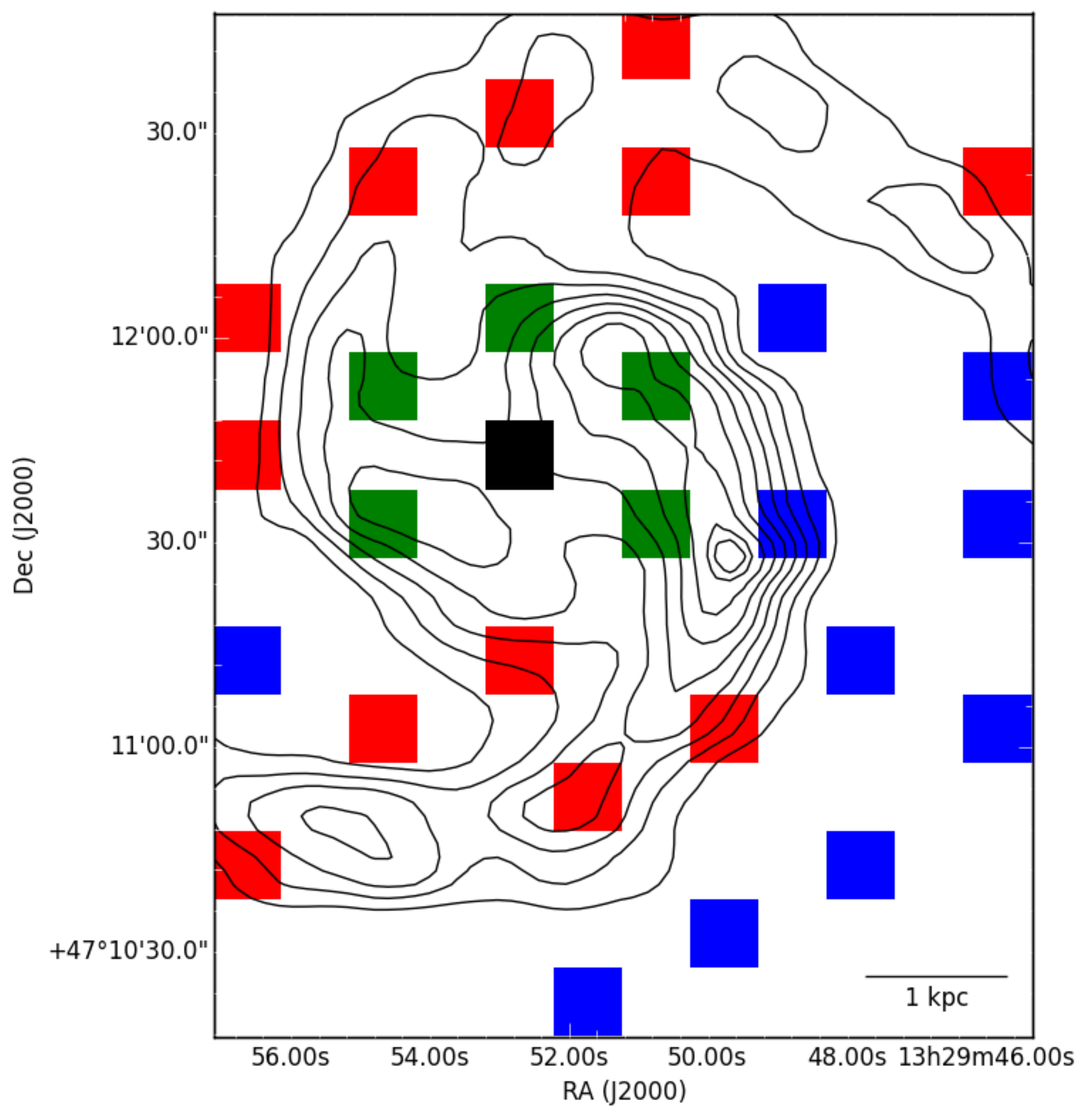}
	\end{array}$
	\caption{\emph{Left:} Schematic of M51 as defined in figure 6
          of \protect\cite{parkin2013} with \CO \ $J=2-1$ contours
          overlaid. \emph{Right:} The same regions of M51  defined for
          the pixels of our FTS maps with the same \CO \ $J=2-1$ contours overlaid. In both figures, the colors correspond to the nucleus (black), centre (green), arm (red) and inter-arm(blue) regions of M51. }
	\label{regionsMap}
\end{figure}

One of the aims of the VNGS is to investigate any regional variations
in the interstellar medium of the galaxies which we resolve.  
%Some of
%the galaxies in our sample that have been observed with the FTS are
%not resolved (e.g. Arp 220, \citealt{rangwala2011}). 
In the case of
M51, we resolve the central $\sim 2'$ at a beam size of $40''$. We
investigate the regional variations in the physical state and heating
mechanisms  of the molecular gas by assigning each pixel in our FTS
map into one of four regions (see Figure \ref{regionsMap}): the
nucleus, centre, arm and inter-arm. We use the same region definitions
  as \cite{parkin2013} to facilitate comparison with their results.

\subsubsection{Line Fitting}

We wrote a custom line fitting routine in HIPE to fit all of the
detected atomic and molecular transitions; 
%in every pixel of our cube. 
a list of detected transitions is shown in Table
\ref{lineFlux}. The intrinsic line profile of the FTS is a Sinc
function with a fixed full-width half-maximum of $1.4305 \unit{GHz}$
in high resolution mode, which corresponds to $280 - 450 \unit{km
  s^{-1}}$ for the SSW, and $440 - 970 \unit{km \ s^{-1}}$ for the SLW
in velocity space\footnote{SPIRE Handbook. Available at
  \url{http://herschel.esac.esa.int/Docs/SPIRE/html/spire_om.html}. Accessed
  July 14th, 2015}. The maximum measured line width from observations
from the James Clerk Maxwell Telescope (JCMT) of $\mol{CO}$ $J=3-2$ convolved to a beam size of $40''$ (see Section \ref{groundCO}) is only $\sim 50 \unit{km \ s^{-1}}$ in the FTS field of view, less than the intrinsic line width of the instrument.  Therefore, we do not resolve the line width in our observations.

\begin{figure*} % FTS Maps
	\centering
	$\begin{array}{@{\hspace{-0.2in}}c@{\hspace{0in}}c@{\hspace{0in}}c@{\hspace{0in}}c}
		\includegraphics[width=0.37\linewidth]{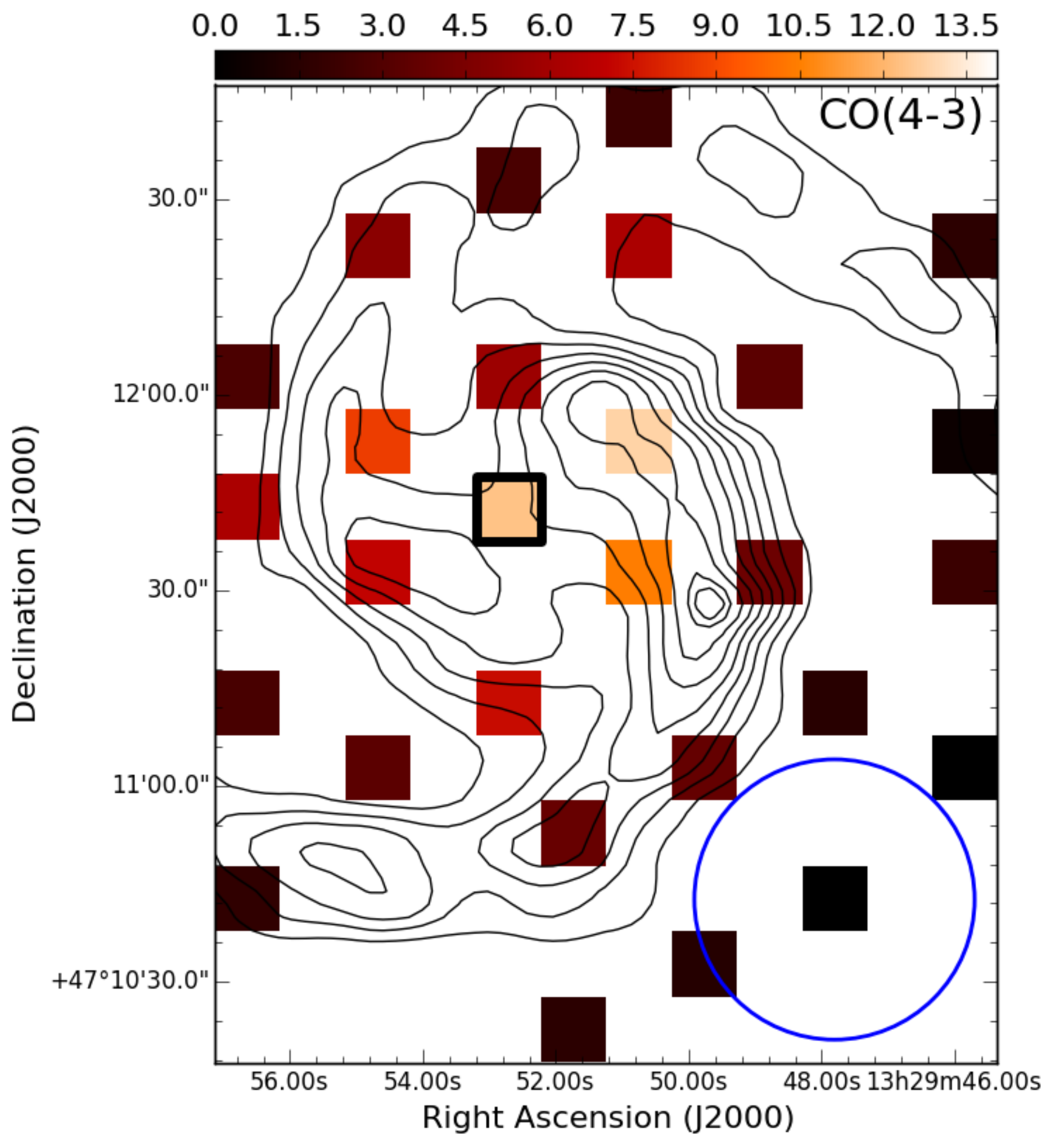} 
		\includegraphics[width=0.37\linewidth]{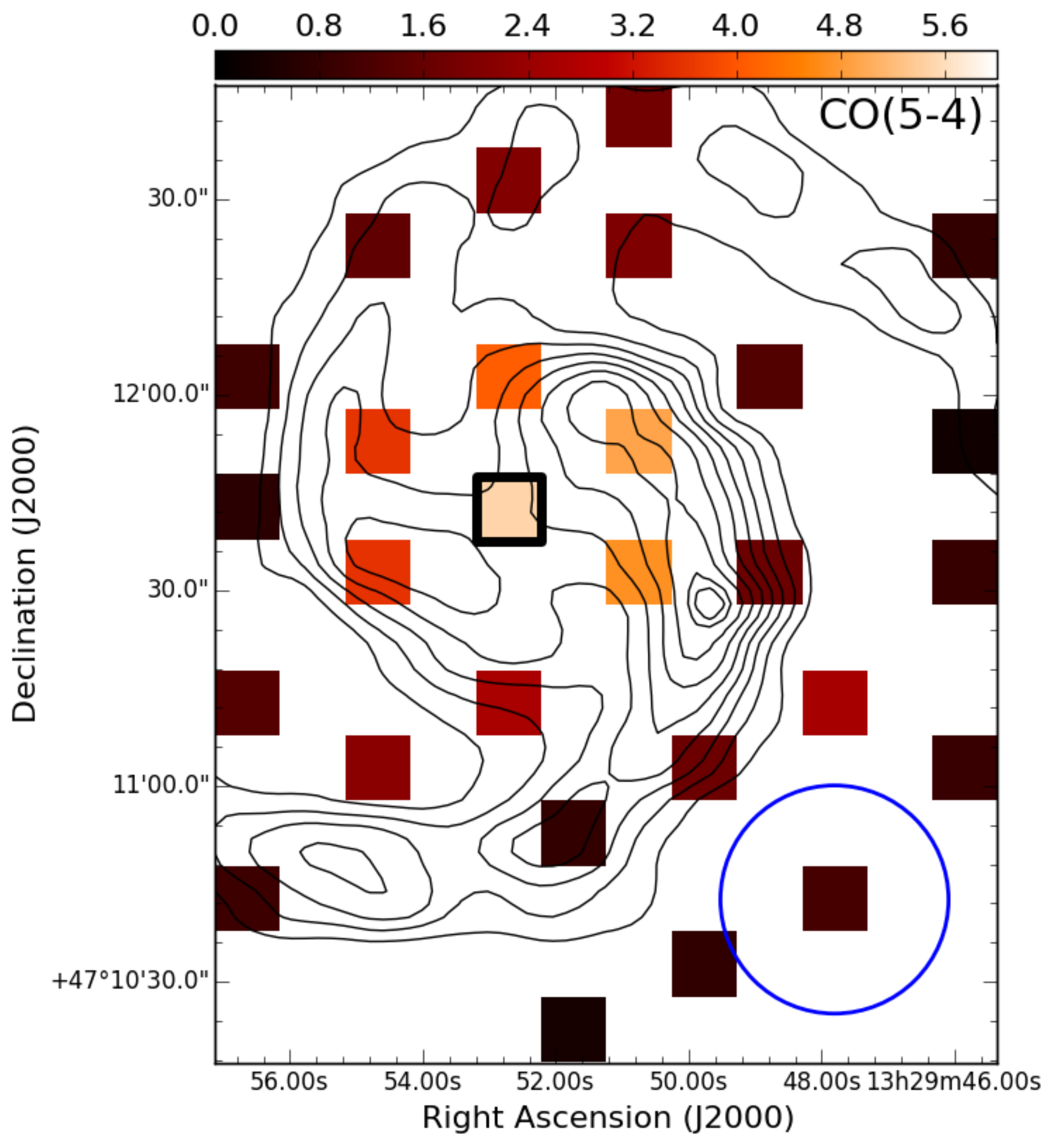} 
		\\
		\includegraphics[width=0.37\linewidth]{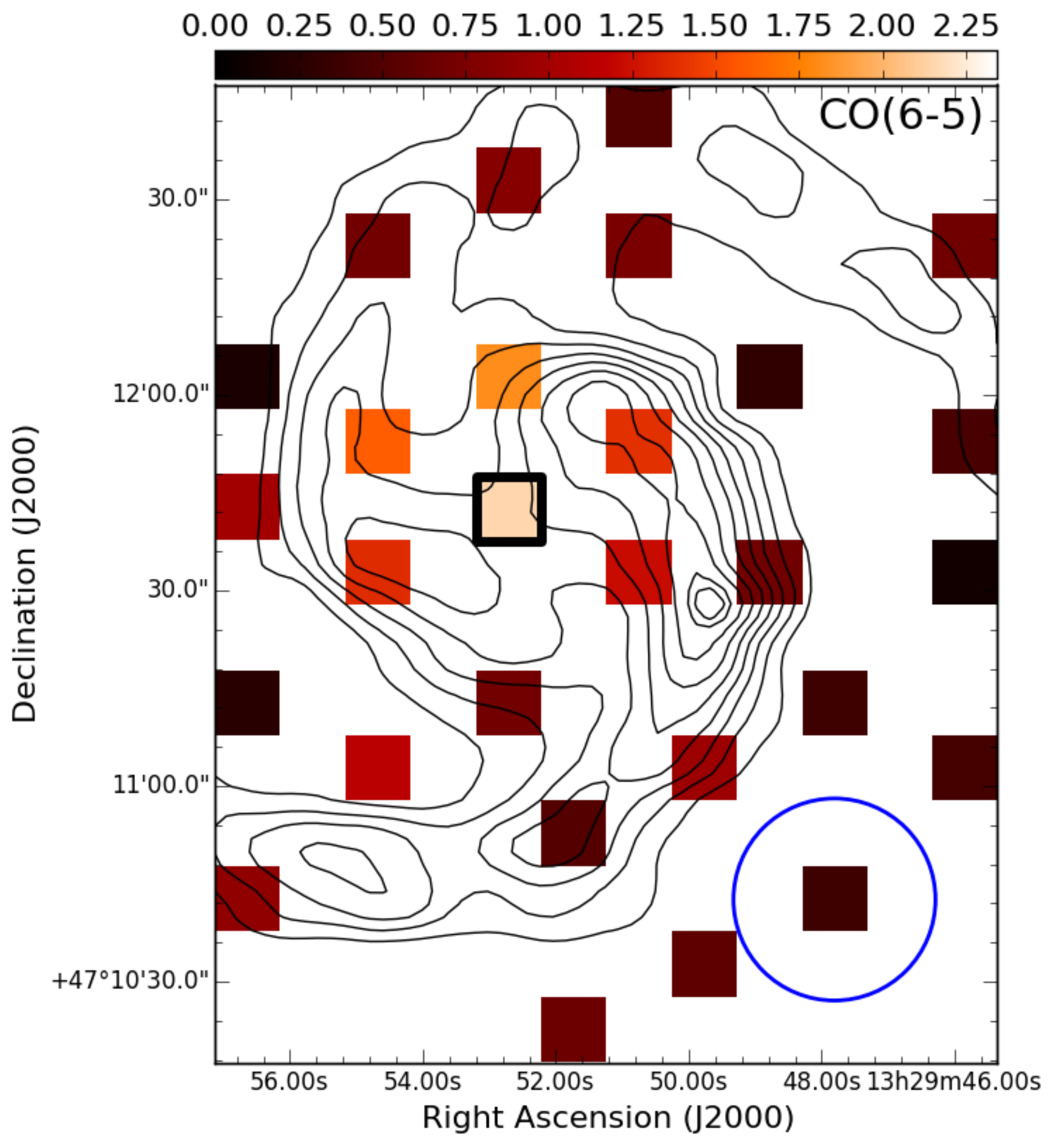} 
		\includegraphics[width=0.37\linewidth]{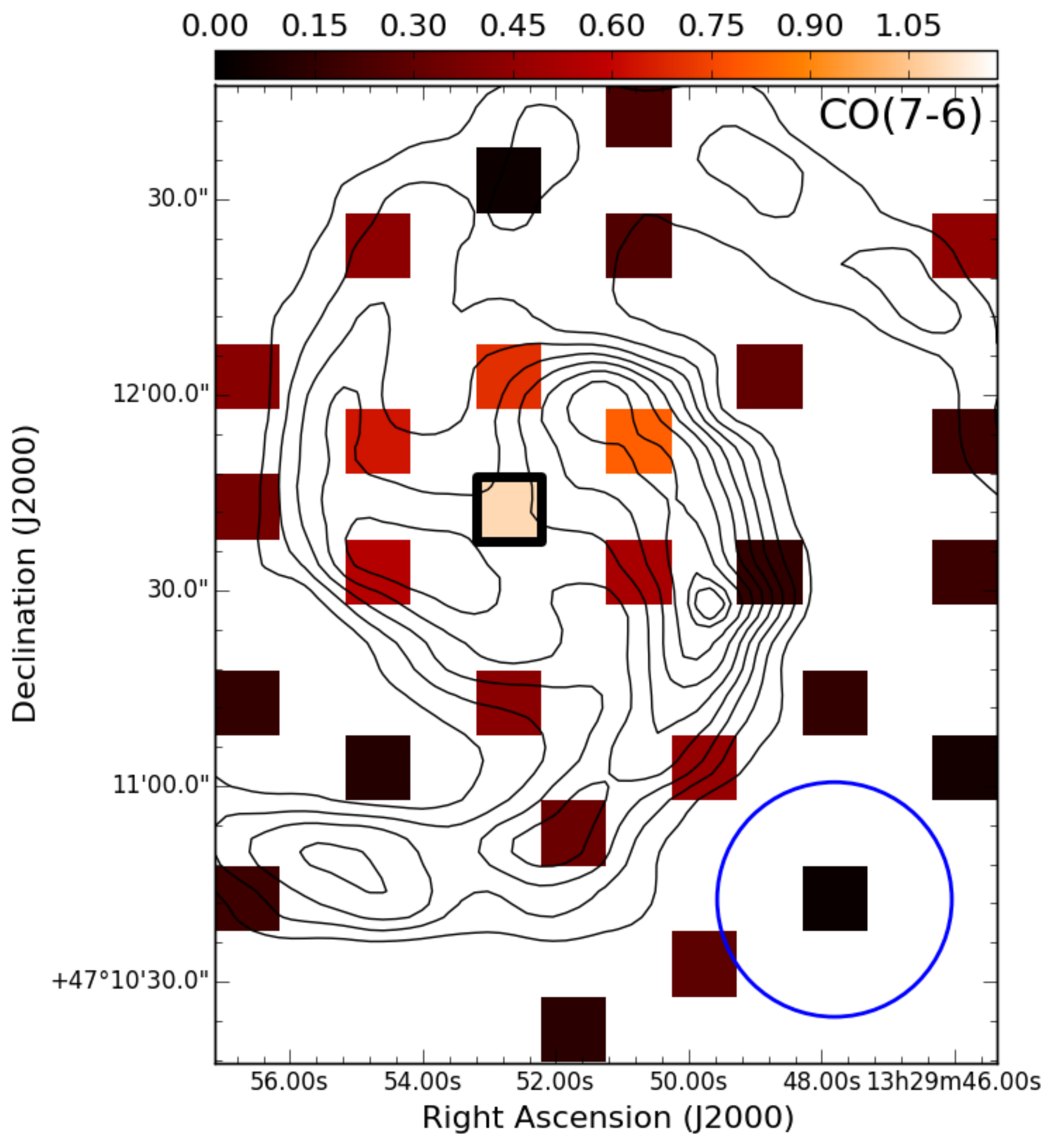} 
	\end{array}$
\caption[]{\CO \ integrated intensity maps in units of $\mol{K
    \ km \ s^{-1}}$ from the Herschel SPIRE-FTS for the $J=4-3$ (top
  left), $J=5-4$ (top right), and $J=6-5$ (bottom left), $J=7-6$
  (bottom right) transitions. The \CO \ $J=2-1$ contours from the IRAM
  $30\unit{m}$ telescope at a beam size of $13''$ are overlaid
  \citep{leroy2009}.  The \CO \ integrated intensity  maps shown
  here have been corrected for the semi-extended nature of the source
  (see Section \ref{FTSdatared}). The native FTS beam size at the
  observed frequencies is shown as a blue circle in the bottom-right
  corner for each of the \CO \ maps, while the centre of M51 is
  denoted by a black box. The semi-extended corrected maps have
  a $40''$ gaussian beam, which is approximately the size of the \CO \ $J=7-6$ beam.}
\label{COFTSImg}
\end{figure*}

\begin{figure*} % FTS Maps
	\centering
	$\begin{array}{@{\hspace{-0.2in}}c@{\hspace{0in}}c@{\hspace{0in}}c}
		\includegraphics[width=0.37\linewidth]{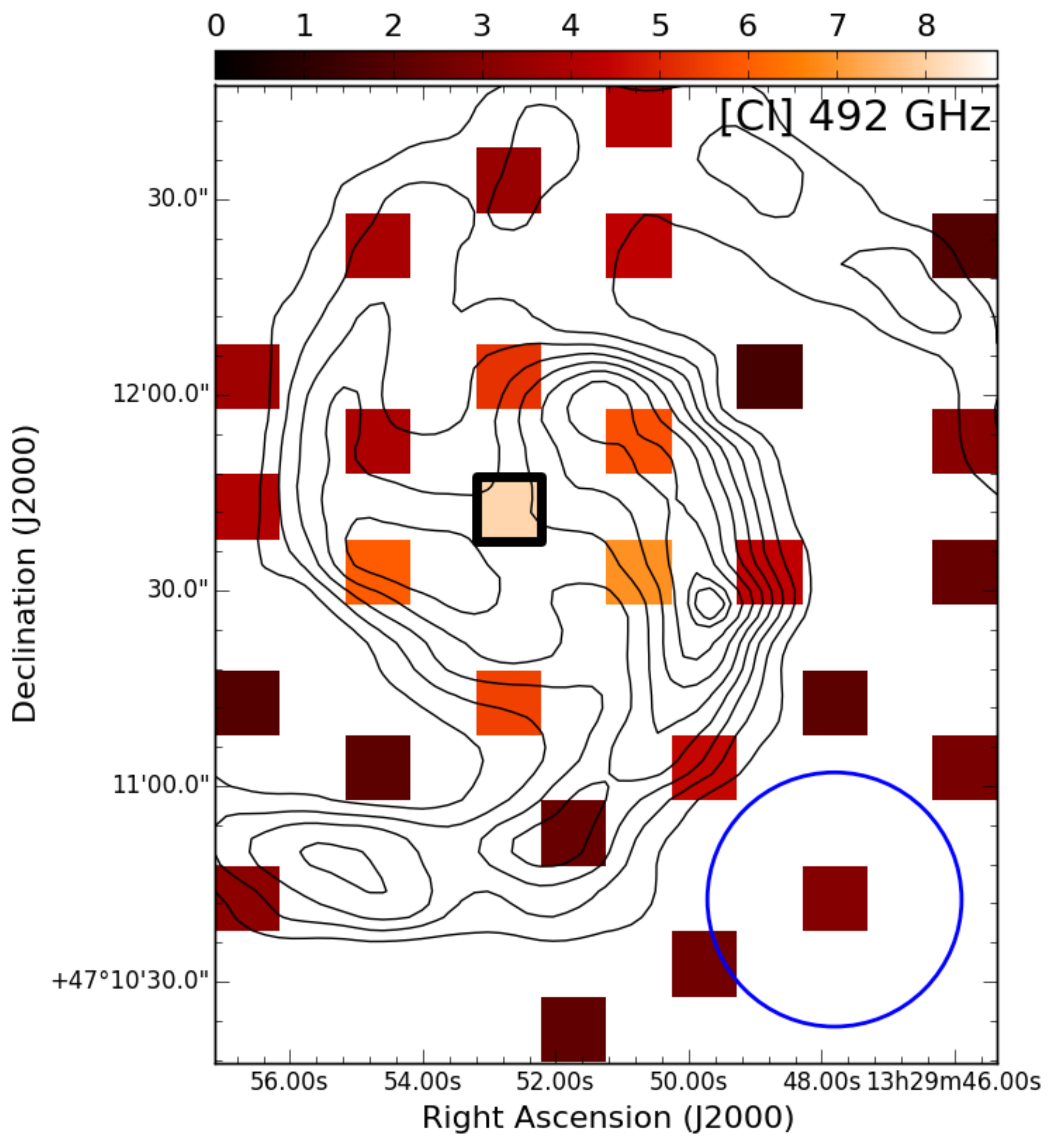} 
		\includegraphics[width=0.37\linewidth]{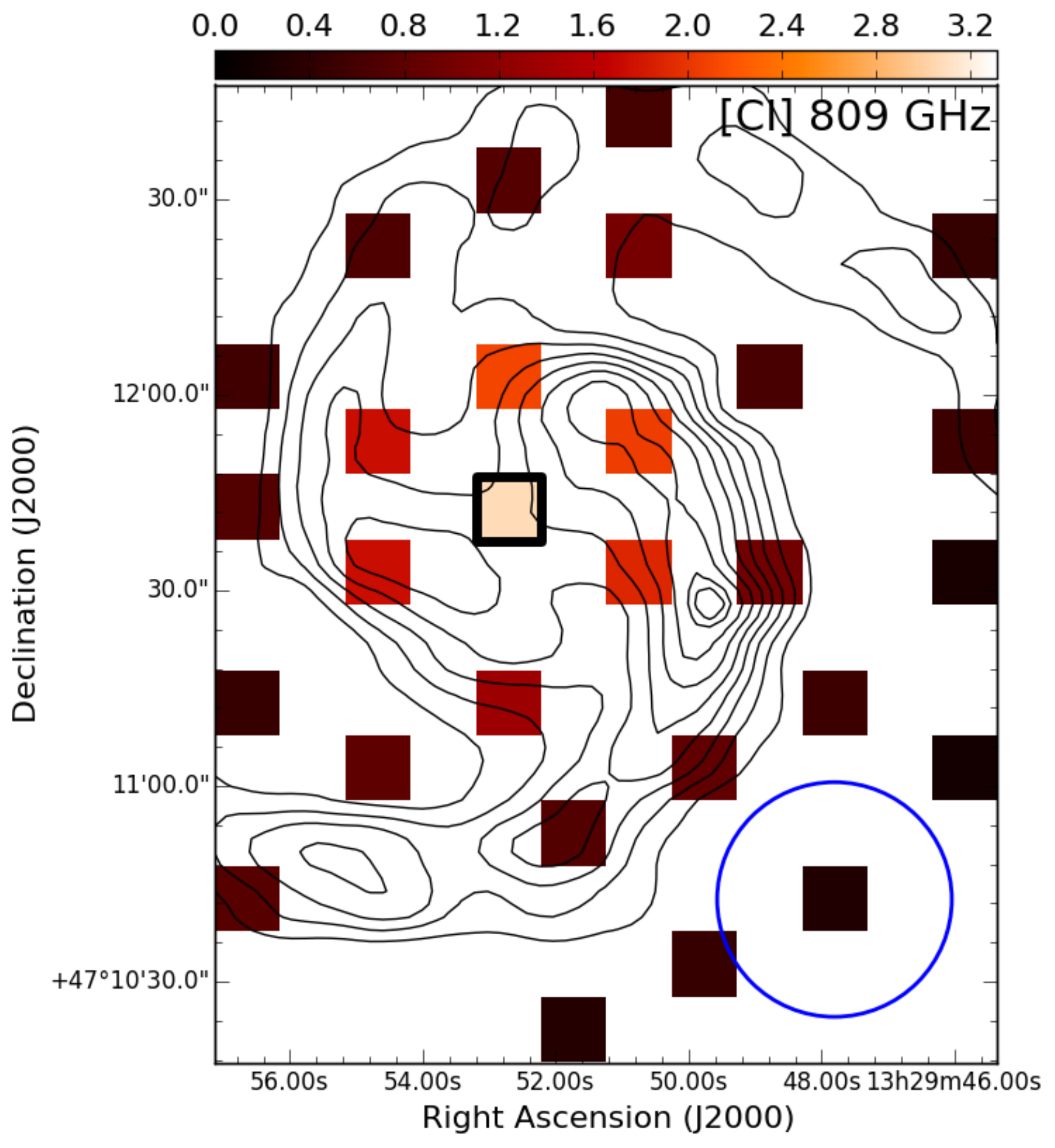} 
	\end{array}$
\caption[]{\CI \  integrated intensity  maps in units of $\mol{K
    \ km \ s^{-1}}$ from the Herschel SPIRE-FTS observations of M51 for the $^3\mol{P}_1 - ^3\mol{P}_0$ transition at $492 \unit{GHz}$ (left) and the $^3\mol{P}_2 - ^3\mol{P}_1$ transition at $809 \unit{GHz}$ (right) transitions. For more details see Figure \ref{COFTSImg}.}
\label{CIFTSImg}
\end{figure*}

For each pixel in our cubes, the routine fits each of the lines listed
in Table \ref{lineFlux} with a Sinc function using a
Levenberg-Marquardt fitter, keeping the width of the line fixed to
$1.4305 \unit{GHz}$, with the amplitude and centroid varying. The
surrounding $30 \unit{GHz}$ is fit with a quadratic at the same time
in order to account for the continuum emission. With the exception of
the \CO \ $J=7-6$ and \CI \ $J=2-1$ transitions, all of the lines
listed in Table \ref{lineFlux} are fit individually. In the case of
the $\mol{CO}$ $J=7-6$ and $\mol{[CI]}$ $J=2-1$ transitions, both
lines are fit concurrently, each with a Sinc function, along with the
continuum emission. We integrate the resulting Sinc functions to
calculate the total integrated intensity for each line, while the
uncertainty is calculated from the uncertainty in the fitting
parameters. The \CO \ $J=10-9$ 
line is   detected only in the nucleus with a flux of
$0.6 \pm 0.2 \unit{K
  \ km \ s^{-1}}$. The calibration uncertainty of the SPIRE-FTS is
$7$ percent, while we add a total of $10$ percent in quadrature to account for
uncertainties in fitting the baseline, and uncertainties in the
semi-extended source correction. The resulting maps for \CO \ and \CI
\ are shown in Figures \ref{COFTSImg} and \ref{CIFTSImg}, respectively.   

\input{\TabPath/LineFluxAuto.tex}

For each region, we perform an unweighted average for each \CO \ and \CI \ transition, including all pixels where all of the transitions are detected with a signal-to-noise ratio (SNR) $> 1$. We opt for an unweighted average as the fluxes within an individual region are similar (Figure \ref{measSLED}). With the exception of the \CO \ $J=8-7$ line, all of the transitions listed in Table \ref{lineFlux} are detected with a SNR $>3$ for all pixels in the nucleus and centre regions. For the arm/inter-arm region, only a single pixel satisfies the same SNR $>3$. As such, we choose a SNR $>1$ in order to explore the parameter space in the arm and inter-arm regions of M51 (Figure \ref{measSLED}). Even with this low SNR cut, few pixels are included from the arm and inter-arm regions (see Figure \ref{measSLED}), while the included pixels all contain a portion of both regions within the $40''$ beam. Therefore, we combine the arm and inter-arm regions into a single arm/inter-arm region. Finally, we combine the pixels included in the nucleus, centre and arm/inter-arm regions into a single ``All'' region.

\subsection{Ancillary Data}

\subsubsection{Ground based \CO} \label{groundCO}

We supplemented our FTS observations of M51 using previously published
\CO \ $J=1-0$ to $J=3-2$ maps from ground based instruments. M51 was
observed in \CO \ $J=1-0$ and $^{13}$\CO \ $J=1-0$ using the Institut
de Radioastronomie Millim\'etrique (IRAM) $30\unit{m}$ telescope as
part of the PAWS program \citep{schinnerer2013}. 
%The $30 \unit{m}$ observations were used to fill in the short
%spacings for the interferometric observations using the PdBI. We use
%only the single-dish observations of \CO \ $J=1-0$ and $^{13}$\CO
%\ $J=1-0$ 
As we are interested only in the very large scales ($\sim
40\unit{pc}$) in M51, we use the single-dish \CO \ $J=1-0$ and $^{13}$\CO \ $J=1-0$ data cubes published in \cite{pety2013}\footnote{Downloaded from \url{http://www.mpia.de/PAWS/PAWS/Data.html}}, with a beam size of $22.5''$.
%We acquired previous observations of in M51 from the Institut de
%Radioastronomie Millim\'etrique (IRAM) $30\unit{m}$
%telescope\footnote{http://www.mpia-hd.mpg.de/HERACLES/Overview.html}. 
M51 was observed in \CO \ $J=2-1$ with the IRAM
  $30\unit{m}$ as part of the HERA CO Line Extragalactic
Survey\footnote{http://www.mpia-hd.mpg.de/HERACLES/Overview.html}
\citep{schuster2007, leroy2009}.  The publicly available data cube has
been smoothed to $13''$, the smallest beam size of all the \CO \ maps
used in this work.  
Finally, M51 was observed in \CO \ $J=3-2$ with $15 ''$ resolution by
\cite{vlahakis2013} using the HARP-B instrument on the JCMT. We
  use the data cube released
as part of the JCMT Nearby Galaxies Legacy
Survey\footnote{http://www.physics.mcmaster.ca/$\sim$wilson/www\_xfer/NGLS/}
(NGLS,
\citealt{wilson2012}).

\begin{figure*}
	\centering
	\includegraphics[width=\linewidth]{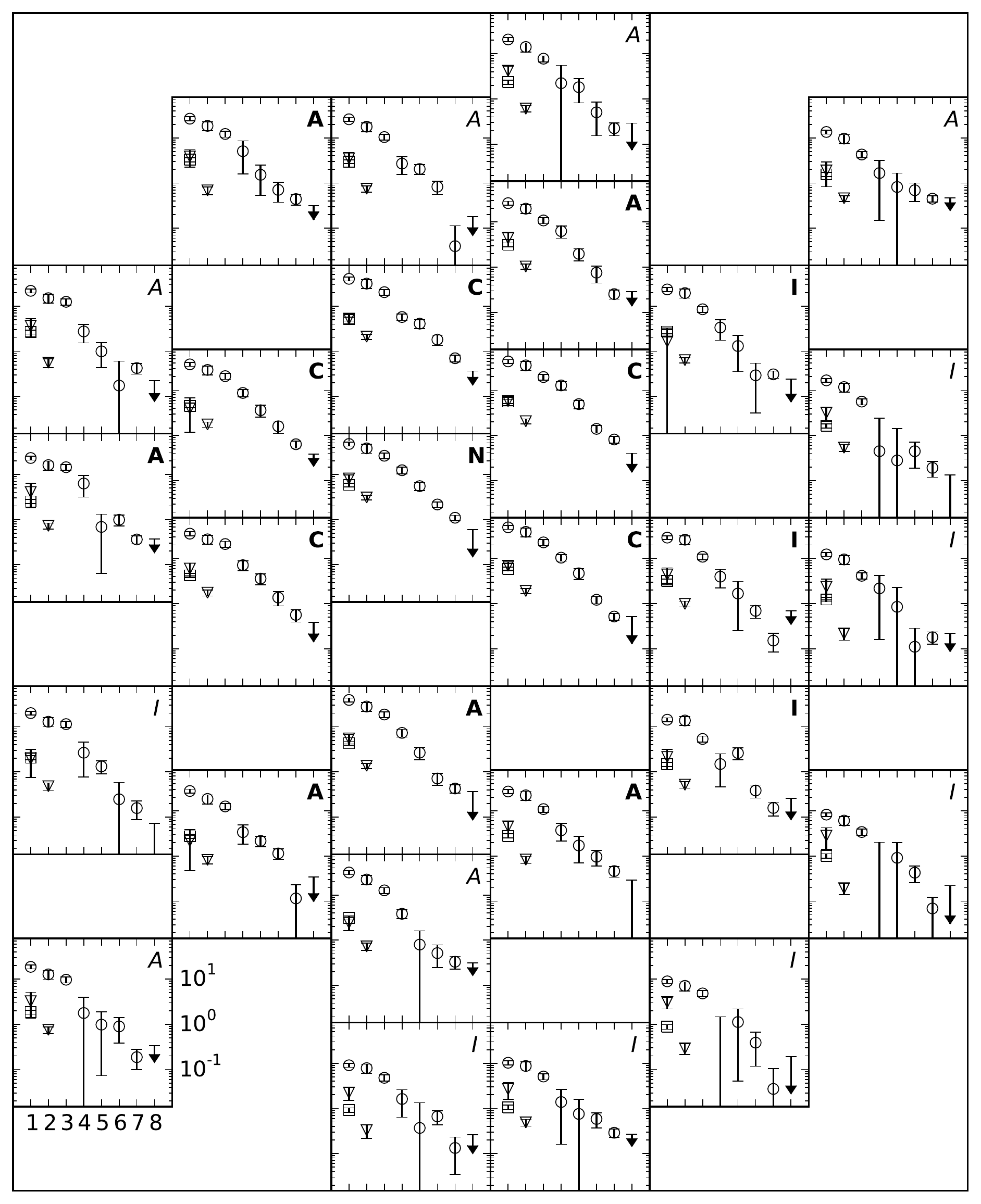} 
	\caption[]{Measured \CO \ (circles), \CI \  (triangles), and
          $^{13}$\CO (squares) spectral line energy distribution in
          units of $\mol{K \ km \ s^{-1}}$ for each pixel in our FTS
          data cube. The y-axis
          corresponds to the integrated intensity in units of $\mol{K
            \ km \ s^{-1}}$ and the x-axis corresponds to
          $J_{up}$. The vertical and horizontal scales are the same for each box and are shown
          in the lower-left box. The \CO \ $J=1-0$ to $J=3-2$ and $^{13}$\CO
          \ $J=1-0$ transitions are from ground based instruments (see
          Section \ref{groundCO}). The error bars shown here do not
          include calibration uncertainties (see Table \ref{lineFlux}
          for the calibration uncertainties). The letter in the upper
          right corner indicates the pixel's region,  where ``N''
          corresponds to the nucleus, ``C'' the centre, ``A'' the arm
          and ``I'' the inter-arm regions. Only the bolded letters are included in our region averages.  } 
	\label{measSLED}
\end{figure*}

All four of the ground-based \CO \ transitions were reduced in the same manner using the \verb+Starlink+ software package \citep{currie2008} and a similar method as for the \CO \ $J=3-2$ observations of NGC 4038/39 in \cite{schirm2014}. First, we convolved the data cubes to a $40''$ gaussian beam using the \verb+gausmooth+ command. We then smoothed the cubes with a top hat with a width 2.5 times the half-power beam-width of the $40''$ beam and smoothed to a velocity width of $20 \unit{km/s}$. Using the \verb+clumpfind+ command, we identified regions of emission above $2 \sigma$ in our smoothed cube to create a mask which we then used to create moment 0, 1 and 2 maps, corresponding the intensity-weighted integrated intensity, velocity and line-width maps, respectively, from our original, $40''$ HPBW, data cubes. We estimated the noise in our moment 0 maps using the emission free channels from our $40''$ HPBW data cubes. Finally, we re-gridded the ground based data using the \verb+wcsalign+ task in \verb+Starlink+, using our \CO \ $J=4-3$ integrated intensity map as a reference. 
The resulting spectral line energy distributions (SLEDs) for \CO \ and
\CI \ are shown in Figures \ref{measSLED} (individual pixels) and
\ref{twoCompMapSLED} (region averages).

\subsubsection{Infrared data} \label{IRdata}

The total infrared luminosity, $L_{\mol{TIR}}$, can be combined
  with the \CO~ line strengths to provide useful constraints on photon dominated region (PDR) models.
%We estimate $L_{\mol{FIR}}$ ($\lambda = 40 - 500 \unit{\mu m}$) using
%a similar method to \cite{parkin2013}. 
We calculate the total infrared luminosity ($L_{\mol{TIR}}$, $\lambda = 3 - 1100 \unit{\mu m}$) using the following empirical relation from \cite{galametz2013}

\begin{equation}
    L_{\mol{TIR}} = c_{24}\nu_{24} L_{24} + c_{70}\nu_{70} L_{70} + c_{160}\nu_{160} L_{160}
\end{equation}

\noindent where the subscripts $24$, $70$, and $160$ correspond to the
$24 \unit{\mu m}$, $70 \unit{\mu m}$, and $160 \unit{\mu m}$
maps respectively, while $c_{24} = 2.133 \pm 0.095$,
$c_{70} = 0.681 \pm 0.028$ and $c_{160} = 1.125 \pm 0.010$. We use the
\emph{Spitzer Space Telescope} MIPS $24 \unit{\mu m}$  map
reprocessed by \cite{bendo2012}, and the \emph{Herschel} PACS $70
\unit{\mu m}$ and $160 \unit{\mu m}$  maps from
\cite{mentuchcooper2012}.  We beam-match and align all the maps 
using appropriate kernels from \cite{aniano2011}
%Aniano et al. (2011)
before calculating the $L_{\mol{TIR}}$ map. We
then beam-match and align the $L_{\mol{TIR}}$ map to our FTS
observations using gaussian kernels with the \emph{imsmooth} command
to beam-match, and the \emph{imregrid} command to regrid, in the Common Astronomy Software Package (CASA) version 4.2.1. %We use a ratio of $L_{\mol{TIR}}/L_{\mol{FIR}} = 1.3$ to calculate our final FIR map as in \cite{parkin2013}, consistent with the value calculated for a sample of normal galaxies ($L_{\mol{TIR}} < 10^{11} L_\odot$)  \citep{graciacarpio2008}.
The calculation of $L_{\mol{TIR}}$ using the $24 \unit{\mu m}$, $70
\unit{\mu m}$, and $160 \unit{\mu m}$ photometric waveband provides a
reliable estimate of $L_{\mol{TIR}}$ to within $25 $ percent \citep{galametz2013}. %\cite{dale2001} calculate the FIR luminosity in a sample of normal star-forming galaxies over a smaller wavelength range than we have ($\lambda = 42 - 122 \unit{\mu m}$), and find that in the nearby universe ($z \sim 0$), the ratio of $L_{\mol{TIR}}/L_{\mol{FIR}} \sim 2 - 3$ for their definition. Given their results cover a smaller wavelength range, and that by definition, $L_{\mol{TIR}}/L_{\mol{FIR}} > 1$, we estimate that our measured values for $L_{\mol{FIR}}$ are accurate to within a factor of 2. 

% Start with the SLED

%M51 was previously observed in \CO $J=1-0$ using the Nobeyama Radio Observatory (NRO) $45\unit{m}$ telescope (e.g. see \citealt{kuno2007}). We acquired the \CO $J=1-0$ data cube published in \cite{kuno2007} as part of the Nobeyama CO Atlas of Nearby Spiral Galaxies\footnote{Downloaded from \url{http://www.nro.nao.ac.jp/~nro45mrt/html/COatlas/}}, with a beam size of $15''$. We reduce this data using the \verb+Starlink+ software package \citep{currie2008} using the same method as used for the \CO $J=3-2$ data (see Section \ref{CO32}). We convolve the data cube to a $40''$ gaussian beam using the \verb+gausmooth+ command before collapsing the cube using the \verb+clumpfind+.

%\subsubsection{PACS spectroscopy}

%Convolve each PACS image using CASA imsmooth command with a gaussian. Original beam sizes are listed in Parkin et al. 2013. \citep{parkin2013}

%% file: LineFluxAuto.tex
\begin{table*} 
\centering 
\begin{minipage}{\linewidth} 
\caption{Line integrated intensity measurements} 
\begin{tabular}{@{}l l c c c c c c} 
\hline 
Species & Transition & Rest frequency & \multicolumn{4}{c}{Average
  measured  integrated intensity\footnotemark[1]} & Calibration  \\ 
& & ($\unit{GHz}$) & \multicolumn{4}{c}{($\unit{K \ km \ s^{-1}}$)} &
uncertainty ($\%$) \\ 
& & & Nucleus & Centre & Arm/inter-arm & All & \\ 
\hline 
\CO~(ancillary)& $J = 1 - 0$& 115.27 & $47.2 \pm 0.2$& $41.2 \pm 0.1$& $26.2 \pm 0.1$& $33.1 \pm 0.1$& $10.0$\footnotemark[2] \\ 
& $J = 2 - 1$& 230.54 & $37.8 \pm 0.2$& $32.2 \pm 0.2$& $20.0 \pm 0.2$& $25.6 \pm 0.2$& $20.0$\footnotemark[3] \\ 
& $J = 3 - 2$& 345.80 & $25.9 \pm 0.2$& $20.9 \pm 0.2$& $11.7 \pm 0.2$& $16.0 \pm 0.2$& $15.0$\footnotemark[4] \\ 
\hline 
\CO~(FTS) & $J = 4 - 3$& 461.04 & $12 \pm 1$& $9 \pm 1$& $< 4.6 \pm 2.1$& $7 \pm 2$& $12.2$ \\ 
& $J = 5 - 4$& 576.27 & $5.5 \pm 0.9$& $4.2 \pm 0.9$& $< 1.8 \pm 0.9$& $2.9 \pm 0.9$& $12.2$ \\ 
& $J = 6 - 5$& 691.47 & $2.2 \pm 0.1$& $1.5 \pm 0.4$& $< 0.7 \pm 0.2$& $1.1 \pm 0.3$& $12.2$ \\ 
& $J = 7 - 6$& 806.65 & $1.10 \pm 0.06$& $0.6 \pm 0.1$& $0.28 \pm 0.07$& $0.47 \pm 0.08$& $12.2$ \\ 
& $J = 8 - 7$& 921.80 & $< 0.4 \pm 0.2$& $< 0.3 \pm 0.1$& $< 0.3 \pm 0.1$& $< 0.3 \pm 0.1$& $12.2$ \\ 
\hline 
$^{13}$\CO~(ancillary)& $J = 1 - 0$& 110.20 & $5.87 \pm 0.06$& $5.09 \pm 0.07$& $2.92 \pm 0.06$& $3.91 \pm 0.06$& $10.0$\footnotemark[2] \\ 
\hline 
\CI~(FTS)& $J = 1 - 0$& 492.16 & $8 \pm 1$& $6 \pm 2$& $< 3.5 \pm 1.6$& $< 4.6 \pm 1.6$& $12.2$ \\ 
& $J = 2 - 1$& 809.34 & $3.08 \pm 0.06$& $1.9 \pm 0.1$& $0.85 \pm 0.08$& $1.39 \pm 0.09$& $12.2$ \\ 
\hline 
\label{lineFlux} 
\end{tabular} 

 \medskip 
\footnotemark[1]Quoted uncertainties are measurement uncertainties only \\ 
\footnotemark[2]\cite{kramer2008} \\ 
\footnotemark[3]\cite{leroy2009} \\ 
\footnotemark[4]\cite{vlahakis2013} \\ 
\end{minipage} 
\end{table*}

%% file: RadTrans.tex
\section{Non-LTE Excitation Analysis} \label{radTransSect}

 We use the methods presented in \cite{kamenetzky2014} to perform a non-LTE excitation analysis to determine the physical
state of the molecular gas. % across the entire observed region. 
Here, we present the highlights of the method used, along with any
differences from the previous work. 
We use the non-LTE excitation code RADEX \citep{vandertak2007} along
with a Bayesian likelihood code \citep{kamenetzky2014} to determine
the kinetic temperature 
(\Tkin), molecular gas density (\nH), area filling factor (\FF) and
\CO \ column density per unit line width (\NCO).
 In this work, we also include 
the \CI / \CO \ column density ratio 
%\CI \ column density per unit line width 
(\NCI/\NCO) as one of the fitted parameters. 
The range of each parameter used in the models is given in Table \ref{gridparameter}.
We compare our measured fluxes to the  RADEX models to calculate
the likelihood distribution for each of the physical parameters. 
The
code determines the %most probable value (``1DMax'') and 
 median and 1$\sigma$ range for each of the physical parameters
   from the  marginalized likelihood distribution,  along with
the most probable  ``best-fit'' (``4DMax'') solution from the combined
multi-dimensional likelihood distribution. 
%for the kinetic temperature, molecular gas density, \CO \ column
%density, area filling factor,
% and \CI / \CO \ column density ratio.  
By using Bayesian
inference, we are able to include priors on the physical parameters
based upon the physical characteristics of the observed region.

\input{\TabPath/GridParam.tex}

\subsection{Priors}

We introduce three priors into our modeling: the length
along the line of sight, the total molecular gas mass and 
the optical depth. These are the same priors used in
\cite{schirm2014}; however the derivation of some of the physical
parameters differs.  The physical parameters used to
calculate the priors are given in Table \ref{radtab}. Note that the
line width, which is used to calculate the total column density \NCO,
is taken as the average from the \CO \ $J=3-2$ moment 2 map 
for all the pixels included in the regional averages.
%Here, we describe the three priors used, along with how we derive some of
%the physical parameters. 

\input{\TabPath/ModelConstraints.tex}

The first prior is on the total length ($L$) of the \CO \ and \CI
\ emitting regions along the line of sight. This prior limits the
column density, area filling factor, and molecular gas density such that

\begin{equation}
	\frac{N_{\mol{CO}}}{\sqrt{\Phi_A}x_{\mol{CO}} n(\mol{H_2})} \le L
\end{equation}

\noindent As with all grand design spiral galaxies, the molecular gas
in M51 resides predominantly in a disk. We derive our length
prior based upon measurements of the scale height of this
disk. \cite{pety2013} calculated the scale height for their extended
and compact components to be $\sim 190 - 250 \unit{pc}$ and $\sim 10 -
40 \unit{pc}$, respectively (figure 17 of their work). Their extended
component corresponds to a warm, diffuse component, which we do not
include in our models. (We  discuss the implications of
including a diffuse component in Section \ref{diffGas}.) We use the
scale height of their compact component and adopt  a length prior of
160 pc, a factor of 4 times the maximum scale height ($40\unit{pc}$)
derived for the compact component.
%: a factor of 2 to account for both above and below the disk midplane, and a factor of 2 to account for the molecular gas beyond the scale height.

The a second prior is on the total mass of molecular gas in a single
beam. In previous publications (e.g. \citealt{rangwala2011,
  kamenetzky2012, schirm2014}), we used the  dynamical mass to
limit the total molecular gas mass within a single beam. This
assumption was sensible for systems contained entirely within a single
FTS beam, such as Arp 220 \citep{rangwala2011}. However, in the case
of galaxies which span multiple beams, such as NGC 4038/39
\citep{schirm2014} or M51, it is more difficult to  determine the
dynamical mass  per beam. Instead, we calculate an upper limit to the
molecular gas mass for each pixel using the $\mol{CO}$ $J=1-0$ map
along with an $\alpha_{\mol{CO}}$ value of $9 \unit{M_\odot \ pc^{-2}
  \ (K \ km \ s^{-1})^{-1}}$. This value for the conversion factor
corresponds to the largest value for $\alpha_{\mol{CO}}$  seen in the
Milky Way   and other Local Group Galaxies ($\alpha_{\mol{CO}} \sim 3 - 9 \unit{M_\odot \ pc^{-2} \ (K \ km \ s^{-1})^{-1}}$, \citealt{leroy2011}). 
The mass prior for each region is the average from all the pixels included in the region average. 

The mass prior places a limit on the beam-averaged column density ($\left<N_{\mol{CO}}\right> =  N_{\mol{CO}} \Phi_A$)

 \begin{equation}
	N_{\mol{CO}} \Phi_A < \frac{M_{mol} x_{\mol{CO}}}{\mu m_{\mol{H_2}} A_{\mol{CO}} }
\end{equation}
 
 \noindent where $M_{mol}$ is our derived mass from the $\mol{CO}$
 $J=1-0$ map, $x_{\mol{CO}}$ is the $\mol{CO}$ abundance relative to
 $\mol{H_2}$, $\mu$ is the mean molecular weight, $m_{\mol{H_2}}$ is the mass of
 the H$_2$ molecule, and $A_{\mol{CO}}$ is the area of the $\mol{CO}$ emitting region, which is the area covered by one beam at the distance of M51.

The third prior limits the optical depth of each line such that $-0.9
< \tau < 100$. An optical depth $<0$ is indicative of a maser, and we
do not expect \CO \ or \CI \ masers to contribute appreciably to the
emission on the observed size scales 
  (the limit of -0.9 allows for computational error). Furthermore, the line intensities calculated by RADEX become more uncertain the further the optical depth drops below $0$. In addition, \cite{vandertak2007} suggest limiting the optical depth to an upper limit of $100$, as the one-zone approximation implied by its escape probability formalism breaks down at optical depths greater than $\tau > 100$.

\subsection{Non-LTE Excitation Modelling}\label{nonLTESectM51}

\CO,  \CI \ and $^{13}$\CO \ are all tracers of molecular gas; all
three species are excited via collisions with \molH. In the classic
slab-geometry model of a photon dominated region (PDR) by
\cite{tielens1985}, \CI \ arises from the surfaces of molecular
clouds, while \CO \ does not begin to form until deeper into the cloud. In this model, some of the molecular gas is ``\CO \ dark'' \citep{wolfire2010}. However, there is strong evidence that \CI \ and \CO \ are co-spatial \citep{papadopoulos2004}, as supported by observations of the Orion molecular cloud \citep{plume1999, ikeda2002, shimajiri2013} and more recent simulations of molecular clouds (e.g. \citealt{offner2014}, \citealt{gaches2014}). Furthermore, \CI \ may be less sensitive to temperature than \CO \ \citep{offner2014}, and so may help constrain the density.

We fit a two-component model to our measured \CO \ and \CI \ intensities
in the nucleus, centre, and arm/inter-arm regions of M51. (A
single-component model fit, which does not produce a physically
realistic solution, is discussed in Appendix \ref{singCompSec}.) We
include the \CO \ transitions from $J=1-0$ to $J=7-6$, while leaving
the $J=8-7$ transition as an upper limit. The total uncertainty used
is the line fitting and calibration uncertainties added in
quadrature. The molecular gas in M51 is unlikely to populate two
distinct components in terms of the physical state of the gas, so 
our two-component fit will represent an
average of the state of all the molecular gas within the 3 distinct
regions. We are therefore investigating the bulk properties of the
molecular gas in the three regions. (For an extensive
discussion of one- and two-component modelling, see \citealt{kamenetzky2014}.)

\begin{figure} 
	\centering
%	\includegraphics[width=\linewidth]{twoComp_regionSLED_cold\coldComp_warm\warmComp .pdf} 
% ** xxx ** new figure 6 from Julia inserted
	\includegraphics[width=\linewidth]{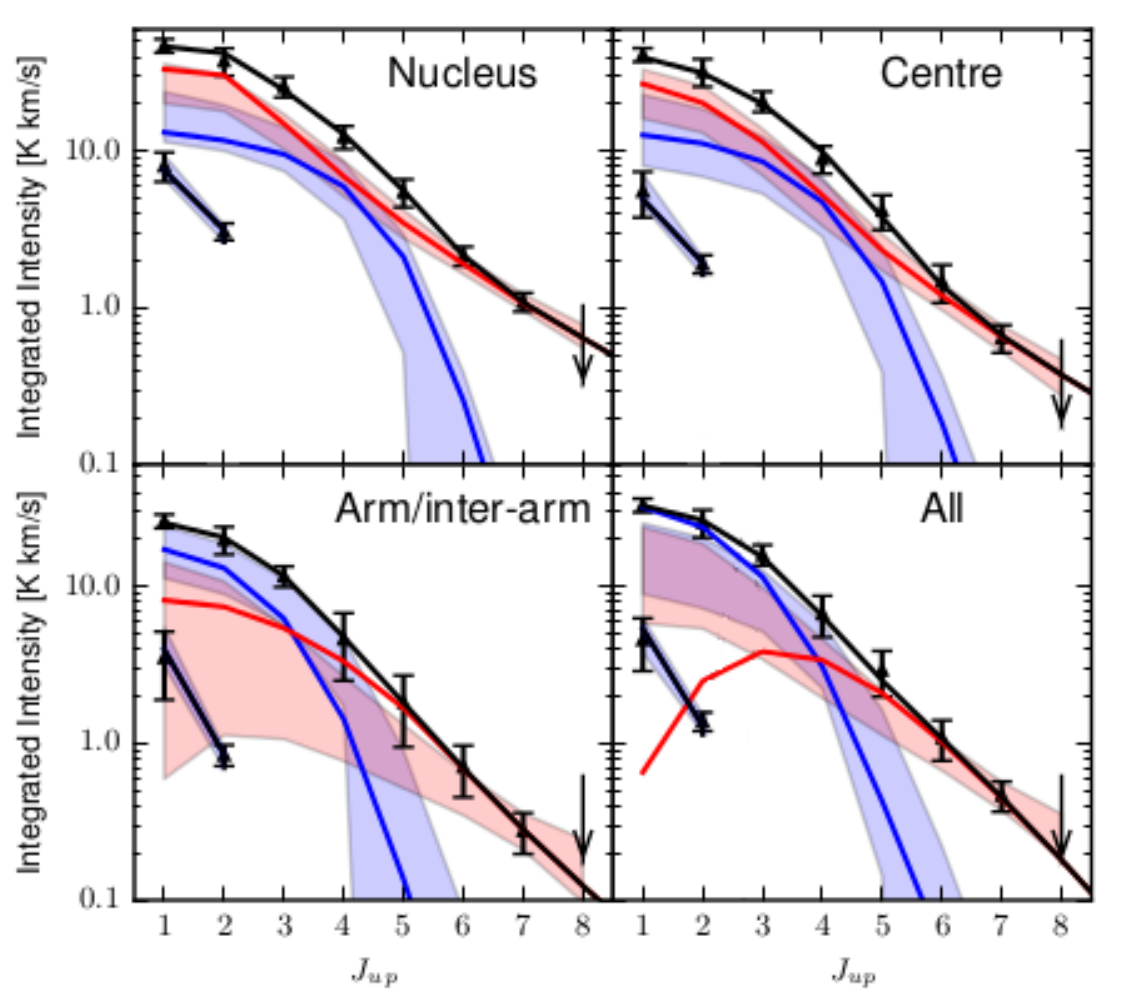} 
	\caption[]{Measured and best-fit SLEDs for the two-component fit for the nucleus
(top-left),  centre (top-right), and arm/inter-arm (bottom-left)
regions, and for all the regions combined (bottom right). The measured
\CO \ SLED (8 points) and \CI \ SLED (2 points) are shown by the black
triangles with error bars. The cold component and warm component fits
to the \CO
\ emission are shown by the blue and red lines, respectively,
while the total calculated \CO \ emission is shown by the solid black
line. The cold component fit to the \CI \ emission is indicated by a
solid black line. For the cold and warm component fits, the coloured shaded
regions indicate the 1$\sigma$ uncertainty region of the fits.}
	\label{twoCompMapSLED}
\end{figure}

\begin{figure} 
	\centering
	\includegraphics[width = \linewidth]{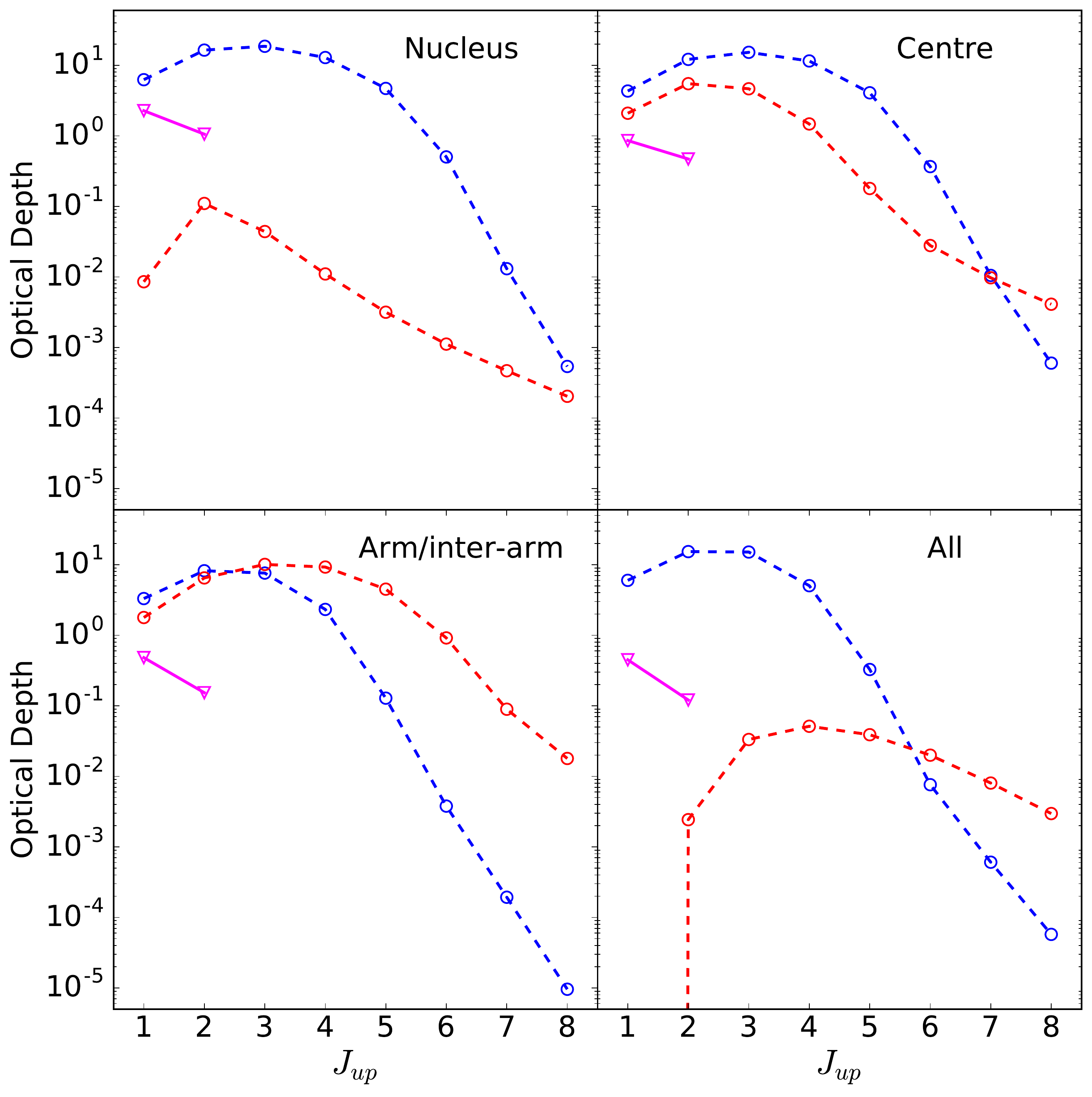} 
	\caption[]{Best-fit line optical depths for the two-component fit for the nucleus (top-left),  centre (top-right), and arm/inter-arm (bottom-left) regions, and for all the regions combined (bottom right). The cold component and warm component \CO \ optical depths are shown by the blue and red dashed lines and circles, respectively. The \CI \ optical depths are indicated by the solid magenta line and triangles.}
	\label{twoCompMapTau}
\end{figure}

Our two-component fit to the molecular gas consists of a ``cold''
component which can dominate the lower-$J$ \CO \ transitions, and a
``warm'' component which dominates the upper-$J$ \CO \ transitions. We
include the \CI \ in the cold component model only. 
 Our initial tests fitting both
components simultaneously while constraining the temperature
of the cold component to be  $<100$ K resulted in a
bimodal temperature distribution, with high probabilities at the upper
(100 K) and lower (10 K) ends of the range and a minimum probablity around
30 K. 
Therefore, to isolate better the cold component from the warmer component, we
reran the fits while constraining the 
temperature of the cold component to be $< 10^{1.5}$ K. 
The derived physical parameters are given
in Table \ref{twoCompTable}; we give both the median value for each
parameter as well as the ``best fit'' value, which is the set of
values in the multi-dimensional parameter space with the highest probability.
The resulting measured and calculated SLEDs are shown in Figure
\ref{twoCompMapSLED}, while the calculated optical depths are shown in
Figure \ref{twoCompMapTau}. The $1\sigma$ ranges are shown in
Figure \ref{fourSquareCut} and
an example of the probability distributions for four of the fit
parameters is given in Appendix \ref{ParamDistns}.

\input{\TabPath/M51\coldComp_twoCompRadexTable.tex}

%%% One thing to consider is whether we put everything on one figure, instead of splitting up case 1 and case 2

% Start with the SLED

\begin{figure*}
	\centering
	\includegraphics[width = 0.9\linewidth]{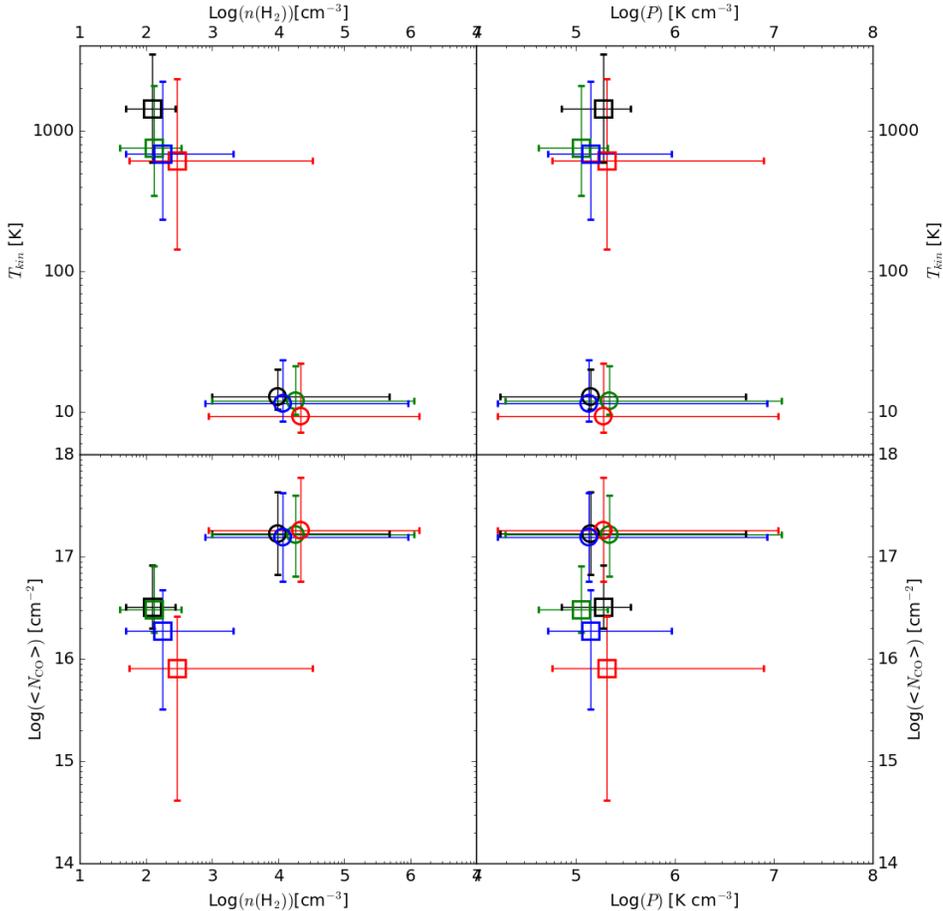}
	\caption[]{Derived physical parameters for the cold
          (\emph{circles}) and warm (\emph{squares}) components of the
          multi-component RADEX model averaged over the nucleus
          (\emph{black}), centre (\emph{green}), and arm/inter-arm (\emph{red}) regions of M51 (see Figure \ref{regionsMap} and Section \ref{radTransSect} for more details), and for the average of all the regions combined (\emph{blue}).  The symbols are plotted at the location of the median solution for each parameter.   \CO \ is included in both the cold and warm components, while \CI \ is included only in the cold component. The error bars correspond to the $1\sigma$ range of the combined likelihood distribution of each region for the kinetic temperature (\emph{top row}), beam-averaged column density (\emph{bottom row}), molecular gas density (\emph{left column}), and pressure (\emph{right column}).}  
	\label{fourSquareCut}
\end{figure*}

%-Would mean three components

%-Could use $\mol{[CI]}$ LTE temperature to fix the temperature of the cold component and see what a warm component would give us.

%-Would mean cold component from which SF is being fuelled exists

%-If we assume optically thick GMCs, then $T_{kin} \sim T_{obs}$. The GMCs measured in \cite{colombo2014} have observed temperatures in \CO \ $J=1-0$ $<10\unit{K}$, implying that the molecular gas is indeed cold. This likely breaks our degeneracy 

%\section{Discussion} \label{discuss}

%% file: GridParam.tex
\begin{table*}
	\centering
	\begin{minipage}{\linewidth}
		\caption{RADEX grid parameters}
		\begin{tabular}{@{}lcl@{}}
			\hline
			Parameter & Range & Units  \\
			\hline
			Kinetic Temperature,  cold component
                        ($T_{kin}$, Cold)  & $10^{0.5} - 10^{1.5}$ & $\unit{[K]}$ \\
			Kinetic Temperature,  warm component
                        ($T_{kin}$, Warm)  & $10^{0.7} - 10^{3.8}$ & $\unit{[K]}$ \\
			H$_2$ density ($n(\mol{H_2}$)) &
                        $10^{1.0}-10^{7.0}$ & $\unit{[cm^{-3}]}$  \\
			Area filling factor ($\Phi_A$) &
                        $10^{-5.0}-1$ & ...  \\
			\CO~column density per unit line width
                        ($N_{\mol{CO}}/\Delta V$)\footnote{Column
                          density is calculated per unit linewidth,
                          while the linewidth is held fixed at $1
                          \unit{km \ s^{-1}}$ in the calculations
                          (see text).}   & $10^{12.0}-10^{18.0}$ &
                        $\unit{[cm^{-2} \ (km \ s^{-1})^{-1}]}$ \\
			\CI/\CO~column density ratio
                        ($N_{\mol{[CI]}}/N_{\mol{CO}}$) & $10^{-2.0} -
                        10^{2.0}$ & ... \\
			Line width ($\Delta V$)   & $1.0$ &
                        $\unit{[km \                        s^{-1}]}$ \\
			\hline
		\end{tabular}
		\label{gridparameter}
	\end{minipage}
\end{table*}

%\begin{deluxetable}{lcccc} %%TABLE4
%\tablecolumns{3}
%\tablewidth{0pt}
%\tabletypesize{\scriptsize}
%\tablecaption{RADEX grid parameters}
%\tablehead{\colhead{Parameter} & \colhead{Range} & \colhead{\# of Points}}
%\startdata
%$T_{kin} \unit{[K]}$ & $10^{0.7} - 10^{3.8}$ & 71\\
%$n(\mol{H_2}) \unit{[cm^{-3}]}$ & $10^{1.0}-10^{7.0}$ & 71 \\
%$\Phi_A$ & $10^{-5.0}-1$ & 71 \\
%$N_{\mol{CO}}/\Delta V \unit{[cm^{-2}]}$ & $10^{12.0}-10^{18.0}$ & 81\\
%$N_{\mol{[CI]}}/N_{\mol{CO}}$ & $10^{-2.0} - 10^{2.0}$ & 20 \\
%$\Delta V \unit{[km \ s^{-1}]} $ & $1.0$ &
%\enddata
%\tablecomments{\bfseries{Column density is calculated per unit linewidth, while the linewidth is held fixed at $1 \unit{km \ s^{-1}}$ in the grid calculations (see text).}}
%\label{gridparameter}
%\end{deluxetable}

%% file: ModelConstraints.tex
\begin{table}
	\centering
	\begin{minipage}{\linewidth}
		\caption{Non-LTE Model Constraints}
		\begin{tabular}{@{}lcc@{}}
			\hline
  			Parameter & Value & Units \\
			\hline
			$\mol{CO}$ abundance ($x_{\mol{CO}}$) &$3\times 10^{-4}$ & ... \\
			Mean molecular weight ($\mu$) & $1.5$ & ...\\
			Angular size scale &$48.9$ &$\unit{pc}/''$ \\
			Source size & $40$&$''$ \\
			Length  ($L$)& $\le 160$ & $\unit{pc}$ \\
			\hline
		\end{tabular}
		\label{radtab}
	\end{minipage}
\end{table}

%\begin{deluxetable}{lcc} 
%\tablecolumns{3}
%\tablewidth{0pt}
%\tabletypesize{\scriptsize}
%\tablecaption{Model Constraints}
%\tablehead{\colhead{Parameter} & \colhead{Value} & \colhead{Units}}
%\startdata
%$\mol{CO}$ abundance ($x_{\mol{CO}}$) &$3\times 10^{-4}$ & ... \\
%Mean molecular weight ($\mu$) & $1.5$ & ...\\
%Angular size scale &$48.9$ &$\unit{pc}/''$ \\
%Source size & $40$&$''$ \\
%Length  ($L$)& $\le 160$ & $\unit{pc}$ \\
%Line width ($\Delta v$)\tablenotemark{a} &  & $\unit{km/s}$
%\enddata
%\tablenotetext{a}{The line width is derived for each pixel from the \CO \ $J=3-2$ moment 2 map. }
%\label{radtab}
%\end{deluxetable}

%% file: M51_7downjcut4_CIcold_n9_twoCompRadexTable.tex
\begin{table*} 
\centering 
\begin{minipage}{\linewidth} 
\caption{Results from two-component non-LTE excitation analysis} 
%\begin{tabular}{@{}l l l l l l l r} 
\begin{tabular}{@{}l l l l l l } 
\hline 
 &  \multicolumn{4}{c}{Median ($-1\sigma \rightarrow +1\sigma$; best fit)} & \\ 
Parameter\footnotemark[1]  & Nucleus & Centre%\footnotemark[1] 
& Arm/inter-arm%\footnotemark[1] 
& All \\ %& Units \\ 
\hline 
Cold component: \\
%\hline 
$\mol{Log}(T_{\mol{kin}})$& $ 1.1 \ (1.0 \rightarrow 1.3; 1.3) $ & $ 1.1
\ (1.0 \rightarrow 1.3; 1.4) $ & $ 1.0 \ (0.9 \rightarrow 1.4; 1.3) $ & $ 1.1 \ (0.9 \rightarrow
1.4; 1.5)
$ \\ % &$\unit{[\mol{Log}(K)]}$\\ 
$\mol{Log}(n(\mol{H_2}))$& $ 4.0 \ (3.0 \rightarrow 5.7; 3.6) $ & $ 4.3
\ (3.0 \rightarrow 6.1; 3.4) $ & $ 4.3 \ (2.9 \rightarrow 6.1; 3.0) $ & $ 4.1 \ (2.9 \rightarrow
6.0; 2.4) $ \\%&$\unit{[\mol{Log}(cm^{-3})]}$\\ 
$\mol{Log}(N_{\mol{CO}})$& $ 19.1 \ (18.6 \rightarrow 19.5; 19.1) $ & $ 19.0
\ (18.6 \rightarrow 19.4; 19.0) $ & $ 18.8 \ (18.3 \rightarrow 19.2; 18.4) $ & $ 18.9
\ (18.4 \rightarrow 19.3; 18.8) $ \\%&$\unit{[\mol{Log}(cm^{-2})]}$\\ 
$\mol{Log}(\Phi_A)$& $ -1.8 \ (-2.1 \rightarrow -1.6; -2.1) $ & $
-1.8 \ (-2.1 \rightarrow -1.5; -2.2) $ & $ -1.5 \ (-1.8 \rightarrow
-1.1; -1.7) $ & $ -1.6 \ (-2.0 \rightarrow -1.4; -1.5) $ \\%&$\unit{[...]}$\\ 
$\mol{Log}(<N_{\mol{CO}}>)$& $ 17.2 \ (16.8 \rightarrow 17.6; 17.0) $ & $
17.2 \ (16.8 \rightarrow 17.6; 16.8) $ & $ 17.3 \ (16.8 \rightarrow 17.8; 16.7) $ & $
17.2 \ (16.8 \rightarrow 17.6; 17.3) $ \\%&$\unit{[\mol{Log}(cm^{-2})]}$\\ 
$\mol{Log}(P)$& $ 5.2 \ (4.2 \rightarrow 6.7; 4.9) $ & $ 5.3 \ (4.3 \rightarrow 7.1;
4.7) $ & $ 5.3 \ (4.2 \rightarrow 7.0; 4.3) $ & $ 5.1 \ (4.2 \rightarrow 6.9; 3.9) $ \\%&$\unit{[\mol{Log}(K \ cm^{-3})]}$\\ 
% ** CI updated to be N[CI]/N(CO)
$\mol{Log}(N_{\mol{[CI]}}/N_{\mol{CO}})$& $ 1.2 \ (0.4 \rightarrow
1.7; 0.4) $ &
$ 1.0 \ (0.1 \rightarrow 1.6; 0.2) $ & $ 1.0 \ (-0.1 \rightarrow
1.6; 0.2) $ & $ 0.9 \ (0.0 \rightarrow 1.6; -0.2) $ \\%&$\unit{[...]}$\\ 
\hline 
Warm component: \\
%\hline 
$\mol{Log}(T_{\mol{kin}})$& $ 3.2 \ (2.8 \rightarrow 3.5; 3.0) $ & $ 2.9
\ (2.5 \rightarrow 3.3; 2.9) $ & $ 2.8 \ (2.2 \rightarrow 3.4; 2.4) $
& $ 2.8 \ (2.4 \rightarrow 
3.4; 2.1) $ \\%&$\unit{[\mol{Log}(K)]}$\\ 
$\mol{Log}(n(\mol{H_2}))$& $ 2.1 \ (1.7 \rightarrow 2.4; 2.4) $ & $ 2.1
\ (1.6 \rightarrow 2.5; 1.9) $ & $ 2.5 \ (1.7 \rightarrow 4.5; 2.4) $
& $ 2.3 \ (1.7 \rightarrow 
3.3; 4.1) $ \\%&$\unit{[\mol{Log}(cm^{-3})]}$\\ 
$\mol{Log}(N_{\mol{CO}})$& $ 17.9 \ (16.9 \rightarrow 18.7; 16.5) $ & $ 17.8
\ (16.9 \rightarrow 18.6; 18.4) $ & $ 17.6 \ (16.3 \rightarrow 18.6; 18.8) $ & $ 17.7
\ (16.6 \rightarrow 18.6; 16.6) $ \\%&$\unit{[\mol{Log}(cm^{-2})]}$\\ 
$\mol{Log}(\Phi_A)$& $ -1.4 \ (-1.8 \rightarrow -0.5; -0.1) $ & $
-1.3 \ (-1.8 \rightarrow -0.6; -1.6) $ & $ -1.7 \ (-3.0 \rightarrow
-0.7; -2.4) $ & $ -1.4 \ (-2.2 \rightarrow -0.6; -1.4) $ \\%&$\unit{[...]}$\\ 
$\mol{Log}(<N_{\mol{CO}}>)$& $ 16.5 \ (16.3 \rightarrow 16.9; 16.4) $ & $
16.5 \ (16.3 \rightarrow 16.9; 16.8) $ & $ 15.9 \ (14.6 \rightarrow 16.4; 16.4) $ & $
16.3 \ (15.5 \rightarrow 16.7; 15.2) $ \\%&$\unit{[\mol{Log}(cm^{-2})]}$\\ 
$\mol{Log}(P)$& $ 5.3 \ (4.9 \rightarrow 5.6; 5.4) $ & $ 5.1 \ (4.6 \rightarrow 5.3;
4.7) $ & $ 5.3 \ (4.8 \rightarrow 6.9; 4.8) $ & $ 5.1 \ (4.7 \rightarrow 6.0; 6.2) $ \\%&$\unit{[\mol{Log}(K \ cm^{-3})]}$\\ 

 \hline\label{twoCompTable}\end{tabular} 

 \medskip 
\footnotemark[1] Units for the parameters are as follows: 
$(T_{\mol{kin}})$: K; 
$(n(\mol{H_2}))$: cm$^{-3}$; 
$(N_{\mol{CO}})$, 
$(<N_{\mol{CO}}>)$: cm$^{-2}$; 
$(P)$: K cm$^{-3}$; 
$(\Phi_A)$, $(N_{\mol{[CI]}}/N_{\mol{CO}})$: dimensionless.
\end{minipage} 
\end{table*}

%% file: PDR.tex
\section{Photon Dominated Regions}\label{pdrSect}

PDRs are regions of molecular gas illuminated by FUV radiation ($6.20 \unit{eV} < E_{phot} < 13.6 \unit{eV}$, \citealt{tielens1985}).  While FUV photons are typically not the right energy to dissociate molecular hydrogen,  this radiation can have a significant effect on the chemistry and heating of the illuminated region. Indeed, the FUV radiation will liberate electrons from dust grains through the photoelectric effect, which in turn will heat the molecular gas through collisions.

\begin{figure*}
	\centering
	\includegraphics[width = 0.9\linewidth]{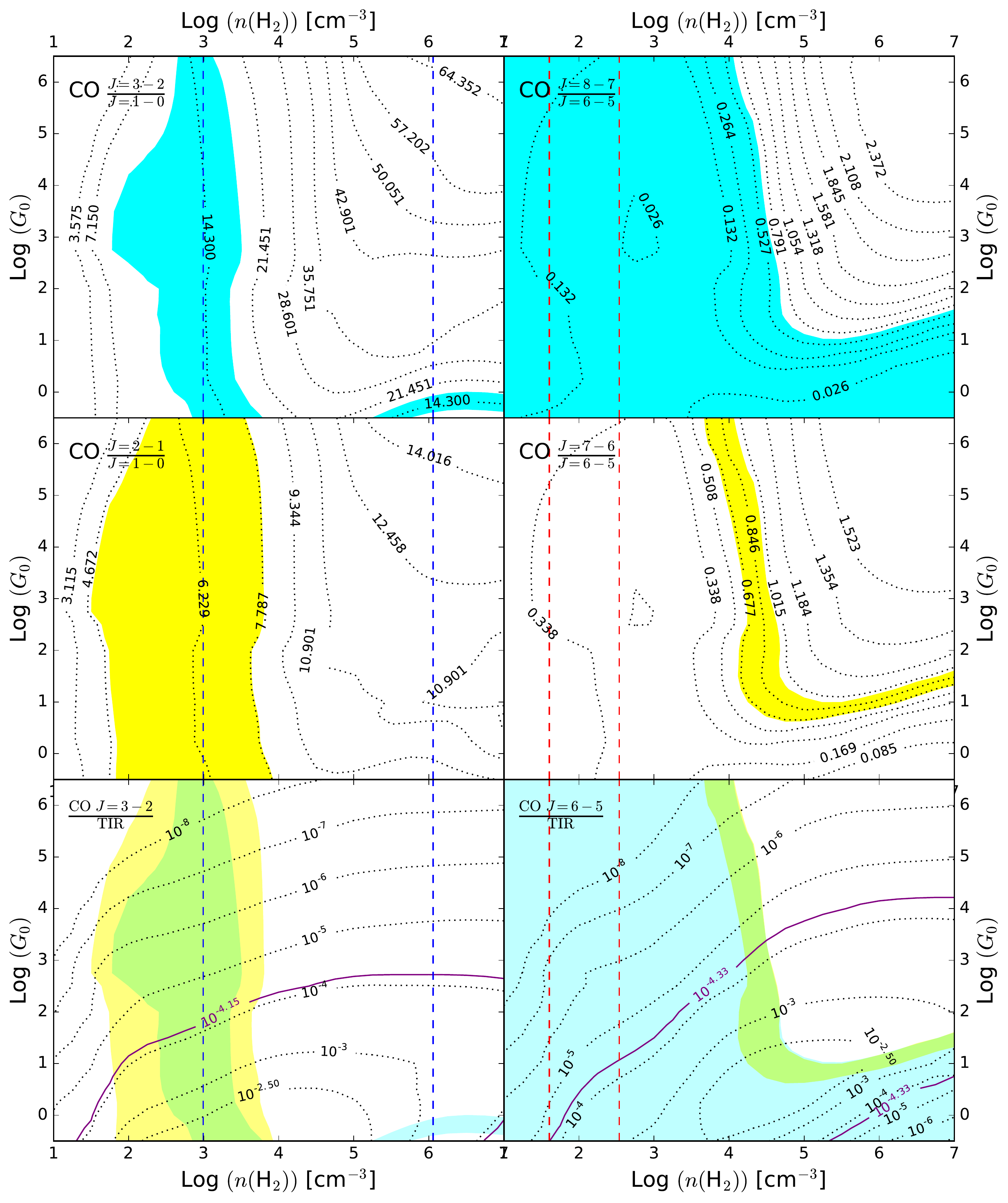}
	\caption[]{
% **  new figure 9 inserted with TIR instead of FIR in label for
% bottom 2 panels
PDR model line ratios for \CO \ $\frac{J=3-2}{J=1-0}$ (\emph{top-left}), $\frac{J=2-1}{J=1-0}$ (\emph{middle-left}), $\frac{J=8-7}{J=6-5}$ (\emph{top-right}), and $\frac{J=7-6}{J=6-5}$ (\emph{middle-right}) for the nucleus of M51 in units of $\mol{W \ m^{-2}}$.  Note that the ratio of $\frac{J=8-7}{J=6-5}$ is an upper limit. In the top two rows, the dotted contours correspond to constant \CO \ line ratios. The blue (\emph{top-row}) and yellow (\emph{middle-row}) shaded regions correspond to the uncertainty in the measured line ratio for the nucleus.  In the bottom row, the dotted contours correspond to constant value of $\frac{L_{\mol{CO} \ J=  3-2}}{L_{\mol{TIR}}}$ (\emph{bottom-left}) and $\frac{L_{\mol{CO} \ J = 6-5}}{L_{\mol{TIR}}}$ (\emph{bottom-right}), while the solid purple line is the measured ratio for each. The shaded regions in the bottom-row panels correspond to the same \CO \ line ratios of the two panels directly above, while the green region indicates where the two line ratios overlap. The blue (\emph{left-column}) and red (\emph{right-column}) dashed vertical lines correspond to the cold and warm component $1\sigma$ ranges for the densities from the two-component RADEX solutions. }  
	\label{pdrMiddle}
\end{figure*}

Using a PDR model grid (\citealt{hollenbach2012} and M. Wolfire,
private communication), we model the ratio of \CO \ transitions
(e.g. \CO \ $J=3-2$/$J=2-1$, etc.) for the nucleus, centre and
arm/inter-arm regions of M51, along with the combination of all three
regions. The model uses the molecular gas density ($n(\molH)$) and FUV
field strength ($G_0$) in units of the Habing field (FUV flux $ = 1.3
\times 10^{-4} G_0 \unit{ergs \ cm^{-2} \ s^{-1} \ sr^{-1}}$). The
model grid spans a large range of density ($n(\mol{H_2})=10^{1.0}
\unit{cm^{-3}}$ to $10^{7.0} \unit{cm^{-3}}$) and FUV field strengths
($G_0 = 10^{-0.5}$ to $10^{6.5}$) to calculate the \CO \ fluxes for
the transitions from $J=1-0$ up to $J=29-28$. We  show the \CO
\ model grid along with the line ratios measured   for the nucleus of M51 in Figure \ref{pdrMiddle}.
We also show the cold component density range calculated from our
two-component RADEX fit overlaid on the low $J$ line ratios 
(Figure \ref{pdrMiddle}, left column) and the warm
component density range overlaid on the high $J$ line ratios (Figure \ref{pdrMiddle}, right column).

%We include the cold and warm component densities calculated from our two-component RADEX fit to the low $J$ (Figure \ref{pdrMiddle}, left column) and high $J$ (Figure \ref{pdrMiddle}, right column) \CO \ transitions, respectively. 

\begin{figure}
	\centering
	\includegraphics[width = \linewidth]{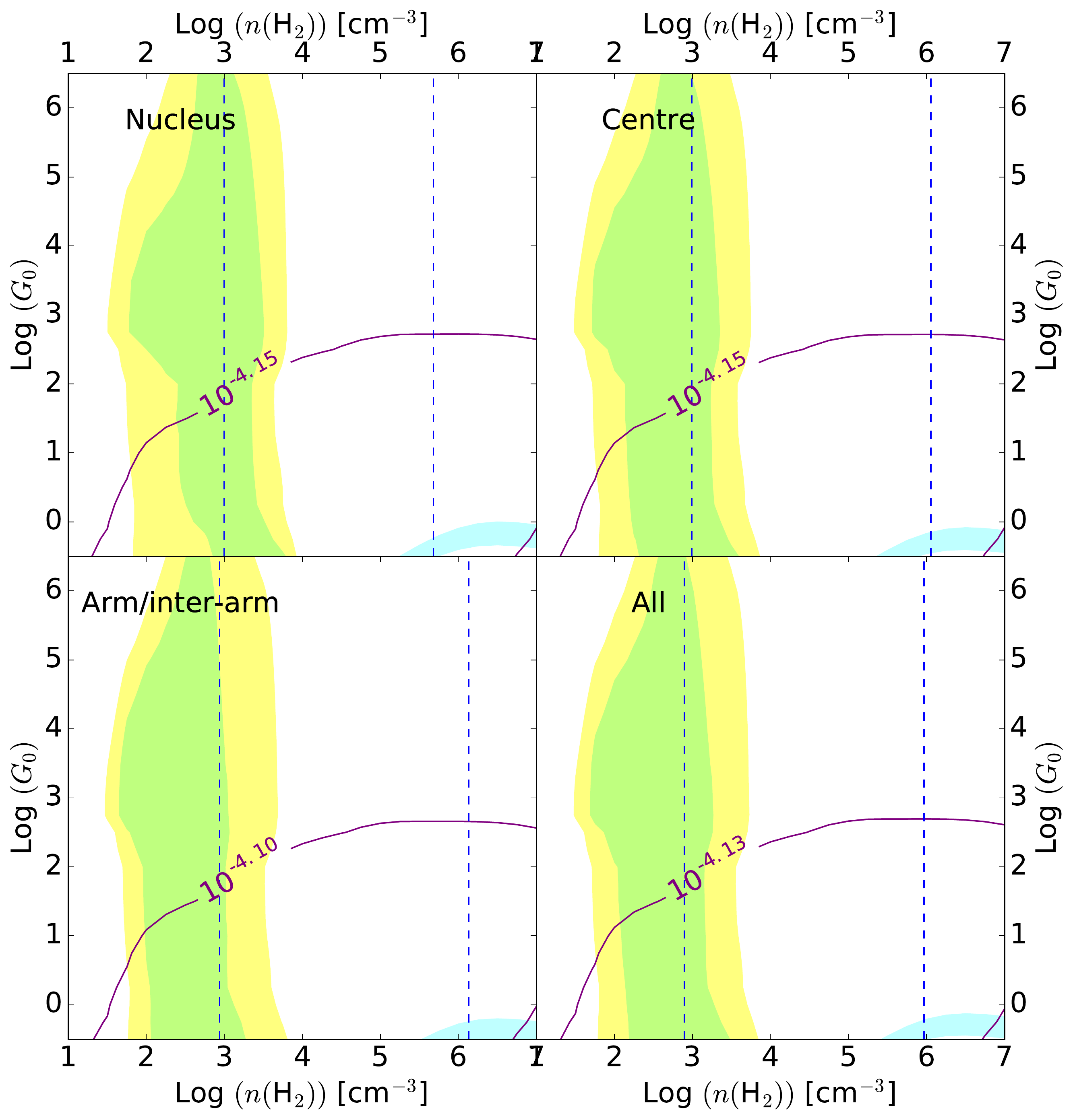}
	\caption[]{Same as the bottom-left panel of Figure \ref{pdrMiddle} except for the cold-component PDR solutions for the nucleus (top-left), centre (top-right), and arm/inter-arm (bottom-left) regions of M51, and for all four regions combined (bottom-right). Note that the line ratios are calculated in units of $\mol{W~ m^{-2}}$. }  
	\label{coldPDR}
\end{figure}

\begin{figure}
	\centering
	\includegraphics[width = \linewidth]{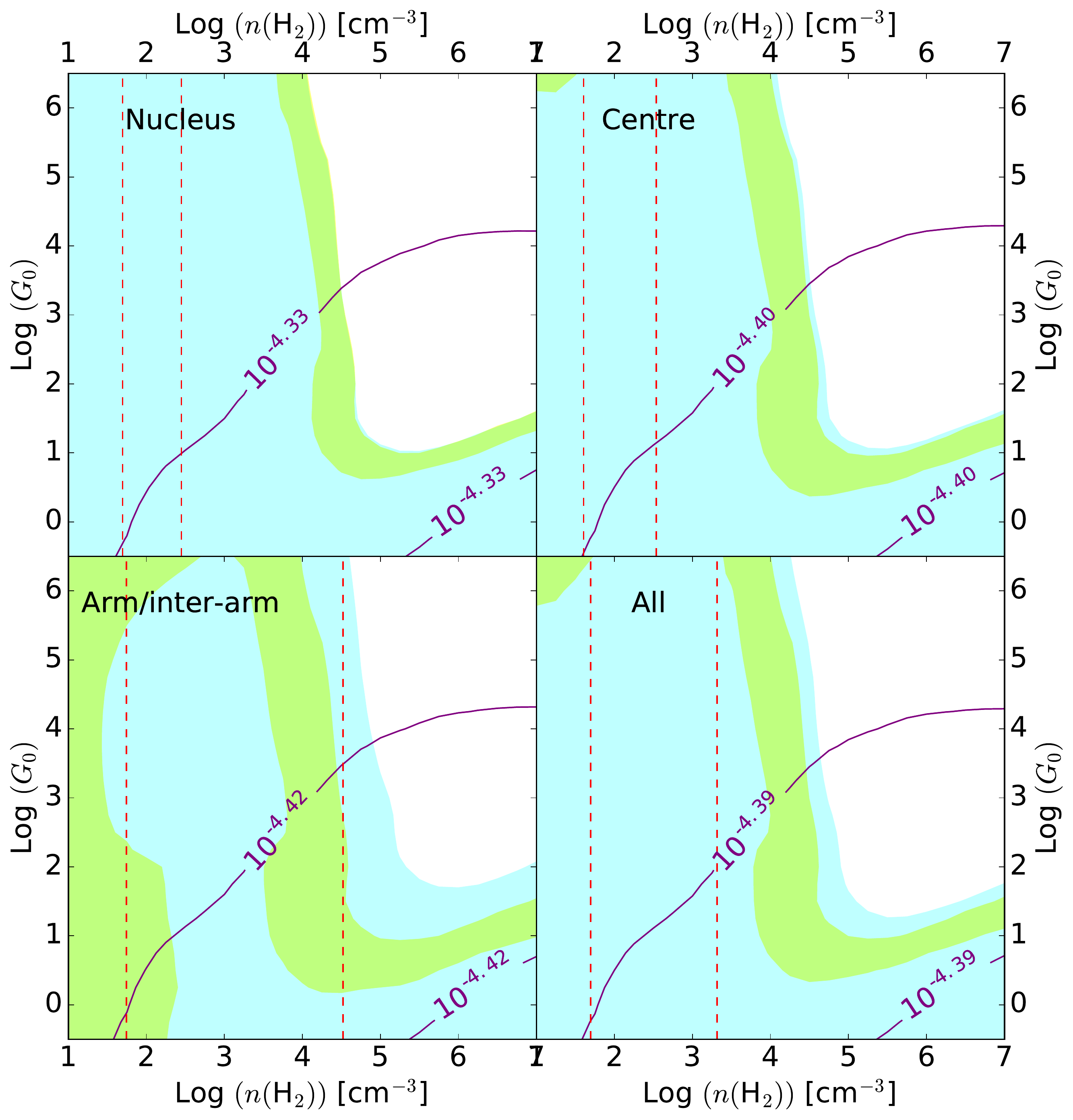}
	\caption[]{Same as the bottom-right panel of Figure \ref{pdrMiddle} except for the warm-component PDR solutions for the nucleus (top-left), centre (top-right), and arm/inter-arm (bottom-left) regions of M51, and for all four regions combined (bottom-right). Note that the line ratios are calculated in units of $\mol{W~ m^{-2}}$.}  
	\label{warmPDR}
\end{figure}

The \CO \ ratios alone are unable to constrain both the density and FUV field strength (Figure \ref{pdrMiddle}).  The ratio of $L_{\mol{CO}}/L_{\mol{TIR}}$ provides an upper limit to the FUV field strength. 
We calculate a grid of the ratio of $L_{\mol{CO} J = 3-2}/L_{\mol{TIR}}$ and $L_{\mol{CO} J = 6-5}/L_{\mol{TIR}}$ by estimating $L_{\mol{TIR}}$ as twice the FUV field strength \citep{kaufman1999}.  We show the resulting grid in the bottom panel of Figure \ref{pdrMiddle}, along with the measured ratio for the nucleus of M51.  We correct the measured \CO-to-TIR line ratios by multiplying the ratio by 2 to account for the fact that the measured \CO \ emission from the PDR model is only from one side of the PDR, while the TIR emission as estimated is assumed to originate from both sides of the PDR.
We show the measured $L_{\mol{CO} J = 3-2}/L_{\mol{TIR}}$ ratios for the remaining regions in Figure \ref{coldPDR}, and the $L_{\mol{CO}(6-5)}/L_{\mol{TIR}}$ ratios in Figure \ref{warmPDR}.

For the low-J \CO \ line ratio PDR model, the \CO \ line ratios
coupled with the ratio of $L_{\mol{CO} J = 3-2}/L_{\mol{TIR}}$
constrain the field strength to $G_0 < 10^{2}$ in all the modelled
regions, while the PDR model density agrees with the low end of the
range of density from our  non-LTE analysis ($n(\mol{H_2})
\sim 10^{3} - 10^{3.5} \unit{cm^{-3}}$, Figure \ref{coldPDR}). 
For the
high-J \CO \ line ratio PDR model, the \CO \ line ratios limit the
density to $n(\mol{H_2}) \gtrsim 10^{4} \unit{cm^{-3}}$, while the
ratio of  $L_{\mol{CO} J = 6-5}/L_{\mol{TIR}}$ limits the field
strength to $G_0 \lesssim 10^3$ for all the modelled regions (Figure
\ref{warmPDR}). 
 For the nucleus and centre regions, these densities are significantly
higher than the densities calculated from our non-LTE
analysis ($n(\mol{H_2}) \sim 10^{2.5} - 10^{3.5} \unit{cm^{-3}}$) and
so these PDR models do not provide a good fit to the data. For the
arm/interarm region, both low 
($n(\mol{H_2}) \sim 10^{2} \unit{cm^{-3}}$) and high
($n(\mol{H_2}) \sim 10^{4} \unit{cm^{-3}}$) density solutions exist;
tighter constraints on the density are required to distinguish between
the two solutions.

%The high density calculated from our radiative transfer analysis ($n(\mol{H_2}) \sim 10^{4.75} - 10^{6.5} \unit{cm^{-3}}$) constrains the field strength in the high-J \CO \ PDR to $G_0 \sim 10^{1} - 10^{2}$. 
% 
%In order to recover a smaller $\frac{\mol{CO} J =6-5}{TIR}$ line
%ratio consistent with the $\frac{\mol{CO} J =8-7}{\mol{CO} J =7-6}$ and $\frac{\mol{CO} J =8-7}{\mol{CO} J =6-5}$ line ratios and the density constraints, we would need only $\sim 1 \%$ of the TIR flux comes from our ``warm'' PDRs. Since only a very small fraction of the TIR flux needs to originate in the warm component, the effect on our ``cold'' PDR models of dividing the TIR flux between the cold and warm components is negligible.

% ** previous sentences were moved from discussion

%% file: Discussion.tex
\section{Discussion} \label{discSec}

\subsection{Physical state of the molecular gas}

\subsubsection{Regional variations}

The physical state of both the cold and warm molecular gas can vary
significantly from source to source, depending upon the environment in which the molecular gas resides. %Within an individual galaxy, the environment surrounding the molecular gas can vary. In Section \ref{radTransSect}, we investigated any differences in the physical state of the molecular gas for the nucleus, centre and arm/inter-arm regions of M51. 
The results for the cold and warm components for the nucleus, centre
and arm/inter-arm regions of M51 are shown in Table \ref{twoCompTable}
and Figure \ref{fourSquareCut}. For each of the components, no
significant differences between regions are seen among any of
the physical parameters. 
%across the 3 regions as all of the $1\sigma$ ranges overlap considerably. %This suggests that either our observations are not sensitive to the differences in the physical states of the molecular gas, or that there are no appreciable changes to the physical state across the observed region. %zz
The density of the cold molecular gas in all 3 regions
($n(\mol{H_2}) \sim 10^{4} \unit{cm^{-3}}$) is typical for
GMCs in the Milky Way (e.g. \citealt{tielens2005}), but is
  uncertain by three orders of magnitude (Table
\ref{twoCompTable}).  The temperature of the cold component in all
3 regions is very similar to  the typically quoted value of
$\sim 10 \unit{K}$ for Milky Way GMCs.
%; however there may be contamination from a warmer,
%diffuse component (see Section \ref{diffGas}). 
Given the $1\sigma$
ranges for the beam-averaged column densities in Table
\ref{twoCompTable}, 
%covers half an order of magnitude, while in the
%arm/inter-arm region, it spans an order of
%magnitude. Any 
any differences 
%in the amount of molecular gas contained within a beam 
would need to be greater than a factor of $\sim 5$ 
($\sim 10$ for the arm/inter-arm region) 
to be seen in our results. 
Similarly for the warm component, we detect no differences between the
temperature, density or beam-averaged column density across the 3
regions. 
%In the case of the temperature, the solutions suggest that 
The warm molecular gas has a temperature of $\sim 1000
\unit{K}$. 
In the nucleus and centre, the density of the warm component is
  quite low ($n(\mol{H_2}) \sim 10^{1.6}- 10^{2.5} \unit{cm^{-3}}$), while for the
  arm/inter-arm region the density is not very well constrained. 
%Unlike in previous results (e.g. \citealt{schirm2014}), the temperatures of the cold and warm component are not as distinct, which may suggest we are tracing two different density components of the molecular gas as opposed to two temperature components.

The similarities in the nucleus and centre of M51 are likely due to
contamination of the beam at the nucleus position by emission from the centre
region. \cite{parkin2013} used data with a 
$12''$ beam and defined the nucleus to be a square $\sim 12''$ across.
%modelled PDRs in M51 using transitions of $\mol{[CII]}$, $\mol{[OI]}$
%and $\mol{[NII]}$, along with the infrared flux. In their work, the
%nucleus consisted of a square $\sim 12''$ across with a beam size of
%$12''$. They found that the density recovered in their PDR models for
%the nucleus ($\sim 10^{3.5} - 10^{4.25}$) and centre ($\sim 10^{2.5}
%- 10^{4.0}$) overlapped, while the surface temperatures for the
%nucleus ($240 - 475 \unit{K}$) and centre ($170-680$) overlap
%considerably. 
Given the $40''$ beam of our observations, our
nucleus region will be somewhat contaminated by emission from the
centre region, and vice versa. 
%Overall, our results suggest that the beam-averaged properties of the
%molecular gas in M51 do not vary by more than factors of a few in the
%case of both the cold and warm component.
%
%The larger beam size coupled with uncertainties in our line measurements contributes to our ability to detect any differences, while the results from \cite{parkin2013} suggest that, if PDRs dominate the \CO \ emission, we would not expect to see significant variations down to a beam size of at least $\sim 12''$. %zz

%%The similarities between the centre and the arm/inter-arm region of M51 are not unexpected: in the PDR models of \cite{parkin2013}, the ranges for the density, field strength and surface temperature across the centre, arm and inter-arm regions overlapped significantly. There was less overlap between the nucleus and the remaining 3 regions modelled by \cite{parkin2013}.  The nucleus region in \cite{parkin2013} covers a square $\sim 12''$ centre on the nucleus (see Figure \ref{regionsMap}, left), while the beam size of their observations were $12''$. In our observations, our nucleus consists of a single beam with a size of $40''$. 

\subsubsection{Comparison to previous studies}

Both \cite{israel2006} and \cite{schinnerer2010} modelled various
ratios of \CO \ and $^{13}$\CO \ in M51. \cite{schinnerer2010} used a
non-LTE analysis to model ratios of $^{12}$\CO
\ $J=1-0$ and $J=2-1$, and $^{13}$\CO \ $J=1-0$ at multiple positions
of the western arm and southern regions of M51 at resolutions of
$2\farcs9$ and $4\farcs5$; however these regions lie beyond our FTS maps. They recovered cold ($14-20 \unit{K}$),
moderately dense ($n(\mol{H_2}) \sim 10^{2} -  10^{2.4}
\unit{cm^{-3}}$) gas.

\cite{israel2006} modelled $^{12}\mol{CO}$ $J=1-0$ to $J=4-3$ and $^{13}\mol{CO}$ $J=1-0$ to $J=3-2$ line ratios at two locations in M51: the centre and in a giant molecular association (GMA) offset from the nucleus $\Delta \alpha = -10'', \Delta \delta = +15''$. Given the small size of the offset and our large beam, both of these positions correspond to our nucleus. Using an LVG model, they fit two components to the \CO \ line ratios, assuming that $[^{12}\mol{CO}]/[^{13}\mol{CO}] = 40$. For the offset GMA, they find a warm ($\sim 100 \unit{K}$), relatively diffuse ($\sim 10^{2.0} \unit{cm^{-3}}$) component, and a warmer ($\sim 150 \unit{K}$), more dense ($\sim 10^{3.0} \unit{cm^{-3}}$) component. %It is important to note that the density of the more diffuse component is at the lower limit of their modelled density space ($10^{2} \unit{cm^{-3}} \le n(\mol{H_2}) \le 10^{5} \unit{cm^{-3}}$), while the temperature of their warm component is at the upper limit of the modelled temperature space ($10 \unit{K} \le T_{kin} \le 150 \unit{K}$). 
For the centre, they find a relatively warm ($\sim 100 - 1000
\unit{K}$), lower density ($\sim 10^{2.0} - 10^{3.0} \unit{cm^{-3}}$)
component, and a cooler ($\sim 20 - 60 \unit{K}$), higher density
($10^{3.0} - 10^{3.5} \unit{cm^{-3}}$) component. 
The cold component
from our two-component fit agrees with their centre results in
  both density and temperature, although our models do not
    constrain the density particularly well.
Our warm component
 fit also agrees with their centre results in 
  both density and temperature. 
%Although the temperature of our warm component agrees at the $1\sigma$ level
%with their warm component, our warm component density 
%is much larger.
Unlike \cite{israel2006}, we have observations of \CO \ beyond the $J=4-3$ transition and so are  able to place significantly tighter
    constraints on the temperature of the warm component.

\cite{brunner2008} probed the warm and hot molecular gas in M51 using
the mid-infrared \molH \ rotational transitions $S(0) - S(5)$ in a
strip. They found that the low-J \molH \ transitions ($S(0) - S(2)$)
trace warm ($\sim 100 - 300 \unit{K}$) molecular gas, 
%similar to our warm component. 
while the high-J \molH \ transitions ($S(2) - S(5)$)
trace hot molecular gas ($\sim 400 - 1000 \unit{K}$). 
The temperature ranges for the warm and hot H$_2$ gas agree quite
well with the warm component from our fit to the CO data.
% with similar temperatures to  the warm component from our CO fits. However,
%the mass of this hot component is small,  with a peak surface density 
%only $\sim 2 \%$ of the peak surface density of the warm \molH \ gas.
% In comparison, our fits to the CO emission suggest that the warm
% component contains $\sim 20\%$ of the total molecular gas mass.

\subsubsection{Diffuse molecular gas} \label{diffGas}

As part of the PAWS collaboration,
% to study the molecular gas in M51,
\cite{pety2013}  mapped 
%used a combination of the Plateau de Bure Interferometer (PdBI) and
%the Institut de Radioastronomy Millim\'etrique (IRAM) 30m telescope
%to map 
the \CO \ $J=1-0$ emission in M51
at arcsecond resolution. By combining the Plateau de Bure
Interferometer (PdBI) interferometric data with the IRAM 30m data, they were
able to correct for the ``missing flux'' from the interferometric
observations. They find that $\sim 50 $ percent of the \CO \ $J=1-0$ emission is from  molecular gas located in a thick, extended disk with a scale height $\sim 200 \unit{pc}$.  \cite{pety2013} argue that this emission originates from a warm ($\sim 50-100 \unit{K}$), diffuse ($\sim 100 - 500 \unit{cm^{-3}}$) molecular component.

\input{\TabPath/M51_compareColdRadexSolutions.tex}

%Some of the emission from the \CO \ $J=1-0$ and $J=2-1$ transitions may in fact be coming from this extended component. 
We investigate the possible effects of this extended \CO
\ emission on our two-component  non-LTE
model by fitting various combinations of the \CO \ and \CI
\ transitions (Table \ref{diffGasTable}). We begin by setting the \CO
\ $J=1-0$ flux to half of the measured flux and fitting it along with
the remaining \CO \ and \CI \ transitions as before
(\emph{half10}). Additionally, to investigate any contributions from
the warm, diffuse component to the \CO \ $J=2-1$ transition, we set
the \CO \ $J=2-1$ transition as an upper limit while fitting half of
the \CO \ $J=1-0$ flux (\emph{half10 2ul}). We compare these solutions
to our original two-component solutions (\emph{all lines}) 
for the centre region
%
%We report the kinetic temperature ($T_{kin}$), molecular gas density
%($n(\mol{H_2})$), beam-averaged column-density ($<N_{\mol{CO}}>$) and
%pressure ($P$) for the centre region from each of the models 
in Table \ref{diffGasTable}.  For the cold component, the $1\sigma$ ranges for the kinetic
temperatures do not shift significantly among the models,
 while the $1\sigma$ ranges for 
the molecular gas density narrow slightly for both the \emph{half10}
and \emph{half10 2ul} solutions compared to the \emph{all lines}
solution.
% ($n(\mol{H_2}) > 10^{3.72} \unit{cm^{-3}}$) lies at the upper limit of the range from the \emph{all lines} solution ($n(\mol{H_2}) < 10^{3.75} \unit{cm^{-3}}$). 
 However, the \emph{half10}
and \emph{half10 2ul} warm component solutions have significantly higher densities,
higher pressures, and lower beam-averaged column densities compared to
the \emph{all lines} solution.

%%The $1\sigma$ ranges for the kinetic temperatures shift to allow lower temperatures; however there is still significant overlap with the \emph{all} solution. The overlap in $1\sigma$ ranges for the \emph{all}, \emph{2up} and \emph{3up} solutions is not surprising: the \emph{2up} and \emph{3up} allow for many of the same solutions as the \emph{all} solution, while also allowing for solutions where not all of the \CO \ $J=1-0$ and/or \CO \ $J=2-1$ flux is recovered. In addition, the \emph{half10} and \emph{half10 2ul} solutions show many similarities, as expected. All of the solutions allow for higher densities than the initial solution (\emph{all}), although only for the \emph{half10} and \emph{half10 2ul} do the $1\sigma$ ranges not overlap with the original solution.

The  changes in the warm component density for both the \emph{half10} and \emph{half10
  2ul} solutions suggest that the diffuse molecular gas component
from \cite{pety2013} contributes to the \CO \ $J=1-0$ emission, and
possibly the \CO \ $J=2-1$ emission. However, determining the physical
characteristics of this component using a  non-LTE  model, such
as the one presented in Section \ref{radTransSect}, would be
difficult. The \CO \ $J=1-0$ emission from the diffuse, extended
component is subthermally excited and there would be even less
contribution from the diffuse component to higher $J$ \CO
\ transitions. While \cite{pety2013} use the ratio of
$\frac{^{12}\mol{CO}}{^{13}\mol{CO}}$ to argue for the existence of
this diffuse component, including only a single $^{13}\mol{CO}$
transition in a non-LTE  analysis requires assuming a relative abundance ratio of $^{12}$\CO \ and $^{13}$\CO \ in the diffuse component. 

The existence of an extended, diffuse component does not preclude the
diffuse molecular gas, at least in part, being contained within
GMCs. Diffuse GMCs exhibiting high ratios of
$\frac{^{12}\mol{CO}}{^{13}\mol{CO}}$ have been observed at high
latitudes within our own Galaxy \citep{blitz1984}. In this scenario, the limited sensitivity of the interferometer along with the unresolved nature of the brightest clumps of the diffuse GMCs would lead to the interferometer filtering out these diffuse GMCs \citep{pety2013}. 

%%The question remains as to where this molecular gas resides: while the ratio of $^{12}$\CO \ and $^{13}$\CO \ suggest that the molecular gas is diffuse, it can still reside within diffuse GMCs at larger scale heights, such as those found in the Milky Way \citethis. \cite{wilson1994} found a similar diffuse component in M33, a smaller, nearby spiral galaxy, by observing line ratios of $^{12}$\CO \ and $^{13}$\CO

The critical densities of the \CI \ $J=1-0$ and $J=2-1$ transitions are $n_{10} \sim 500 \unit{cm^{-3}}$ and $n_{21} \sim 10^{3} \unit{cm^{-3}}$, respectively \citep{papadopoulos2004}. These are comparable to the critical density of \CO \ $J=1-0$ ($n_{cr} \sim 1.1 \times 10^{3} \unit{cm^{-3}}$), which indicates that \CI \ could also, at least in part, be tracing a diffuse molecular component. High-resolution observations of \CI \ in galaxies like M51 coupled with single dish observations, as for \CO \ in \cite{pety2013}, may be useful in constraining the physical state of this diffuse gas. Combining such observations with interferometric observations of dense gas tracers, such as $\mol{HCN}$ and $\mol{HCO+}$ \ would allow us to discriminate between dense and diffuse GMCs. Finally, high-sensitivity, flux-recovered observations of a combination of these molecular gas-tracing species using the Atacama Large Millimeter Array (ALMA) would help differentiate between a truly extended component or a collection of diffuse GMCs. Unfortunately, M51 itself is not a viable target for ALMA due to its high declination.

\subsection{PDR modelling}\label{subsec-PDR}

Our comparison of the PDR and non-LTE models suggests that
the warm component of the molecular gas in the nucleus and centre of
M51 is unlikely to reside
primarily in simple PDRs. 
The LINER nucleus \citep{satyapal2004} may produce a small X-ray
  dominated region that could affect the excitation in the vicinity of
  the nucleus. On larger scales,
\citet{kazandjian2015} has shown that a relatively small
amount of mechanical heating from supernovae and stellar winds can have
significant effects on the temperature of the molecular
gas. The presence of active star formation
in M51 certainly suggests that supernovae will be present in the
disk. However, we cannot easily estimate the mechanical heating
due to supernovae and stellar winds for M51 as the necessary supernova
rate data do not exist. 
The strong spiral density wave in M51 may also create shocks that
  are themselves a potential source of mechanical heating.
We also note that our efforts to estimate the effect of the diffuse
molecular component proposed by \cite{pety2013} result in models with
higher warm component densities that {\it would} be consistent with
PDR models.
We therefore focus our discussion in this
section on the cold component of the molecular gas, for which the PDR
models are able to reproduce the CO data.

Both \cite{kramer2005} and \cite{parkin2013} previously modelled PDRs in M51. \cite{kramer2005} modelled various ratios of $\mol{[CII]}(158\unit{\mu m})$, $\mol{[OI]} (\unit{63\mu m})$, \CI \ $J=1-0$, \CO \ $J=1-0$ and \CO \ $J=3-2$ using the PDR models from \cite{kaufman1999}. They found that the best fit solution to their line ratios at all 3 pointings was consistent with density of $n(\mol{H_2}) \sim 4 \times 10^{4} \unit{cm^{-3}}$ and a field strength of $18 < G_0 < 32$ in an $80''$ beam. 
While their value of $G_0$ is comparable to the results for our
  cold component, their value for the density is an order of magnitude
  larger than our result.

\cite{parkin2013} used the $\mol{[CII]}/\mol{[OI]}63\unit{\mu m}$ and $(\mol{[CII]}+\mol{[OI]}63\unit{\mu m})/\mol{TIR}$ ratios along with the PDR models from \cite{kaufman1999} and \cite{kaufman2006} to constrain the density and field strength in the nucleus, centre, arm and inter-arm regions of M51 (see Figure \ref{regionsMap}). They corrected the $\mol{[CII]}$ emission for the ionized gas fraction and $\mol{[OI]}(63 \unit{\mu m})$ for orientation effects due to the plane-parallel slab-nature of the models (see Section 4.1 of \citealt{parkin2013} for details). Their results for the nucleus ($n(\mol{H_2}) = 10^{3.75} - 10^{4.0} \unit{cm^{-3}}$, $G_0 = 10^{3.25} - 10^{3.75}$) do not agree with  our ``cold'' PDR solution, falling above the range allowed by our \CO \ line ratios and our $\mol{CO}/\mol{TIR}$ ratios. 
Their density for the centre ($n(\mol{H_2}) = 10^{3.0} - 10^{3.25} \unit{cm^{-3}}$) agrees with our ``cold'' PDR solution density, while their field strength ($G_0 = 10^{2.75} - 10^{3.0}$) exceeds what is allowed by our ratio of $\mol{CO}/\mol{TIR}$ combined with various \CO \ line ratios (Figure \ref{coldPDR}). Their results for both the arm and inter-arm regions are similar: the ranges reported for the density ($n(\mol{H_2}) = 10^{2.75} - 10^{3.0} \unit{cm^{-3}}$)  agree with our results, while the FUV field strength ($G_0 = 10^{2.25} - 10^{2.5}$) is larger than our allowed solutions.
 
% *** MOVE TO PDR SECTION???
%In order to recover a $\frac{\mol{CO} J =6-5}{TIR}$ line ratio consistent with the $\frac{\mol{CO} J =8-7}{\mol{CO} J =7-6}$ and $\frac{\mol{CO} J =8-7}{\mol{CO} J =6-5}$ line ratios, we would require that only $\sim 1 \%$ of the TIR flux comes from our ``warm'' PDRs. If we assume the remaining TIR flux arises from our ``cold'' PDRs, our results for the ``cold'' PDR do not change appreciably. 
%{this paragraph needs to go somewhere more sensible}

Our ``cold'' PDR results suggest a density of $n(\mol{H_2} \sim 10^{3} \unit{cm^{-3}}$ and FUV field strength of $G_0 < 10^{2}$. In the \cite{kaufman1999} and \cite{kaufman2006} PDR models, this would correspond to $\frac{\mol{[CII]}}{\mol{[OI]}(63 \unit{\mu m})} \gtrsim 1.7$ and $\frac{\mol{[CII]} + \mol{[OI]}(63 \unit{\mu m})}{TIR} \gtrsim 1.3 \times 10^{-2}$. The average values of these ratios in each of the nucleus, centre, arm and inter-arm regions from \cite{parkin2013} vary between $\frac{\mol{[CII]}}{\mol{[OI]}(63 \unit{\mu m})} \sim 0.2 - 1.2 $ and $\frac{\mol{[CII]} + \mol{[OI]}(63 \unit{\mu m})}{TIR} \sim (5.0 - 8.1) \times 10^{-3}$. For both ratios, the values for our ``cold'' PDR solution are a factor of $\sim 1.5$ greater than the values measured in \cite{parkin2013}.
Smoothing the 
\cite{parkin2013} data to our $40''$ beam and remeasuring these 
ratios does not change the average values significantly.
%
%We investigate the effects of our larger beam on our PDR results by convolving both the $\mol{[CII]}$ and $\mol{[OI]}(63\unit{\mu m})$ maps from \cite{parkin2013} to our $40''$ gaussian beam, and re-gridding the images to our SPIRE-FTS maps. Using these new maps, we find that the ratios vary between $\frac{\mol{[CII]}}{\mol{[OI]}(63 \unit{\mu m})} \sim 0.6 - 1.3$ and $\frac{\mol{[CII]} + \mol{[OI]}(63 \unit{\mu m})}{TIR} \sim (5.6 - 8.4) \times 10^{-3}$ across the region observed with SPIRE-FTS. These ratios are consistent with those from \cite{parkin2013}, suggesting that the beam sizes are not the cause of the differences in solutions.
%
However, including
%The ranges for the measured ratios reported in \cite{parkin2013} do not account for 
the $\sim 30 $ percent calibration uncertainty of the PACS spectrometer\footnote{PACS ObserverÕs Manual is available for
  download from the ESA Herschel Science Centre.} leads to a range of
$\frac{\mol{[CII]} + \mol{[OI]}(63 \unit{\mu m})}{TIR} \sim (3.5 -
10.5 \times 10^{-3})$, which agrees within $50$ percent with
%$\frac{\mol{[CII]} + \mol{[OI]}(63 \unit{\mu m})}{TIR} \gtrsim 1.3
%\times 10^{-2}$, 
the range necessary to reproduce our results using the
\cite{kaufman2006} PDR models. 
%The PACS calibration uncertainty would
                               %have a greater effect on the ratio of
                               %$\frac{\mol{[CII]}}{\mol{[OI]}(63
                               %\unit{\mu m})}$, corresponding to an
                               %uncertainty of $\sim 40\%$ in the line
                               %ratio. 
Accounting for the calibration uncertainty leads to a range of $ \sim 0.4 - 1.8$ for the ratio of $\frac{\mol{[CII]}}{\mol{[OI]}(63 \unit{\mu m})}$, which agrees within uncertainties with the model value of $>1.7$.  

We attribute the remaining disagreement to the simplified geometries
assumed by the PDR models, which can lead to differences in the model
results \citep{roellig2007}. Both the model used in \cite{parkin2013}
and the model used here assume a semi-infinite slab of gas illuminated
by an FUV field. In reality, PDRs are clumpy media which are affected
by other physical processes concurrently. Mechanical heating, for
example, can have a significant effect on the atomic and ion line
ratios, %reported by \cite{parkin2013}, 
as well as on the overall shape of the \CO \ SLED \citep{kazandjian2015}. Combining the measurements from {\cite{parkin2013} and our measured \CO \ SLED, along with observations of other molecular gas tracers such as $\mol{HCN}$ and $\mol{HCO+}$ would be necessary to quantify the contributions of mechanical heating to the PDRs in M51.

%%In our ``cold'' PDR, the value of $G_0$ is largely bound by the ratio of $\frac{\mol{CO} \ J=3-2}{TIR}$. In order to recover a value of $G_0 \sim 2.75$, which is a typical value from \cite{parkin2013}, we would need a value of $\frac{\mol{CO} \ J=3-2}{TIR}$ an order of magnitude smaller than what we have measured (see Figure \ref{pdrMiddle}). It is unlikely that uncertainties in our measurements of TIR could account for more than a factor of 2, and so to account for the remaining differences, the \CO \ flux from PDRs would have to be $5-10$ times smaller than the total \CO \ flux. The remaining \CO \ emission would then arise from molecular gas which is either shielded completely from the incident FUV radiation, or in a region where no significant sources of FUV radiation is present. 

\subsection{Comparison to other VNGS galaxies} \label{compVNGS}

M51 is the 6th galaxy in the VNGS sample for which the analysis of the
SPIRE-FTS observations of \CO \ have been published, and is the first
normal, quiescent spiral galaxy from the sample.  Only a
single-component was fit to the \CO \ SLED for M83 \citep{wu2014},
while for NGC 1068 \citep{spinoglio2012}, the extended (star-forming
ring) and compact (circumnuclear disk) \CO \ emission were fit
separately, taking advantage of the varying beam size of the SLW and
SSW. In the case of Arp 220 \citep{rangwala2011}, M82
\citep{panuzzo2010, kamenetzky2012} and NGC 4038/39
\citep{schirm2014}, a two-component fit was performed for the \CO,
$^{13}$\CO (M82) and \CI (NGC 4038/39) emission. 
Arp 220, M82 and NGC 4038/39 are all examples of either an interaction or an ongoing merger. Arp 220 ($D = 77 \unit{Mpc}$, \citealt{scoville1997}) is an ultra luminous infrared galaxy and is an advanced merger between two galaxies. M82 ($D = 3.4 \unit{Mpc}$, \citealt{dalcanton2009}) is a starburst galaxy \citep{yun1993} whose increased star formation rate is due to a recent interaction with the nearby M81. Finally, NGC 4038/39 ($D = 22 \unit{Mpc}$, \citealt{schweizer2008}) is an ongoing merger between two gas rich spiral galaxies.

We compare our two-component fit for the centre region  of M51 to
those of Arp 220, M82 and NGC 4038/39 (Table \ref{compareTable} and
Figure \ref{compareFigure}). We also compare our results to the mean values for the two-component fits of
\cite{kamenetzky2014} for 17 galaxies, including the 5 galaxies from
the VNGS.
For NGC 4038/39, we distinguish between
the region where the two merging gas disks overlap (the ``overlap
region'') and the nucleus of NGC 4038 (hereafter NGC 4038). For 
 NGC 4038, the physical size corresponding to the beam of
the observations ($\sim 43''$) is $\sim 4.6 \unit{kpc}$. In
comparison, our analysis of M51 covers the central $\sim 5 \unit{kpc}$
($\sim 100''$). Note that for Arp 220, the \CO \ emission is
point-like within the FTS beam \citep{rangwala2011} and the actual
column density is reported in Table
\ref{compareTable}. (\citet{rangwala2011} only report best fit values
for the parameters for Arp 220 (rather than medians); however, the
probability distributions for the Arp 220 fits are quite narrow and so
the difference between median and best fit is likely small for this galaxy.)

\input{\TabPath/M51_compareVNGSGals.tex}

\begin{figure}
	\centering
	\includegraphics[width = \linewidth]{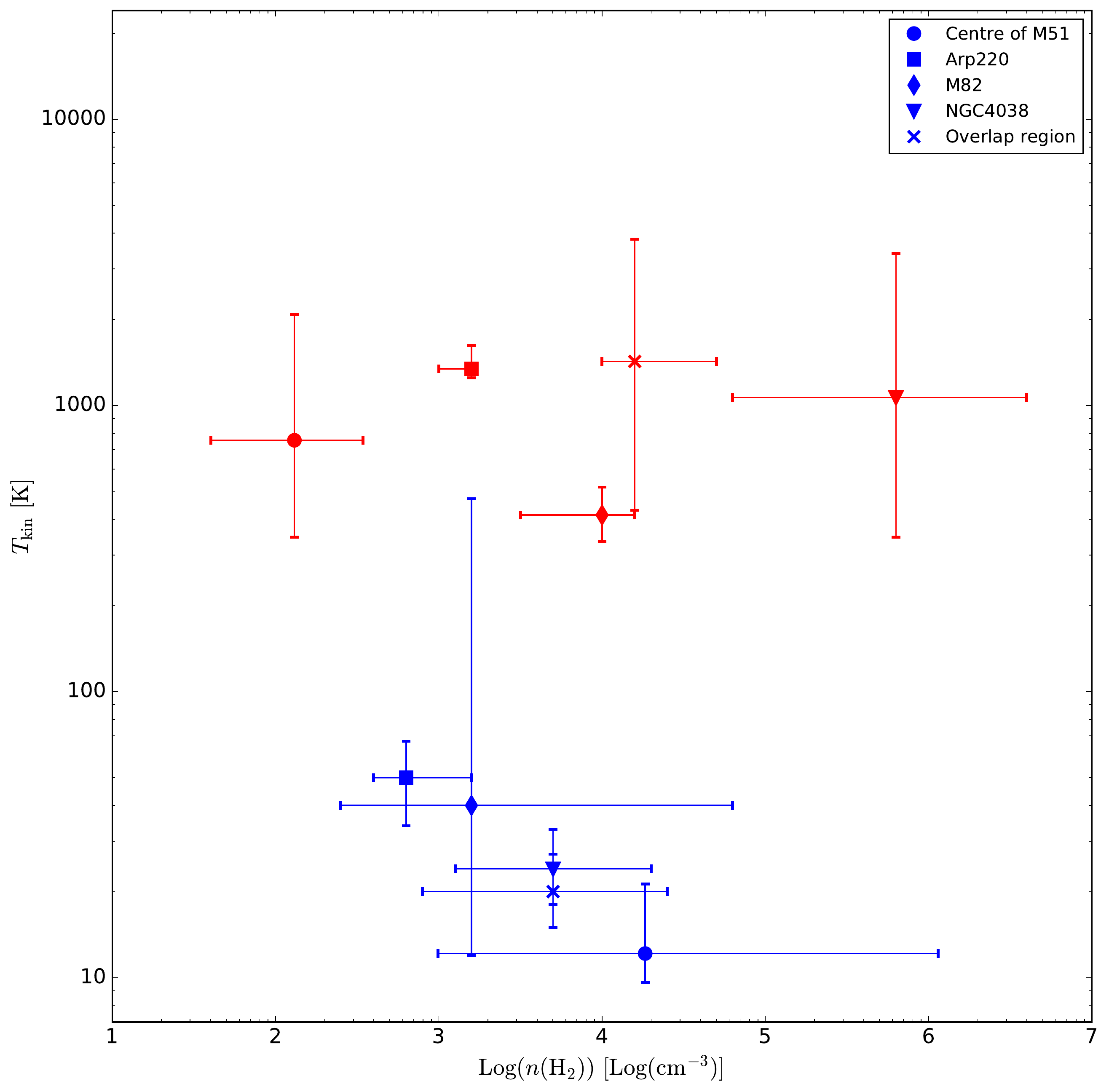}
	\caption[
% *** xxx new version of Figure 12 needed
Comparison of VNGS galaxies]{ Comparison of results 
           from the Very Nearby Galaxies Survey. The error
          bars represent the $1\sigma$ ranges reported in Table
          \ref{compareTable}  and the symbols are
          placed at the median of the distribution.   
%, while the data points correspond to where the error bars intersect. 
The values for the temperature (y-axis) and molecular gas density (x-axis) are the non-LTE excitation analysis results from this work (centre of M51), \cite{rangwala2011} (Arp 220), \cite{kamenetzky2012} (M82), and \cite{schirm2014} (NGC 4038 and the overlap region). The cold and warm components are represented by the blue and red data points, respectively. }  
	\label{compareFigure}
\end{figure}

The characteristics of the cold component for the centre region of M51
are very similar to the two regions in NGC 4038/39.  Aside from the large
temperature range for M82 ($T_{kin} = 12-472 \unit{K}$), the cold
molecular gas in M51 is also similar to that of M82. The temperature
of the cold molecular gas in Arp 220 ($T_{kin} = 34-67 \unit{K}$) is
 above the temperature range we adopted in fitting the cold
  component in M51.
 The cold
component fits to the 17 galaxies from \cite{kamenetzky2014} cover a
similar range in temperature ($T_{kin} = 12-250 \unit{K}$).

The warm component of M51 shows more differences than the cold
component when compared to the other galaxies, with a lower
 density ($n(\mol{H_2}) = 10^{1.6} - 10^{2.5} \unit{cm^{-3}}$) than any
of the other 3 systems 
($10^{3} < n(\mol{H_2}) < 10^{6.6} \unit{cm^{-3}}$).
%Furthermore, the density of the molecular gas ($n(\mol{H_2}) > 10^{4.87} \unit{cm^{-3}}$) is higher than in M82 ($n(\mol{H_2}) < 10^{3.20} \unit{cm^{-3}}$), Arp 220 ($n(\mol{H_2}) < 10^{4.20} \unit{cm^{-3}}$) or the overlap region ($n(\mol{H_2}) < 10^{4.70} \unit{cm^{-3}}$), while it is comparable to NGC 4038.  
The warm
component fits from \cite{kamenetzky2014} have  a similar
temperature  range to M51, but densities 
($10^{2.3} < n(\mol{H_2}) < 10^{4.9} \unit{cm^{-3}}$)
that only just overlap the M51 solution. 
%, again quite different from what we find for M51.

The molecular gas mass is proportional to the beam size multiplied by
the beam-averaged column density. Since the beam-size in any one
system will be the same for the cold and warm components, we can
calculate the warm gas mass as a fraction of the cold gas mass using
the beam-averaged column densities.  In M51, the 1$\sigma$ range for
the warm gas mass is 7-65 percent (mean 20 percent)
of the cold gas mass for the nucleus and centre regions, and 0.2-30 percent
(mean 3 percent) 
for the arm/interarm region.
Interestingly, these fractions are comparable to the warm gas mass
fractions of $\sim 10$ percent in Arp
220 \citep{rangwala2011} and $\sim 3$ percent in M82 \citep{kamenetzky2012} but are significantly larger
than the warm gas mass fraction of $\sim 0.1 $ percent in NGC 4038 and the
overlap region \citep{schirm2014}.  The average ratio of warm to
cold gas mass in the \cite{kamenetzky2014} sample is 12 percent, with a
range from 0.4 percent (for NGC~4038) to 40 percent. 

The global star formation rates in Arp 220 ($\sim 200 \unit{M_\odot
  \ yr^{-1}}$, \citealt{barcos-munoz2015}) and M82 ($\sim 10
\unit{M_\odot \ yr^{-1}}$, \citealt{yun1993}) are larger than the
global star formation rates in M51 ($\sim 2.6 \unit{M_\odot
  \ yr^{-1}}$  \citealt{schuster2007})  and NGC 4038/39 ($\gtrsim 5 \unit{M_\odot \ yr^{-1}}$, \citealt{stanford1990}) by factors of $2-80$.
In addition, the gas and star formation rate surface
  densities are significantly higher in Arp 220
  \citep{rangwala2011} and M82 \citep{kamenetzky2012}, where
  the active regions are roughly 1 kpc in size, than in M51 and the
  Antennae, where the activity is spread over the entire disk.
%Arp 220 SFR $\sim 200 \unit{M_\odot \ yr^{-1}}$ \citep{barcos-munoz2015}
%M82 $\sim 10 \unit{M_\odot \ yr^{-1}}$  \citep{yun1993}
%NGC 4038/39 $\grtsim 5 \unit{M_\odot \ yr^{-1}}$ \citep{stanford1990}
%M51 $\sim 2.6 \unit{M_\odot \ yr^{-1}}$  \citep{schuster2007}
The primary heating source for the warm molecular gas in Arp 220
\citep{rangwala2011} and M82 \citep{kamenetzky2012} was found to be
mechanical heating due primarily to supernova and stellar winds
\citep{maloney1999}.
%, while it is unlikely that this warm gas resides primarily in PDRs. 
The higher warm gas mass fraction in Arp 220 and M82 compared to NGC 4038/39 was attributed to an increase in the efficiency by which energy from supernova and stellar winds is injected as thermal energy into the molecular gas \citep{schirm2014}. 
 It is unclear whether stellar feedback can also explain
  the high warm gas mass fraction in M51.
Perhaps shocks produced by the strong spiral density wave in M51
  can provide
  an efficient source of turbulent energy.

 The overlap region of  NGC 4038/39 provides us with a comparison
of the effects of an early stage major merger to the less pronounced
interaction seen in M51. The most striking differences are in the
density and pressure of the warm molecular gas, which are both roughly
two orders of magnitude smaller in M51 compared to the overlap region
of NGC 4038/9. 
%where the temperature in the centre region of M51 ($T_{kin} \sim 44 - 183 \unit{K}$) is $2-7$ times less than the lower limit of the temperature in NGC 4038 ($T_{kin} \gtrsim 350 \unit{K}$). 
%In NGC 4038 and the overlap region, \cite{schirm2014} were unable to
%rule out PDRs as a possible source of heating, while evidence
%suggests the supernova and stellar winds contribute significantly to
%the heating. 
In NGC 4038/39, supernova and stellar winds were found to be
sufficient to heat the warm molecular gas \citep{schirm2014}. 
For the densest gas,  there  is also evidence    that  PDRs  with  at
  least 5 percent mechanical heating  contribute to
  the overall heating budget 
\citep{schirm2016}. In M51, there is strong evidence to suggest that
PDRs are fundamental to the molecular gas heating (e.g. see
\citealt{roussel2007} and \citealt{parkin2013}),  although PDR
models may not be sufficient to explain the heating of the warm
component (\S\ref{subsec-PDR}). Both PDRs and mechanical heating, however, are tied to the star formation rate as both require the formation of O and B stars which are relatively short lived. 
A relatively small amount of mechanical heating compared to
heating due to PDRs ($\sim 1$ percent) is able to produce a significant
effect on the temperature of the molecular gas
(e.g. \citealt{kazandjian2012}). In NGC 4038/39, the turbulent motion
due to both the ongoing merger and stellar feedback 
% supernova   and stellar winds 
should exceed this minimum threshold \citep{schirm2014}.  
%The mechanical
                                %heating due to supernova and stellar
                                %winds has not been calculated for M51
                                %as suitable supernova rate data do
                                %not exist. However, we would expect
                                %the contributions of both turbulent
                                %motion and star formation feedback
                                %would have a larger effect on the
                                %overall heating of the molecular gas
                                %in NGC 4038/39 than in M51. 
 It is possible that similar effects are also at work in M51,
  which is itself an interacting system.
Using more sophisticated PDR models which include contributions from mechanical heating, along with dense gas tracers such as $\mol{HCN}$ and $\mol{HCO+}$, may allow us to calculate the contributions to the total gas heating from PDRs and mechanical heating in M51, NGC 4038/39, and other systems.

%% file: M51_compareColdRadexSolutions.tex
\begin{table*} 
\centering 
\begin{minipage}{\linewidth} 
\caption{Including diffuse gas in radiative transfer solutions for the centre region} 
\begin{tabular}{@{}l c c c c l} 
\hline 
& &  \multicolumn{4}{c}{Parameter} \\ 
& & \multicolumn{4}{c}{Median ($1\sigma$ range)} \\ 
& &$\mol{Log}(n(\mol{H_2}))$ &$\mol{Log}(T_{\mol{kin}})$ &$\mol{Log}(<N_{\mol{CO}}>)$ &$\mol{Log}(P)$ \\ 
Component & Solution &$\unit{[\mol{Log}(cm^{-3})]}$ &$\unit{[\mol{Log}(K)]}$ &$\unit{[\mol{Log}(cm^{-2})]}$ &$\unit{[\mol{Log}(K \ cm^{-2})]}$ \\ 
\hline 
Cold & \emph{all lines}& $ 4.3 \ (3.0 - 6.1) $ & $ 1.1 \ (1.0 - 1.3) $ & $ 17.2 \ (16.8 - 17.6) $ & $ 5.3 \ (4.3 - 7.1) $  \\ 
& \emph{half10}\footnotemark[1]& $ 3.9 \ (3.1 - 5.4) $ & $ 1.2 \ (1.0 - 1.4) $ & $ 16.6 \ (16.3 - 17.3) $ & $ 5.2 \ (4.3 - 6.5) $  \\ 
& \emph{half10 2ul}\footnotemark[2]& $ 4.0 \ (3.1 - 5.7) $ & $ 1.2 \ (1.0 - 1.4) $ & $ 17.0 \ (16.5 - 17.6) $ & $ 5.2 \ (4.4 - 6.8) $  \\ 
\hline
Warm & \emph{all lines}& $ 2.1 \ (1.6 - 2.5) $ & $ 2.9 \ (2.5 - 3.3) $ & $ 16.5 \ (16.3 - 16.9) $ & $ 5.1 \ (4.6 - 5.3) $  \\ 
& \emph{half10}\footnotemark[1]& $ 3.2 \ (2.8 - 4.0) $ & $ 2.4 \ (2.1 - 2.9) $ & $ 15.9 \ (15.3 - 16.1) $ & $ 5.7 \ (5.5 - 6.3) $  \\ 
& \emph{half10 2ul}\footnotemark[2]& $ 3.2 \ (2.7 - 4.4) $ & $ 2.5 \ (2.0 - 3.1) $ & $ 15.7 \ (15.0 - 16.1) $ & $ 5.8 \ (5.5 - 6.8) $  \\ 

 \hline\label{diffGasTable}\end{tabular} 

 \medskip 
\footnotemark[1] \CO \ $J=1-0$ set to half its observed value; see text.\hfill\break
\footnotemark[2] \CO \ $J=1-0$ set to half its observed value and \CO
\ $J=2-1$ treated as an upper limit; see text.
\end{minipage} 
\end{table*}

%% file: M51_compareVNGSGals.tex
\begin{table*} 
\centering 
\begin{minipage}{\linewidth} 
\caption{Comparing radiative transfer solutions} 
\begin{tabular}{@{}l l c c c c l} 
\hline 
 &  & \multicolumn{4}{c}{Parameter} & \\ 
 & & \multicolumn{4}{c}{Median ($1\sigma$ range)} & Reference \\ 
& &$\mol{Log}(n(\mol{H_2}))$& $T_{\mol{kin}}$& $\mol{Log}(<N_{\mol{CO}}>)$& $\mol{Log}(P)$ \\ 
 Comp. & Source &$\unit{[\mol{Log}(cm^{-3})]}$ & $\unit{[K]}$ & $\unit{[\mol{Log}(cm^{-2})]}$ & $\unit{[\mol{Log}(K \ cm^{-2})]}$ \\ 
\hline 
Cold & Centre of M51& $ 4.3 \ (3.0 - 6.1) $ & $ 12 \ (10 - 21) $ & $ 17.2 \ (16.8 - 17.6) $ & $ 5.3 \ (4.3 - 7.1) $ & This work\\ 
&Arp220\footnotemark[1] & $ 2.8 \ (2.6 - 3.2) $ & $ 50 \ (34 - 67) $ & $ 20.3 \ (19.9 - 20.3)$ & $ 4.5 \ (4.5 - 4.8) $ & \cite{rangwala2011}\\ 
&M82& $ 3.2 \ (2.4 - 4.8) $ & $ 40 \ (12 - 472) $ & $ 18.2 \ (17.6 - 18.8) $ & $ 5.1 \ (4.6 - 5.8) $ & \cite{kamenetzky2012}\\ 
&NGC4038& $ 3.7 \ (3.1 - 4.3) $ & $ 24 \ (18 - 33) $ & $ 17.1 \ (16.6 - 17.6) $ & $ 5.0 \ (4.5 - 5.6) $ & \cite{schirm2014}\\ 
&Overlap region& $ 3.7 \ (2.9 - 4.4) $ & $ 20 \ (15 - 27) $ & $ 17.4 \ (16.8 - 17.9) $ & $ 4.9 \ (4.3 - 5.6) $ & \cite{schirm2014}\\ 
\hline 
 Warm & Centre of M51& $ 2.1 \ (1.6 - 2.5) $ & $ 755 \ (347 - 2089) $ & $ 16.5 \ (16.3 - 16.9) $ & $ 5.1 \ (4.6 - 5.3) $ & This work\\ 
&Arp220\footnotemark[1] & $ 3.2 \ (3.0 - 3.2) $ & $ 1343 \ (1247 - 1624) $ & $ 19.4 \ (19.4 - 19.5)$  & $ 6.3 \ (6.2 - 6.4) $ & \cite{rangwala2011}\\ 
&M82& $ 4.0 \ (3.5 - 4.2) $ & $ 414 \ (335 - 518) $ & $ 16.7 \ (16.4 - 17.2) $ & $ 6.6 \ (6.2 - 6.8) $ & \cite{kamenetzky2012}\\ 
&NGC4038& $ 5.8 \ (4.8 - 6.6) $ & $ 1065 \ (347 - 3397) $ & $ 14.4 \ (14.2 - 14.8) $ & $ 9.3 \ (8.2 - 10.2) $ & \cite{schirm2014}\\ 
&Overlap region& $ 4.2 \ (4.0 - 4.7) $ & $ 1425 \ (430 - 3811) $ & $ 14.6 \ (14.4 - 14.7) $ & $ 7.4 \ (7.3 - 7.7) $ & \cite{schirm2014}\\ 

 \hline\label{compareTable}\end{tabular} 

 \medskip 
\footnotemark[1] ``Best fit'' values and actual $N_{\mol{CO}}$
reported for Arp 220; see text.\end{minipage} 
\end{table*}

%% file: Conclusions.tex
\section{Summary and conclusions} \label{conc}

In this paper, we have presented intermediate-sampled SPIRE-FTS observations of \CO \ from $J=4-3$ to $J=8-7$ and both \CI \ transitions of the central region of M51. We supplemented these observations with ground-based observations of \CO \ $J=1-0$ to $J=3-2$. We separate M51 into 3 regions, the nucleus, centre and arm/inter-arm regions, by performing an unweighted average of the emission for each pixel contained with a region. We also combine all the pixels within the three into a single ``All'' region.

\begin{enumerate}
	\item Using the non-LTE excitation code RADEX along with a
          Bayesian likelihood code, we perform a two-component fit to
          the \CO \ and \CI \ emission in the nucleus, centre and
          arm/inter-arm regions of M51, along with all three regions
          combined. We find that the results do not vary beyond
          $1\sigma$ for all three regions. The results for the nucleus
          and centre regions of M51 consist of a cold component ($T_{kin} \sim
          10-20 \unit{K}$) with a moderate but poorly
          constrained density ($n(\mol{H_2}) = 10^{3} - 10^{6}$) and a
          warm, somewhat lower
          density component ($T_{kin} \sim 300-3000 \unit{K}$,
          $n(\mol{H_2}) = 10^{1.6} - 
          10^{2.5}$). The results for the arm/inter-arm region, and
          for all regions combined are not as well constrained. The
          warm gas mass fraction for the centre of M51 is  $\sim
            20$ percent.

	\item 
\citet{pety2013} argue that $\sim 50$ percent of the CO J=1-0
  emission in M51 arises from  warm diffuse gas.
We investigate the possible effect of this extended
          component by re-running our models with
          the  \CO \ $J=1-0$ flux set to half of the measured flux in our
          non-LTE analysis. The density range of the cold
          molecular gas narrows somewhat in these models, 
          while the warm component
          shifts to significantly higher density and
          pressure. This analysis  is consistent with the
          remaining  \CO \ $J=1-0$ emission originating from a more diffuse, possibly
          extended component of the molecular gas. We suggest that
          this diffuse molecular gas may still arise from
          GMCs. High-resolution, high-sensitivity, flux-recovered
          observations of multiple molecular gas-tracing species, such
          as \CI, $\mol{HCN}$ and $\mol{HCO+}$ along with \CO \ would
          allow us to distinguish between dense and diffuse GMCs.  
	
	\item We compare line ratios of \CO \ along with the
          $\frac{\mol{CO} J = 3-2}{TIR}$ and $\frac{\mol{CO} J =
          6-5}{TIR}$ ratios to a PDR model. Using the densities calculated
          from our non-LTE excitation analysis, our PDR modelling
          suggest a density of $n(\mol{H_2}) \sim 10^3 \unit{cm^{-3}}$
          and a field strength $G_0 < 10^2$  for the cold
          component. Although the warm component in the
            arm/interarm regions is consistent with PDR heating,
            additional heating sources beyond PDRs seem to be required
            for the nucleus and centre regions.
We compare our results to
          previous results \citep{parkin2013} which used various
          atomic line ratios and the total infrared flux to model the
          FUV field strength and gas density, and find that their FUV
          field strength ($G_0 > 10^{2.25}$) is greater than what is
          allowed by our models. We attribute the differences to
          calibration uncertainties in the atomic line ratios used by
          \cite{parkin2013} and  the simple geometry assumed by the PDR models.
	
	\item We compare our two-component model for %the \CO \ SLED of
          the centre region of M51 to similar models %non-LTE analysis results for
          for the ULIRG Arp 220, the starburst galaxy M82 and the on-going
          merger NGC 4038/39. The characteristics of the cold
          component are comparable across all 4 systems, with the
          exception of the temperature of Arp 220, which is slightly
          higher. In the case of the warm component, the density
          in the centre region of M51 ($n(\mol{H_2}) \sim 10^{1.6} - 10^{2.5}$)
          is lower than the other 3 systems ($n(\mol{H_2}) \sim 10^3 -
          10^{6.6}$), while the temperatures are comparable. Interestingly,
          the warm gas mass fraction in M51 is higher than in the NGC
          4038/9 merger and comparable to what is seen in Arp 220 and M82.
We suggest that a more complete multi-phase analysis of the
molecular gas in M51 including both PDR models and mechanical heating
would increase our understanding of this iconic system.

\end{enumerate}

%% file: appendix.tex
\appendix 

\begin{appendices}

\section{Single-component fit} \label{singCompSec}

\begin{figure}
	\centering
	\includegraphics[width = \linewidth]{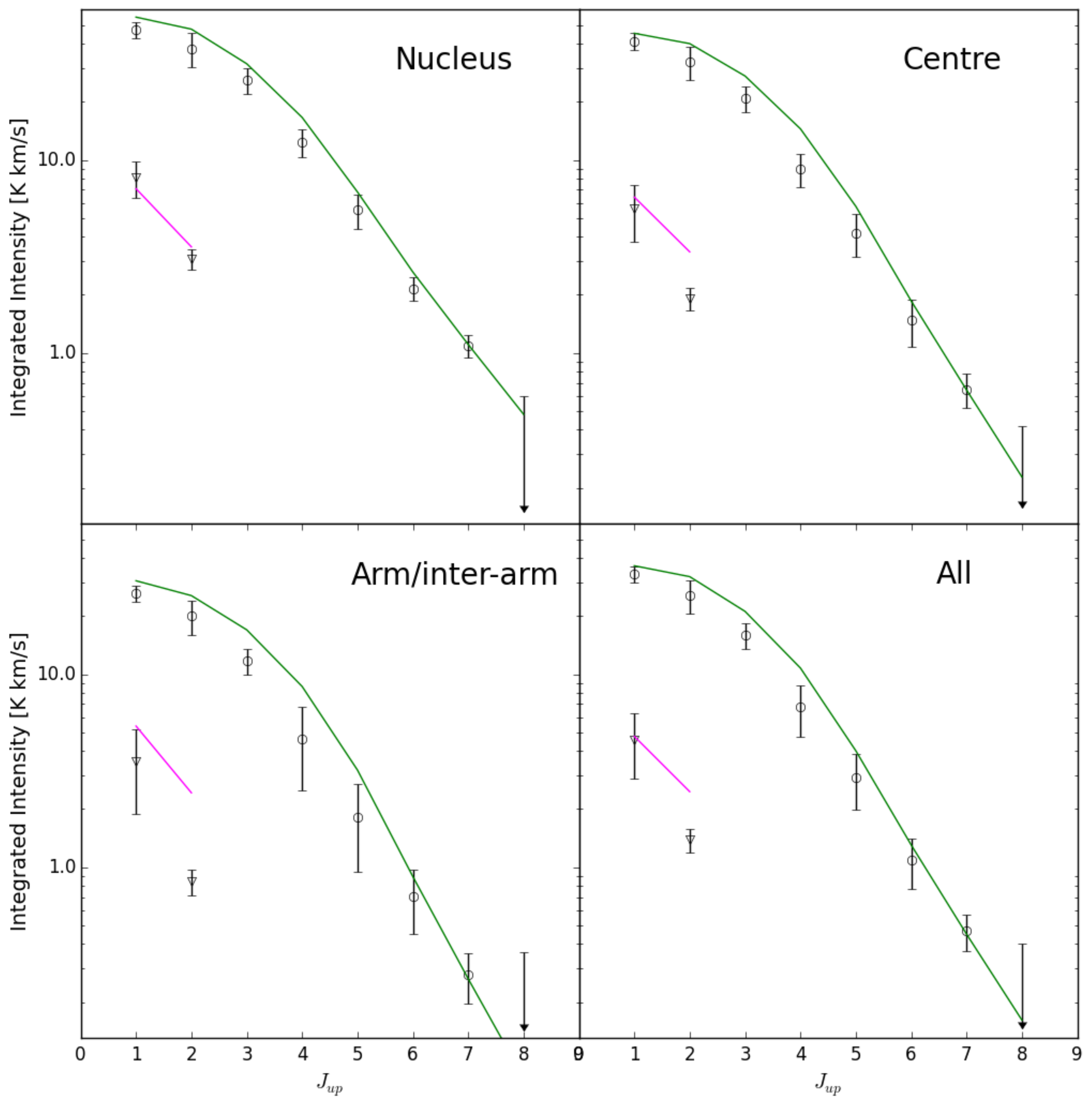}
	\caption[]{Measured and calculated spectral line energy distributions for the single component fit for the nucleus (top-left), centre (top-right), and arm/inter-arm (bottom-left) regions, and for all the regions combined (bottom right). The measured fluxes are shown by black circles (\CO) and triangles (\CI), while the calculated fluxes are shown by the green (\CO) and magenta (\CI) solid lines. }
	\label{SLEDSingle}
\end{figure}

We fit a single-component to the 7 detected \CO \ transitions ($J=1-0$
to $J=7-6$), and both \CI \ transitions, setting the \CO \ $J=8-7$
transition as an upper limit, in the nucleus, centre and arm/inter-arm
regions of M51, along with all 3 regions combined.  We show the
calculated fluxes from the ``best fit'' solution compared to the
measured fluxes in Figure \ref{SLEDSingle}.  We report the best-fit
solution for each of the physical parameters in Table \ref{oneCompTable}, along with the $1\sigma$ ranges for each parameter.

For all of the modelled regions, our single-component fits suggest
that the \CO \ emission is dominated by warm ($\sim
100-300 \unit{K}$), relatively diffuse ($\lesssim 10^{2.4} -
10^{2.8} \unit{cm^{-3}}$) molecular gas. While diffuse, \CO \ emitting
molecular gas has been observed within the Milky Way
(e.g. see \citealt{pety2008} and \citealt{liszt2009}), most of the
star-forming molecular gas is much colder than
$200 \unit{K}$. Furthermore, GMC scale observations of \CO \ $J=1-0$
in M51 by \cite{hughes2013} find that the \CO \ peak brightness
temperature ranges from $T_{mb} = 1-10 \unit{K}$ on spatial scales of
$\sim 50 \unit{pc}$ ($\sim 1''$). Assuming the \CO \ emission fills
the beam, the \CO \ peak brightness temperature corresponds to the
molecular gas temperature. If the molecular gas were at a temperature
of $\sim 200 \unit{K}$, as recovered in our single-component model,
only $\sim 0.5 - 5 $ percent of the $1''$ beam would be filled. Given the typical sizes of GMCs ($10 - 100 \unit{pc}$), our single-component model does not appear likely to represent accurately the bulk of the molecular gas in M51.

%We find that the single component fit recovers a warm ($T_{kin} \sim 100 - 300 \unit{K}$), low density ($n(\mol{H_2}) \sim 10^{2.4} - 10^{2.8} \unit{cm^{-3}}$) gas. In addition, the results for all four region are very similar; the $1\sigma$ ranges for all the physical parameters overlap considerably for all four regions (Figure \ref{fourSquareSingle}), suggesting that the state of the molecular gas is similar in all four regions. 

\input{\TabPath/M51\oneComp_singleRadexTable.tex}

%\begin{figure*}
%	\centering
%	\includegraphics[width = \linewidth]{\FigPath/SLED/OneCompRegion/oneComp_regionTau\oneComp.png}
%	\caption[]{}
%	\label{TauSingle}
%\end{figure*}

\section{Parameter distributions} \label{ParamDistns}

\begin{figure}
	\centering
	\includegraphics[width = \linewidth]{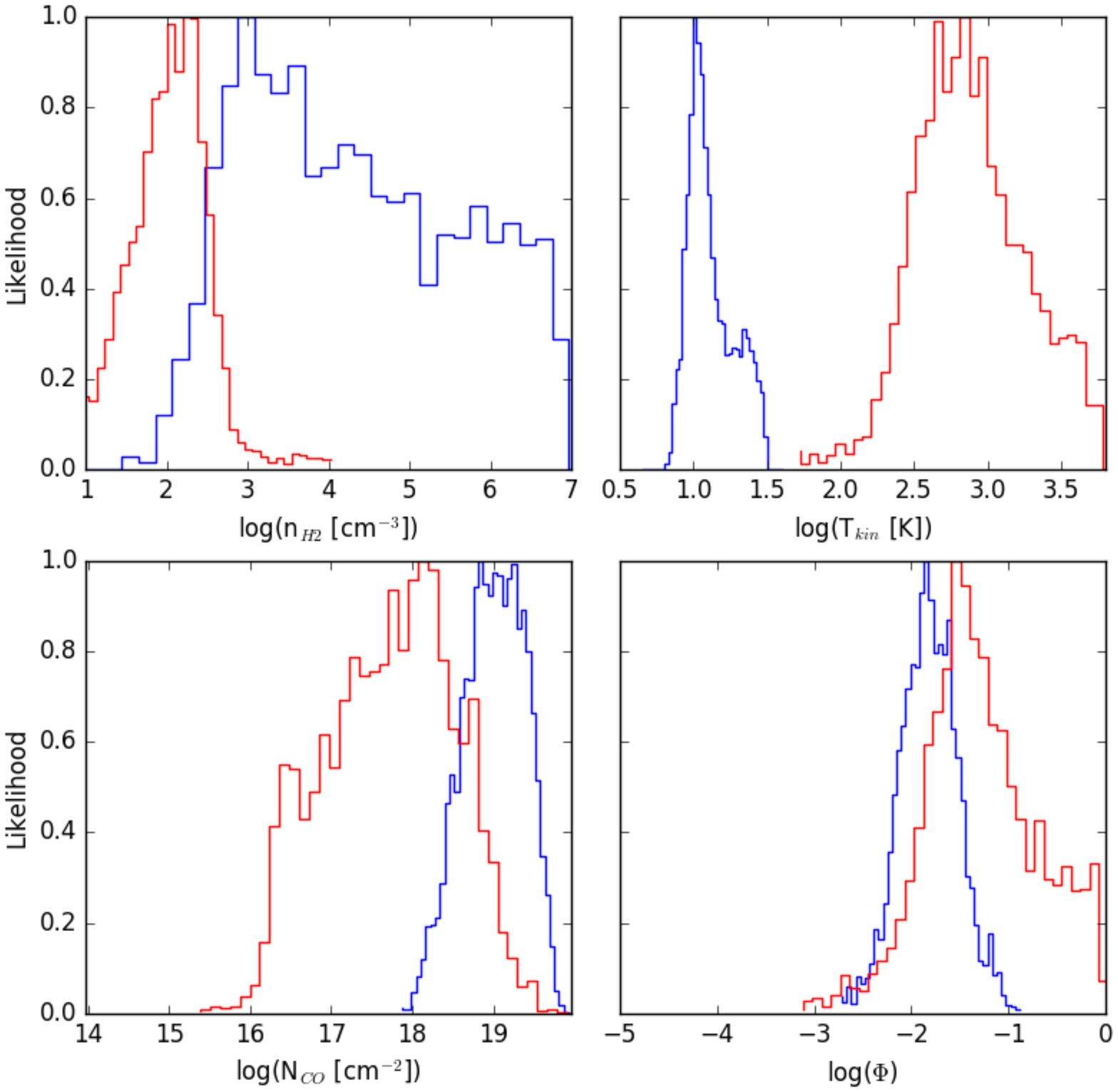}
	\caption[]{Probability distributions of 4
	fitted parameters from the two-component RADEX model for the
	centre region of M51. The cold component distributions are
	shown in blue and the warm component distributions are shown
	in red for molecular
	gas density (\emph{top left}), kinetic temperature (\emph{top right}),
	column density (\emph{bottom left}),  and filling factor
	(\emph{bottom right}).}  
	\label{fourDistnsCentre}
\end{figure}

Figure \ref{fourDistnsCentre} shows the probability
distributions from the two-component Bayesian RADEX
model fits to the centre region of M51. The distributions for the cold
component fit are shown in blue while the distributions for the warm
component fit are shown in red. Many of the distributions are
decidedly asymmetric. In particular, the density distribution for the
cold component fit is very broad (see also
Table  \ref{twoCompTable}). The probability distributions for the fits
to the nucleus and arm/interarm regions as well as to teh average of
all the data 
are very similar to the distributions for the centre region
shown in Figure \ref{fourDistnsCentre}.

\end{appendices}

%% file: M51_all7down_CI_singleRadexTable.tex
\begin{table*} 
\centering 
\begin{minipage}{\linewidth} 
\caption{Results from single-component non-LTE excitation analysis} 
\begin{tabular}{@{}l l l l l  r} 
\hline 
 & \multicolumn{4}{c}{Best fit ($1\sigma$ range)} & \\ 
 Parameter & Nucleus & Centre & Arm/inter-arm & All & Units \\ 
\hline 
 $T_{\mol{kin}}$& $ 234 \ (162 - 320) $ & $ 141 \ (112 - 256) $ & $ 104 \ (90 - 236) $ & $ 141 \ (110 - 277) $ &$\unit{[K]}$\\ 
$\mol{Log}(n(\mol{H_2}))$& $ 2.37 \ (2.36 - 2.74) $ & $ 2.54 \ (2.41 - 2.82) $ & $ 2.46 \ (2.38 - 2.84) $ & $ 2.54 \ (2.36 - 2.82) $ &$\unit{[\mol{Log}(cm^{-3})]}$\\ 
$\mol{Log}(N_{\mol{CO}})$& $ 18.68 \ (17.85 - 18.64) $ & $ 18.70 \ (17.69 - 18.67) $ & $ 18.74 \ (17.05 - 18.53) $ & $ 18.56 \ (17.36 - 18.57) $ &$\unit{[\mol{Log}(cm^{-2})]}$\\ 
$\mol{Log}(\Phi_A)$& $ -1.57 \ (-1.72 - -1.26) $ & $ -1.64 \ (-1.74 - -1.18) $ & $ -1.71 \ (-1.76 - -0.82) $ & $ -1.64 \ (-1.74 - -0.99) $ &$\unit{[...]}$\\ 
$\mol{Log}(<N_{\mol{CO}}>)$& $ 16.92 \ (16.57 - 17.02) $ & $ 17.00 \ (16.49 - 17.01) $ & $ 16.31 \ (16.21 - 16.83) $ & $ 16.51 \ (16.35 - 16.91) $ &$\unit{[\mol{Log}(cm^{-2})]}$\\ 
$\mol{Log}(P)$& $ 5.09 \ (4.71 - 5.13) $ & $ 5.09 \ (4.61 - 5.08) $ & $ 4.96 \ (4.52 - 5.01) $ & $ 5.09 \ (4.58 - 5.08) $ &$\unit{[\mol{Log}(K \ cm^{-2})]}$\\ 

 \hline\end{tabular} 
\label{oneCompTable}\end{minipage} 
\end{table*}

%% file: M51_Schirm.bbl
\begin{thebibliography}{78}
\expandafter\ifx\csname natexlab\endcsname\relax\def\natexlab#1{#1}\fi

\bibitem[{{Aniano} {et~al}\mbox{.}(2011){Aniano}, {Draine}, {Gordon}, \&
  {Sandstrom}}]{aniano2011}
{Aniano} G., {Draine} B.~T., {Gordon} K.~D., {Sandstrom} K., 2011, PASP, 123,
  1218

\bibitem[{{Barcos-Mu{\~n}oz} {et~al}\mbox{.}(2015){Barcos-Mu{\~n}oz}, {Leroy},
  {Evans}, {Privon}, {Armus}, {Condon}, {Mazzarella}, {Meier}, {Momjian},
  {Murphy}, {Ott}, {Reichardt}, {Sakamoto}, {Sanders}, {Schinnerer},
  {Stierwalt}, {Surace}, {Thompson}, \& {Walter}}]{barcos-munoz2015}
{Barcos-Mu{\~n}oz} L. {et~al.}, 2015, \apj, 799, 10

\bibitem[{{Bendo}, {Galliano} \& {Madden}(2012){Bendo}, {Galliano}, \&
  {Madden}}]{bendo2012}
{Bendo} G.~J., {Galliano} F., {Madden} S.~C., 2012, \mnras, 423, 197

\bibitem[{{Blitz}, {Magnani} \& {Mundy}(1984){Blitz}, {Magnani}, \&
  {Mundy}}]{blitz1984}
{Blitz} L., {Magnani} L., {Mundy} L., 1984, \apjl, 282, L9

\bibitem[{{Brunner} {et~al}\mbox{.}(2008){Brunner}, {Sheth}, {Armus},
  {Wolfire}, {Vogel}, {Schinnerer}, {Helou}, {Dufour}, {Smith}, \&
  {Dale}}]{brunner2008}
{Brunner} G. {et~al.}, 2008, \apj, 675, 316

\bibitem[{{Calzetti} {et~al}\mbox{.}(2010){Calzetti}, {Wu}, {Hong},
  {Kennicutt}, {Lee}, {Dale}, {Engelbracht}, {van Zee}, {Draine}, {Hao},
  {Gordon}, {Moustakas}, {Murphy}, {Regan}, {Begum}, {Block}, {Dalcanton},
  {Funes}, {Gil de Paz}, {Johnson}, {Sakai}, {Skillman}, {Walter}, {Weisz},
  {Williams}, \& {Wu}}]{calzetti2010}
{Calzetti} D. {et~al.}, 2010, \apj, 714, 1256

\bibitem[{{Colombo} {et~al}\mbox{.}(2014{\natexlab{a}}){Colombo}, {Hughes},
  {Schinnerer}, {Meidt}, {Leroy}, {Pety}, {Dobbs}, {Garc{\'{\i}}a-Burillo},
  {Dumas}, {Thompson}, {Schuster}, \& {Kramer}}]{colombo2014}
{Colombo} D. {et~al.}, 2014{\natexlab{a}}, \apj, 784, 3

\bibitem[{{Colombo} {et~al}\mbox{.}(2014{\natexlab{b}}){Colombo}, {Meidt},
  {Schinnerer}, {Garc{\'{\i}}a-Burillo}, {Hughes}, {Pety}, {Leroy}, {Dobbs},
  {Dumas}, {Thompson}, {Schuster}, \& {Kramer}}]{colombo2014b}
{Colombo} D. {et~al.}, 2014{\natexlab{b}}, \apj, 784, 4

\bibitem[{{Currie} {et~al}\mbox{.}(2008){Currie}, {Draper}, {Berry}, {Jenness},
  {Cavanagh}, \& {Economou}}]{currie2008}
{Currie} M.~J., {Draper} P.~W., {Berry} D.~S., {Jenness} T., {Cavanagh} B.,
  {Economou} F., 2008, in Astronomical Society of the Pacific Conference
  Series, Vol. 394, Astronomical Data Analysis Software and Systems XVII,
  {Argyle} R.~W., {Bunclark} P.~S., {Lewis} J.~R., eds., p. 650

\bibitem[{{Dalcanton} {et~al}\mbox{.}(2009){Dalcanton}, {Williams}, {Seth},
  {Dolphin}, {Holtzman}, {Rosema}, {Skillman}, {Cole}, {Girardi}, {Gogarten},
  {Karachentsev}, {Olsen}, {Weisz}, {Christensen}, {Freeman}, {Gilbert},
  {Gallart}, {Harris}, {Hodge}, {de Jong}, {Karachentseva}, {Mateo}, {Stetson},
  {Tavarez}, {Zaritsky}, {Governato}, \& {Quinn}}]{dalcanton2009}
{Dalcanton} J.~J. {et~al.}, 2009, \apjs, 183, 67

\bibitem[{{Dobbs} {et~al}\mbox{.}(2010){Dobbs}, {Theis}, {Pringle}, \&
  {Bate}}]{dobbs2010}
{Dobbs} C.~L., {Theis} C., {Pringle} J.~E., {Bate} M.~R., 2010, \mnras, 403,
  625

\bibitem[{{Gaches} {et~al}\mbox{.}(2015){Gaches}, {Offner}, {Rosolowsky}, \&
  {Bisbas}}]{gaches2014}
{Gaches} B.~A.~L., {Offner} S.~S.~R., {Rosolowsky} E.~W., {Bisbas} T.~G., 2015,
  \apj, 799, 235

\bibitem[{{Galametz} {et~al}\mbox{.}(2013){Galametz}, {Kennicutt}, {Calzetti},
  {Aniano}, {Draine}, {Boquien}, {Brandl}, {Croxall}, {Dale}, {Engelbracht},
  {Gordon}, {Groves}, {Hao}, {Helou}, {Hinz}, {Hunt}, {Johnson}, {Li},
  {Murphy}, {Roussel}, {Sandstrom}, {Skibba}, \& {Tabatabaei}}]{galametz2013}
{Galametz} M. {et~al.}, 2013, \mnras, 431, 1956

\bibitem[{{Garcia-Burillo}, {Guelin} \& {Cernicharo}(1993){Garcia-Burillo},
  {Guelin}, \& {Cernicharo}}]{garciaburillo1993}
{Garcia-Burillo} S., {Guelin} M., {Cernicharo} J., 1993, \aap, 274, 123

\bibitem[{{Griffin} {et~al}\mbox{.}(2010){Griffin}, {Abergel}, {Abreu}, {Ade},
  {Andr{\'e}}, {Augueres}, {Babbedge}, {Bae}, {Baillie}, {Baluteau}, {Barlow},
  {Bendo}, {Benielli}, {Bock}, {Bonhomme}, {Brisbin}, {Brockley-Blatt},
  {Caldwell}, {Cara}, {Castro-Rodriguez}, {Cerulli}, {Chanial}, {Chen},
  {Clark}, {Clements}, {Clerc}, {Coker}, {Communal}, {Conversi}, {Cox},
  {Crumb}, {Cunningham}, {Daly}, {Davis}, {de Antoni}, {Delderfield}, {Devin},
  {di Giorgio}, {Didschuns}, {Dohlen}, {Donati}, {Dowell}, {Dowell}, {Duband},
  {Dumaye}, {Emery}, {Ferlet}, {Ferrand}, {Fontignie}, {Fox}, {Franceschini},
  {Frerking}, {Fulton}, {Garcia}, {Gastaud}, {Gear}, {Glenn}, {Goizel},
  {Griffin}, {Grundy}, {Guest}, {Guillemet}, {Hargrave}, {Harwit}, {Hastings},
  {Hatziminaoglou}, {Herman}, {Hinde}, {Hristov}, {Huang}, {Imhof}, {Isaak},
  {Israelsson}, {Ivison}, {Jennings}, {Kiernan}, {King}, {Lange}, {Latter},
  {Laurent}, {Laurent}, {Leeks}, {Lellouch}, {Levenson}, {Li}, {Li},
  {Lilienthal}, {Lim}, {Liu}, {Lu}, {Madden}, {Mainetti}, {Marliani}, {McKay},
  {Mercier}, {Molinari}, {Morris}, {Moseley}, {Mulder}, {Mur}, {Naylor},
  {Nguyen}, {O'Halloran}, {Oliver}, {Olofsson}, {Olofsson}, {Orfei}, {Page},
  {Pain}, {Panuzzo}, {Papageorgiou}, {Parks}, {Parr-Burman}, {Pearce},
  {Pearson}, {P{\'e}rez-Fournon}, {Pinsard}, {Pisano}, {Podosek}, {Pohlen},
  {Polehampton}, {Pouliquen}, {Rigopoulou}, {Rizzo}, {Roseboom}, {Roussel},
  {Rowan-Robinson}, {Rownd}, {Saraceno}, {Sauvage}, {Savage}, {Savini},
  {Sawyer}, {Scharmberg}, {Schmitt}, {Schneider}, {Schulz}, {Schwartz},
  {Shafer}, {Shupe}, {Sibthorpe}, {Sidher}, {Smith}, {Smith}, {Smith},
  {Spencer}, {Stobie}, {Sudiwala}, {Sukhatme}, {Surace}, {Stevens}, {Swinyard},
  {Trichas}, {Tourette}, {Triou}, {Tseng}, {Tucker}, {Turner}, {Vaccari},
  {Valtchanov}, {Vigroux}, {Virique}, {Voellmer}, {Walker}, {Ward}, {Waskett},
  {Weilert}, {Wesson}, {White}, {Whitehouse}, {Wilson}, {Winter}, {Woodcraft},
  {Wright}, {Xu}, {Zavagno}, {Zemcov}, {Zhang}, \& {Zonca}}]{griffin2010}
{Griffin} M.~J. {et~al.}, 2010, \aap, 518, L3

\bibitem[{{Hollenbach} {et~al}\mbox{.}(2012){Hollenbach}, {Kaufman}, {Neufeld},
  {Wolfire}, \& {Goicoechea}}]{hollenbach2012}
{Hollenbach} D., {Kaufman} M.~J., {Neufeld} D., {Wolfire} M., {Goicoechea}
  J.~R., 2012, \apj, 754, 105

\bibitem[{{Hughes} {et~al}\mbox{.}(2013{\natexlab{a}}){Hughes}, {Meidt},
  {Colombo}, {Schinnerer}, {Pety}, {Leroy}, {Dobbs}, {Garc{\'{\i}}a-Burillo},
  {Thompson}, {Dumas}, {Schuster}, \& {Kramer}}]{hughes2013b}
{Hughes} A. {et~al.}, 2013{\natexlab{a}}, \apj, 779, 46

\bibitem[{{Hughes} {et~al}\mbox{.}(2013{\natexlab{b}}){Hughes}, {Meidt},
  {Schinnerer}, {Colombo}, {Pety}, {Leroy}, {Dobbs}, {Garc{\'{\i}}a-Burillo},
  {Thompson}, {Dumas}, {Schuster}, \& {Kramer}}]{hughes2013}
{Hughes} A. {et~al.}, 2013{\natexlab{b}}, \apj, 779, 44

\bibitem[{Hunter(2007)}]{hunter2007}
Hunter J.~D., 2007, Computing In Science \& Engineering, 9, 90

\bibitem[{{Ikeda} {et~al}\mbox{.}(2002){Ikeda}, {Oka}, {Tatematsu}, {Sekimoto},
  \& {Yamamoto}}]{ikeda2002}
{Ikeda} M., {Oka} T., {Tatematsu} K., {Sekimoto} Y., {Yamamoto} S., 2002,
  \apjs, 139, 467

\bibitem[{{Israel}, {Tilanus} \& {Baas}(2006){Israel}, {Tilanus}, \&
  {Baas}}]{israel2006}
{Israel} F.~P., {Tilanus} R.~P.~J., {Baas} F., 2006, \aap, 445, 907

\bibitem[{{Kamenetzky} {et~al}\mbox{.}(2012){Kamenetzky}, {Glenn}, {Rangwala},
  {Maloney}, {Bradford}, {Wilson}, {Bendo}, {Baes}, {Boselli}, {Cooray},
  {Isaak}, {Lebouteiller}, {Madden}, {Panuzzo}, {Schirm}, {Spinoglio}, \&
  {Wu}}]{kamenetzky2012}
{Kamenetzky} J. {et~al.}, 2012, \apj, 753, 70

\bibitem[{{Kamenetzky} {et~al}\mbox{.}(2014){Kamenetzky}, {Rangwala}, {Glenn},
  {Maloney}, \& {Conley}}]{kamenetzky2014}
{Kamenetzky} J., {Rangwala} N., {Glenn} J., {Maloney} P.~R., {Conley} A., 2014,
  \apj, 795, 174

\bibitem[{{Kaufman}, {Wolfire} \& {Hollenbach}(2006){Kaufman}, {Wolfire}, \&
  {Hollenbach}}]{kaufman2006}
{Kaufman} M.~J., {Wolfire} M.~G., {Hollenbach} D.~J., 2006, \apj, 644, 283

\bibitem[{{Kaufman} {et~al}\mbox{.}(1999){Kaufman}, {Wolfire}, {Hollenbach}, \&
  {Luhman}}]{kaufman1999}
{Kaufman} M.~J., {Wolfire} M.~G., {Hollenbach} D.~J., {Luhman} M.~L., 1999,
  \apj, 527, 795

\bibitem[{{Kazandjian} {et~al}\mbox{.}(2012){Kazandjian}, {Meijerink},
  {Pelupessy}, {Israel}, \& {Spaans}}]{kazandjian2012}
{Kazandjian} M.~V., {Meijerink} R., {Pelupessy} I., {Israel} F.~P., {Spaans}
  M., 2012, \aap, 542, A65

\bibitem[{{Kazandjian} {et~al}\mbox{.}(2015){Kazandjian}, {Meijerink},
  {Pelupessy}, {Israel}, \& {Spaans}}]{kazandjian2015}
{Kazandjian} M.~V., {Meijerink} R., {Pelupessy} I., {Israel} F.~P., {Spaans}
  M., 2015, \aap, 574, A127

\bibitem[{{Koda} {et~al}\mbox{.}(2009){Koda}, {Scoville}, {Sawada}, {La Vigne},
  {Vogel}, {Potts}, {Carpenter}, {Corder}, {Wright}, {White}, {Zauderer},
  {Patience}, {Sargent}, {Bock}, {Hawkins}, {Hodges}, {Kemball}, {Lamb},
  {Plambeck}, {Pound}, {Scott}, {Teuben}, \& {Woody}}]{koda2009}
{Koda} J. {et~al.}, 2009, \apjl, 700, L132

\bibitem[{{Koulouridis}(2014)}]{kouloridis2014}
{Koulouridis} E., 2014, \aap, 570, A72

\bibitem[{{Kramer} {et~al}\mbox{.}(2005){Kramer}, {Mookerjea}, {Bayet},
  {Garcia-Burillo}, {Gerin}, {Israel}, {Stutzki}, \& {Wouterloot}}]{kramer2005}
{Kramer} C., {Mookerjea} B., {Bayet} E., {Garcia-Burillo} S., {Gerin} M.,
  {Israel} F.~P., {Stutzki} J., {Wouterloot} J.~G.~A., 2005, \aap, 441, 961

\bibitem[{{Kramer}, {Moreno} \& {Greve}(2008){Kramer}, {Moreno}, \&
  {Greve}}]{kramer2008}
{Kramer} C., {Moreno} R., {Greve} A., 2008, \aap, 482, 359

\bibitem[{{Leroy} {et~al}\mbox{.}(2011){Leroy}, {Bolatto}, {Gordon},
  {Sandstrom}, {Gratier}, {Rosolowsky}, {Engelbracht}, {Mizuno}, {Corbelli},
  {Fukui}, \& {Kawamura}}]{leroy2011}
{Leroy} A.~K. {et~al.}, 2011, \apj, 737, 12

\bibitem[{{Leroy} {et~al}\mbox{.}(2009){Leroy}, {Walter}, {Bigiel}, {Usero},
  {Weiss}, {Brinks}, {de Blok}, {Kennicutt}, {Schuster}, {Kramer},
  {Wiesemeyer}, \& {Roussel}}]{leroy2009}
{Leroy} A.~K. {et~al.}, 2009, \aj, 137, 4670

\bibitem[{{Liszt}, {Pety} \& {Tachihara}(2009){Liszt}, {Pety}, \&
  {Tachihara}}]{liszt2009}
{Liszt} H.~S., {Pety} J., {Tachihara} K., 2009, \aap, 499, 503

\bibitem[{{Maloney}(1999)}]{maloney1999}
{Maloney} P.~R., 1999, \apss, 266, 207

\bibitem[{{Meidt} {et~al}\mbox{.}(2013){Meidt}, {Schinnerer},
  {Garc{\'{\i}}a-Burillo}, {Hughes}, {Colombo}, {Pety}, {Dobbs}, {Schuster},
  {Kramer}, {Leroy}, {Dumas}, \& {Thompson}}]{meidt2013}
{Meidt} S.~E. {et~al.}, 2013, \apj, 779, 45

\bibitem[{{Mentuch Cooper} {et~al}\mbox{.}(2012){Mentuch Cooper}, {Wilson},
  {Foyle}, {Bendo}, {Koda}, {Baes}, {Boquien}, {Boselli}, {Ciesla}, {Cooray},
  {Eales}, {Galametz}, {Lebouteiller}, {Parkin}, {Roussel}, {Sauvage},
  {Spinoglio}, \& {Smith}}]{mentuchcooper2012}
{Mentuch Cooper} E. {et~al.}, 2012, \apj, 755, 165

\bibitem[{{Nakai} {et~al}\mbox{.}(1994){Nakai}, {Kuno}, {Handa}, \&
  {Sofue}}]{nakai1994}
{Nakai} N., {Kuno} N., {Handa} T., {Sofue} Y., 1994, \pasj, 46, 527

\bibitem[{{Naylor} {et~al}\mbox{.}(2010){Naylor}, {Baluteau}, {Barlow},
  {Benielli}, {Ferlet}, {Fulton}, {Griffin}, {Grundy}, {Imhof}, {Jones},
  {King}, {Leeks}, {Lim}, {Lu}, {Makiwa}, {Polehampton}, {Savini}, {Sidher},
  {Spencer}, {Surace}, {Swinyard}, \& {Wesson}}]{naylor2010}
{Naylor} D.~A. {et~al.}, 2010, in Society of Photo-Optical Instrumentation
  Engineers (SPIE) Conference Series, Vol. 7731, Society of Photo-Optical
  Instrumentation Engineers (SPIE) Conference Series, p.~16

\bibitem[{{Nikola} {et~al}\mbox{.}(2001){Nikola}, {Geis}, {Herrmann}, {Madden},
  {Poglitsch}, {Stacey}, \& {Townes}}]{nikola2001}
{Nikola} T., {Geis} N., {Herrmann} F., {Madden} S.~C., {Poglitsch} A., {Stacey}
  G.~J., {Townes} C.~H., 2001, \apj, 561, 203

\bibitem[{{Offner} {et~al}\mbox{.}(2014){Offner}, {Bisbas}, {Bell}, \&
  {Viti}}]{offner2014}
{Offner} S.~S.~R., {Bisbas} T.~G., {Bell} T.~A., {Viti} S., 2014, \mnras, 440,
  L81

\bibitem[{{Panuzzo} {et~al}\mbox{.}(2010){Panuzzo}, {Rangwala}, {Rykala},
  {Isaak}, {Glenn}, {Wilson}, {Auld}, {Baes}, {Barlow}, {Bendo}, {Bock},
  {Boselli}, {Bradford}, {Buat}, {Castro-Rodr{\'{\i}}guez}, {Chanial},
  {Charlot}, {Ciesla}, {Clements}, {Cooray}, {Cormier}, {Cortese}, {Davies},
  {Dwek}, {Eales}, {Elbaz}, {Fulton}, {Galametz}, {Galliano}, {Gear}, {Gomez},
  {Griffin}, {Hony}, {Levenson}, {Lu}, {Madden}, {O'Halloran}, {Okumura},
  {Oliver}, {Page}, {Papageorgiou}, {Parkin}, {P{\'e}rez-Fournon}, {Pohlen},
  {Polehampton}, {Rigby}, {Roussel}, {Sacchi}, {Sauvage}, {Schulz}, {Schirm},
  {Smith}, {Spinoglio}, {Stevens}, {Srinivasan}, {Symeonidis}, {Swinyard},
  {Trichas}, {Vaccari}, {Vigroux}, {Wozniak}, {Wright}, \&
  {Zeilinger}}]{panuzzo2010}
{Panuzzo} P. {et~al.}, 2010, \aap, 518, L37

\bibitem[{{Papadopoulos}, {Thi} \& {Viti}(2004){Papadopoulos}, {Thi}, \&
  {Viti}}]{papadopoulos2004}
{Papadopoulos} P.~P., {Thi} W.-F., {Viti} S., 2004, \mnras, 351, 147

\bibitem[{{Parkin} {et~al}\mbox{.}(2013){Parkin}, {Wilson}, {Schirm}, {Baes},
  {Boquien}, {Boselli}, {Cooray}, {Cormier}, {Foyle}, {Karczewski},
  {Lebouteiller}, {de Looze}, {Madden}, {Roussel}, {Sauvage}, \&
  {Spinoglio}}]{parkin2013}
{Parkin} T.~J. {et~al.}, 2013, \apj, 776, 65

\bibitem[{{Pety}, {Lucas} \& {Liszt}(2008){Pety}, {Lucas}, \&
  {Liszt}}]{pety2008}
{Pety} J., {Lucas} R., {Liszt} H.~S., 2008, \aap, 489, 217

\bibitem[{{Pety} {et~al}\mbox{.}(2013){Pety}, {Schinnerer}, {Leroy}, {Hughes},
  {Meidt}, {Colombo}, {Dumas}, {Garc{\'{\i}}a-Burillo}, {Schuster}, {Kramer},
  {Dobbs}, \& {Thompson}}]{pety2013}
{Pety} J. {et~al.}, 2013, \apj, 779, 43

\bibitem[{{Pilbratt} {et~al}\mbox{.}(2010){Pilbratt}, {Riedinger}, {Passvogel},
  {Crone}, {Doyle}, {Gageur}, {Heras}, {Jewell}, {Metcalfe}, {Ott}, \&
  {Schmidt}}]{pilbratt2010}
{Pilbratt} G.~L. {et~al.}, 2010, \aap, 518, L1

\bibitem[{{Plume} {et~al}\mbox{.}(1999){Plume}, {Jaffe}, {Tatematsu}, {Evans},
  \& {Keene}}]{plume1999}
{Plume} R., {Jaffe} D.~T., {Tatematsu} K., {Evans}, II N.~J., {Keene} J., 1999,
  \apj, 512, 768

\bibitem[{{Poglitsch} {et~al}\mbox{.}(2010){Poglitsch}, {Waelkens}, {Geis},
  {Feuchtgruber}, {Vandenbussche}, {Rodriguez}, {Krause}, {Renotte}, {van
  Hoof}, {Saraceno}, {Cepa}, {Kerschbaum}, {Agn{\`e}se}, {Ali}, {Altieri},
  {Andreani}, {Augueres}, {Balog}, {Barl}, {Bauer}, {Belbachir}, {Benedettini},
  {Billot}, {Boulade}, {Bischof}, {Blommaert}, {Callut}, {Cara}, {Cerulli},
  {Cesarsky}, {Contursi}, {Creten}, {De Meester}, {Doublier}, {Doumayrou},
  {Duband}, {Exter}, {Genzel}, {Gillis}, {Gr{\"o}zinger}, {Henning},
  {Herreros}, {Huygen}, {Inguscio}, {Jakob}, {Jamar}, {Jean}, {de Jong},
  {Katterloher}, {Kiss}, {Klaas}, {Lemke}, {Lutz}, {Madden}, {Marquet},
  {Martignac}, {Mazy}, {Merken}, {Montfort}, {Morbidelli}, {M{\"u}ller},
  {Nielbock}, {Okumura}, {Orfei}, {Ottensamer}, {Pezzuto}, {Popesso},
  {Putzeys}, {Regibo}, {Reveret}, {Royer}, {Sauvage}, {Schreiber}, {Stegmaier},
  {Schmitt}, {Schubert}, {Sturm}, {Thiel}, {Tofani}, {Vavrek}, {Wetzstein},
  {Wieprecht}, \& {Wiezorrek}}]{poglitsch2010}
{Poglitsch} A. {et~al.}, 2010, \aap, 518, L2

\bibitem[{{Rangwala} {et~al}\mbox{.}(2011){Rangwala}, {Maloney}, {Glenn},
  {Wilson}, {Rykala}, {Isaak}, {Baes}, {Bendo}, {Boselli}, {Bradford},
  {Clements}, {Cooray}, {Fulton}, {Imhof}, {Kamenetzky}, {Madden}, {Mentuch},
  {Sacchi}, {Sauvage}, {Schirm}, {Smith}, {Spinoglio}, \&
  {Wolfire}}]{rangwala2011}
{Rangwala} N. {et~al.}, 2011, \apj, 743, 94

\bibitem[{{R{\"o}llig} {et~al}\mbox{.}(2007){R{\"o}llig}, {Abel}, {Bell},
  {Bensch}, {Black}, {Ferland}, {Jonkheid}, {Kamp}, {Kaufman}, {Le Bourlot},
  {Le Petit}, {Meijerink}, {Morata}, {Ossenkopf}, {Roueff}, {Shaw}, {Spaans},
  {Sternberg}, {Stutzki}, {Thi}, {van Dishoeck}, {van Hoof}, {Viti}, \&
  {Wolfire}}]{roellig2007}
{R{\"o}llig} M. {et~al.}, 2007, \aap, 467, 187

\bibitem[{{Rose} \& {Searle}(1982)}]{rose1982}
{Rose} J.~A., {Searle} L., 1982, \apj, 253, 556

\bibitem[{{Roussel} {et~al}\mbox{.}(2007){Roussel}, {Helou}, {Hollenbach},
  {Draine}, {Smith}, {Armus}, {Schinnerer}, {Walter}, {Engelbracht},
  {Thornley}, {Kennicutt}, {Calzetti}, {Dale}, {Murphy}, \&
  {Bot}}]{roussel2007}
{Roussel} H. {et~al.}, 2007, \apj, 669, 959

\bibitem[{{Satyapal}, {Sambruna} \& {Dudik}(2004){Satyapal}, {Sambruna}, \&
  {Dudik}}]{satyapal2004}
{Satyapal} S., {Sambruna} R.~M., {Dudik} R.~P., 2004, \aap, 414, 825

\bibitem[{{Schinnerer} {et~al}\mbox{.}(2013){Schinnerer}, {Meidt}, {Pety},
  {Hughes}, {Colombo}, {Garc{\'{\i}}a-Burillo}, {Schuster}, {Dumas}, {Dobbs},
  {Leroy}, {Kramer}, {Thompson}, \& {Regan}}]{schinnerer2013}
{Schinnerer} E. {et~al.}, 2013, \apj, 779, 42

\bibitem[{{Schinnerer} {et~al}\mbox{.}(2010){Schinnerer}, {Wei{\ss}}, {Aalto},
  \& {Scoville}}]{schinnerer2010}
{Schinnerer} E., {Wei{\ss}} A., {Aalto} S., {Scoville} N.~Z., 2010, \apj, 719,
  1588

\bibitem[{{Schirm} {et~al}\mbox{.}(2016){Schirm}, {Wilson}, {Madden}, \&
  {Clements}}]{schirm2016}
{Schirm} M.~R.~P., {Wilson} C.~D., {Madden} S.~C., {Clements} D.~L., 2016,
  \apj, 823, 87

\bibitem[{{Schirm} {et~al}\mbox{.}(2014){Schirm}, {Wilson}, {Parkin},
  {Kamenetzky}, {Glenn}, {Rangwala}, {Spinoglio}, {Pereira-Santaella}, {Baes},
  {Barlow}, {Clements}, {Cooray}, {De Looze}, {Karczewski}, {Madden},
  {R{\'e}my-Ruyer}, \& {Wu}}]{schirm2014}
{Schirm} M.~R.~P. {et~al.}, 2014, \apj, 781, 101

\bibitem[{{Schuster} {et~al}\mbox{.}(2007){Schuster}, {Kramer}, {Hitschfeld},
  {Garcia-Burillo}, \& {Mookerjea}}]{schuster2007}
{Schuster} K.~F., {Kramer} C., {Hitschfeld} M., {Garcia-Burillo} S.,
  {Mookerjea} B., 2007, \aap, 461, 143

\bibitem[{{Schweizer} {et~al}\mbox{.}(2008){Schweizer}, {Burns}, {Madore},
  {Mager}, {Phillips}, {Freedman}, {Boldt}, {Contreras}, {Folatelli},
  {Gonz{\'a}lez}, {Hamuy}, {Krzeminski}, {Morrell}, {Persson}, {Roth}, \&
  {Stritzinger}}]{schweizer2008}
{Schweizer} F. {et~al.}, 2008, \aj, 136, 1482

\bibitem[{{Scoville} \& {Young}(1983)}]{scoville1983}
{Scoville} N., {Young} J.~S., 1983, \apj, 265, 148

\bibitem[{{Scoville}, {Yun} \& {Bryant}(1997){Scoville}, {Yun}, \&
  {Bryant}}]{scoville1997}
{Scoville} N.~Z., {Yun} M.~S., {Bryant} P.~M., 1997, \apj, 484, 702

\bibitem[{{Shimajiri} {et~al}\mbox{.}(2013){Shimajiri}, {Sakai}, {Tsukagoshi},
  {Kitamura}, {Momose}, {Saito}, {Oshima}, {Kohno}, \&
  {Kawabe}}]{shimajiri2013}
{Shimajiri} Y. {et~al.}, 2013, \apjl, 774, L20

\bibitem[{{Spinoglio} {et~al}\mbox{.}(2012){Spinoglio}, {Pereira-Santaella},
  {Busquet}, {Schirm}, {Wilson}, {Glenn}, {Kamenetzky}, {Rangwala}, {Maloney},
  {Parkin}, {Bendo}, {Madden}, {Wolfire}, {Boselli}, {Cooray}, \&
  {Page}}]{spinoglio2012}
{Spinoglio} L. {et~al.}, 2012, \apj, 758, 108

\bibitem[{{Stanford} {et~al}\mbox{.}(1990){Stanford}, {Sargent}, {Sanders}, \&
  {Scoville}}]{stanford1990}
{Stanford} S.~A., {Sargent} A.~I., {Sanders} D.~B., {Scoville} N.~Z., 1990,
  \apj, 349, 492

\bibitem[{{Swinyard} {et~al}\mbox{.}(2014){Swinyard}, {Polehampton}, {Hopwood},
  {Valtchanov}, {Lu}, {Fulton}, {Benielli}, {Imhof}, {Marchili}, {Baluteau},
  {Bendo}, {Ferlet}, {Griffin}, {Lim}, {Makiwa}, {Naylor}, {Orton},
  {Papageorgiou}, {Pearson}, {Schulz}, {Sidher}, {Spencer}, {van der Wiel}, \&
  {Wu}}]{swinyard2014}
{Swinyard} B.~M. {et~al.}, 2014, \mnras, 440, 3658

\bibitem[{{Tielens}(2005)}]{tielens2005}
{Tielens} A.~G.~G.~M., 2005, {The Physics and Chemistry of the Interstellar
  Medium}

\bibitem[{{Tielens} \& {Hollenbach}(1985)}]{tielens1985}
{Tielens} A.~G.~G.~M., {Hollenbach} D., 1985, \apj, 291, 722

\bibitem[{{Tikhonov}, {Galazutdinova} \& {Tikhonov}(2009){Tikhonov},
  {Galazutdinova}, \& {Tikhonov}}]{tikhonov2009}
{Tikhonov} N.~A., {Galazutdinova} O.~A., {Tikhonov} E.~N., 2009, Astronomy
  Letters, 35, 599

\bibitem[{{Tran}(2001)}]{tran2001}
{Tran} H.~D., 2001, \apjl, 554, L19

\bibitem[{{van der Tak} {et~al}\mbox{.}(2007){van der Tak}, {Black},
  {Sch{\"o}ier}, {Jansen}, \& {van Dishoeck}}]{vandertak2007}
{van der Tak} F.~F.~S., {Black} J.~H., {Sch{\"o}ier} F.~L., {Jansen} D.~J.,
  {van Dishoeck} E.~F., 2007, \aap, 468, 627

\bibitem[{{Vlahakis} {et~al}\mbox{.}(2013){Vlahakis}, {van der Werf}, {Israel},
  \& {Tilanus}}]{vlahakis2013}
{Vlahakis} C., {van der Werf} P., {Israel} F.~P., {Tilanus} R.~P.~J., 2013,
  \mnras, 433, 1837

\bibitem[{{Wilson} {et~al}\mbox{.}(2012){Wilson}, {Warren}, {Israel},
  {Serjeant}, {Attewell}, {Bendo}, {Butner}, {Chanial}, {Clements}, {Golding},
  {Heesen}, {Irwin}, {Leech}, {Matthews}, {M{\"u}hle}, {Mortier}, {Petitpas},
  {S{\'a}nchez-Gallego}, {Sinukoff}, {Shorten}, {Tan}, {Tilanus}, {Usero},
  {Vaccari}, {Wiegert}, {Zhu}, {Alexander}, {Alexander}, {Azimlu}, {Barmby},
  {Brar}, {Bridge}, {Brinks}, {Brooks}, {Coppin}, {C{\^o}t{\'e}},
  {C{\^o}t{\'e}}, {Courteau}, {Davies}, {Eales}, {Fich}, {Hudson}, {Hughes},
  {Ivison}, {Knapen}, {Page}, {Parkin}, {Rigopoulou}, {Rosolowsky}, {Seaquist},
  {Spekkens}, {Tanvir}, {van der Hulst}, {van der Werf}, {Vlahakis}, {Webb},
  {Weferling}, \& {White}}]{wilson2012}
{Wilson} C.~D. {et~al.}, 2012, \mnras, 424, 3050

\bibitem[{{Wolfire}, {Hollenbach} \& {McKee}(2010){Wolfire}, {Hollenbach}, \&
  {McKee}}]{wolfire2010}
{Wolfire} M.~G., {Hollenbach} D., {McKee} C.~F., 2010, \apj, 716, 1191

\bibitem[{{Wu} {et~al}\mbox{.}(2015){Wu}, {Madden}, {Galliano}, {Wilson},
  {Kamenetzky}, {Lee}, {Schirm}, {Hony}, {Lebouteiller}, {Spinoglio},
  {Cormier}, {Glenn}, {Maloney}, {Pereira-Santaella}, {R{\'e}my-Ruyer}, {Baes},
  {Boselli}, {Bournaud}, {De Looze}, {Hughes}, {Panuzzo}, \&
  {Rangwala}}]{wu2014}
{Wu} R. {et~al.}, 2015, \aap, 575, A88

\bibitem[{{Wu} {et~al}\mbox{.}(2013){Wu}, {Polehampton}, {Etxaluze}, {Makiwa},
  {Naylor}, {Salji}, {Swinyard}, {Ferlet}, {van der Wiel}, {Smith}, {Fulton},
  {Griffin}, {Baluteau}, {Benielli}, {Glenn}, {Hopwood}, {Imhof}, {Lim}, {Lu},
  {Panuzzo}, {Pearson}, {Sidher}, \& {Valtchanov}}]{wu2013}
{Wu} R. {et~al.}, 2013, \aap, 556, A116

\bibitem[{{Yun}, {Ho} \& {Lo}(1993){Yun}, {Ho}, \& {Lo}}]{yun1993}
{Yun} M.~S., {Ho} P.~T.~P., {Lo} K.~Y., 1993, \apjl, 411, L17

\bibitem[{{Zaritsky}, {Rix} \& {Rieke}(1993){Zaritsky}, {Rix}, \&
  {Rieke}}]{zartisky1993}
{Zaritsky} D., {Rix} H.-W., {Rieke} M., 1993, \nat, 364, 313

\end{thebibliography}
